\pdfoutput=1

\documentclass[12pt]{article}

\usepackage{amsfonts}
\usepackage{amsmath}
\usepackage{amssymb}
\usepackage{bigints}
\usepackage{booktabs}
\usepackage[nosort]{cite}
\usepackage{color}
\usepackage{dsfont}
\usepackage{float}
\usepackage{framed}
\usepackage{graphicx}
\usepackage{indentfirst}
\usepackage{mathrsfs}
\usepackage{multirow}
\usepackage{pdflscape}
\usepackage{setspace}
\usepackage{subdepth}
\usepackage{subfig}
\usepackage{titlesec}
\usepackage[dotinlabels]{titletoc}
\usepackage{wrapfig}
\usepackage[all]{xy}
\usepackage{young}
\usepackage[vcentermath]{youngtab}
\usepackage{relsize}
\usepackage{stackengine}

\usepackage{hyperref}

\numberwithin{equation}{section}

\usepackage{verbatim}

\newcommand{\be}{\begin{equation}}
\newcommand{\ee}{\end{equation}}

\newcommand{\bml}{\begin{multline}}
\newcommand{\emll}{\end{multline}}

\newcommand{\nn}{\nonumber}

\def\({\left(} \def\){\right)}
\def\[{\left[} \def\]{\right]}

\def\al{\alpha}

\def\mO{\mathcal{O}}

\def\mA{\mathcal{A}}

\def\b{\bar}

\def\lam{\lambda}

\def\d{\partial}

\newcommand{\la}{\langle}
\newcommand{\ra}{\rangle}

\newcommand{\bea}{\begin{eqnarray}}
\newcommand{\eea}{\end{eqnarray}}

\def\mbS{\mathbb{S}}

\newcommand{\bi}{\begin{itemize}}
\newcommand{\ei}{\end{itemize}}

\usepackage[left=2cm,right=2cm,top=2cm,bottom=2cm]{geometry}
\titleformat{\section}{\Large\bfseries}{\thesection.}{4pt}{}
\titlespacing{\section}{0pt}{22pt}{6pt}

\titleformat{\subsection}{\large\bfseries}{\thesubsection.}{4pt}{}
\titlespacing{\subsection}{0pt}{18pt}{6pt}

\titleformat{\subsubsection}{\normalfont\bfseries}{\thesubsubsection.}{4pt}{}
\titlespacing{\subsubsection}{0pt}{16pt}{6pt}

\def\ie{\begin{equation}\begin{aligned}}
\def\fe{\end{aligned}\end{equation}}

\def\tilde{\widetilde}
\def\t{\tilde}

\def\bar{\overline}

\def\d{\partial}

\def\1{{\mathds 1}}

\def\mN{\mathcal{N}}
\def\mA{\mathcal{A}}

\def\mL{\mathcal{L}}

\def\tp{\tilde p}

\DeclareFontShape{OT1}{cmr}{mx}{n}%
    {<->cmr10}{}
\newcommand{\mytitlefont}{\fontseries{mx}\selectfont}
\DeclareMathAlphabet{\titlemath}{OT1}{cmr}{mx}{n}

\usepackage[UKenglish]{datetime}

\begin{document}

\begin{titlepage}

\begin{center}

~\\[1cm]

{\fontsize{20pt}{0pt} \mytitlefont Chaotic scattering of highly excited strings}\\[10pt]

~\\[0.2cm]

{\fontsize{14pt}{0pt} David J.~Gross{\small $^{1}$} and Vladimir Rosenhaus{\small $^{2, 3}$}}

~\\[0.1cm]

\it{$^1$ Kavli Institute for Theoretical Physics}\\ \it{University of California, Santa Barbara, CA 93106}\\[.5cm]

\it{$^2$ School of Natural Sciences, Institute for Advanced Study}\\ \it{1  Einstein Drive, Princeton, NJ 08540}\\[.5cm]

\it{$^3$ Initiative for the Theoretical Sciences*}\\ \it{ The Graduate Center, City University of New York}\\ \it{
 365 Fifth Ave, New York, NY 10016}

~\\[0.6cm]

\end{center}

\noindent 

Motivated by the desire to understand chaos in the $S$-matrix of string theory, we study tree level scattering amplitudes involving highly excited strings. While the amplitudes for scattering of light  strings have been a hallmark of string theory since its early days, scattering of excited strings has been far less studied. Recent results on black hole chaos, combined with the correspondence principle between black holes and strings, suggest that the amplitudes  have a rich structure. We review the procedure by which an excited string is formed by repeatedly scattering photons off of an initial tachyon (the DDF formalism). We compute the scattering amplitude of one arbitrary excited string and any number of tachyons in bosonic string theory. At high energies and high mass excited state these amplitudes are determined by a saddle-point in the integration over the positions of the string vertex operators on the sphere (or the upper half plane), thus yielding a generalization of the ``scattering equations''. We find a compact expression for the amplitude of an excited string decaying into two tachyons, and study its properties for a generic excited string. We find the amplitude is highly erratic as a function of both the precise excited string state and of the tachyon scattering angle relative to its polarization, a sign of chaos.

\vfill

\indent *On leave\\

\newdateformat{UKvardate}{%
 \monthname[\THEMONTH]  \THEDAY,  \THEYEAR}
\UKvardate
\end{titlepage}


\tableofcontents
~\\[1cm]


\section{Introduction}
\vspace{-.03cm}
This will be the first in a series of papers in which we study string theory scattering amplitudes involving highly excited strings. This differs from most previous studies of string scattering, which  either involve only light strings, or compute the decay rate of an excited string. We will be studying exclusive amplitudes involving precisely specified, yet generic, heavy string states. \\[-25pt]
\subsubsection*{Three questions}\vspace{-.15cm}
The motivation for this study stems from three related questions. \\[5pt]
\indent The first question has to do with  black hole chaos:  fairly recently it was recognized that black holes are chaotic \cite{Shenker:2013pqa, Kitaev, PolchinskiC}. The chaos - exponential divergence of trajectories in phase space - is a result of the redshift near the horizon. The geometry of the black hole  sets the Lyaponuv exponent. Chaotic systems are known to have a rich array of properties: are there qualitative or quantitative statements that can be made about black hole chaos that go beyond this elementary diagnostic of Lyaponuv behavior? 

The second question has to do with chaos in quantum field theory. Chaos has been extensively discussed in classical and quantum mechanics. What is the structure of chaos in the more general context of quantum field theory? It was recently proposed \cite{VRchaos} that chaos can be seen in the erratic behavior of scattering amplitudes - physically measurable quantities. What is an analytically tractable example of such a system? 

\begin{figure}
\centering

\includegraphics[width=5in]{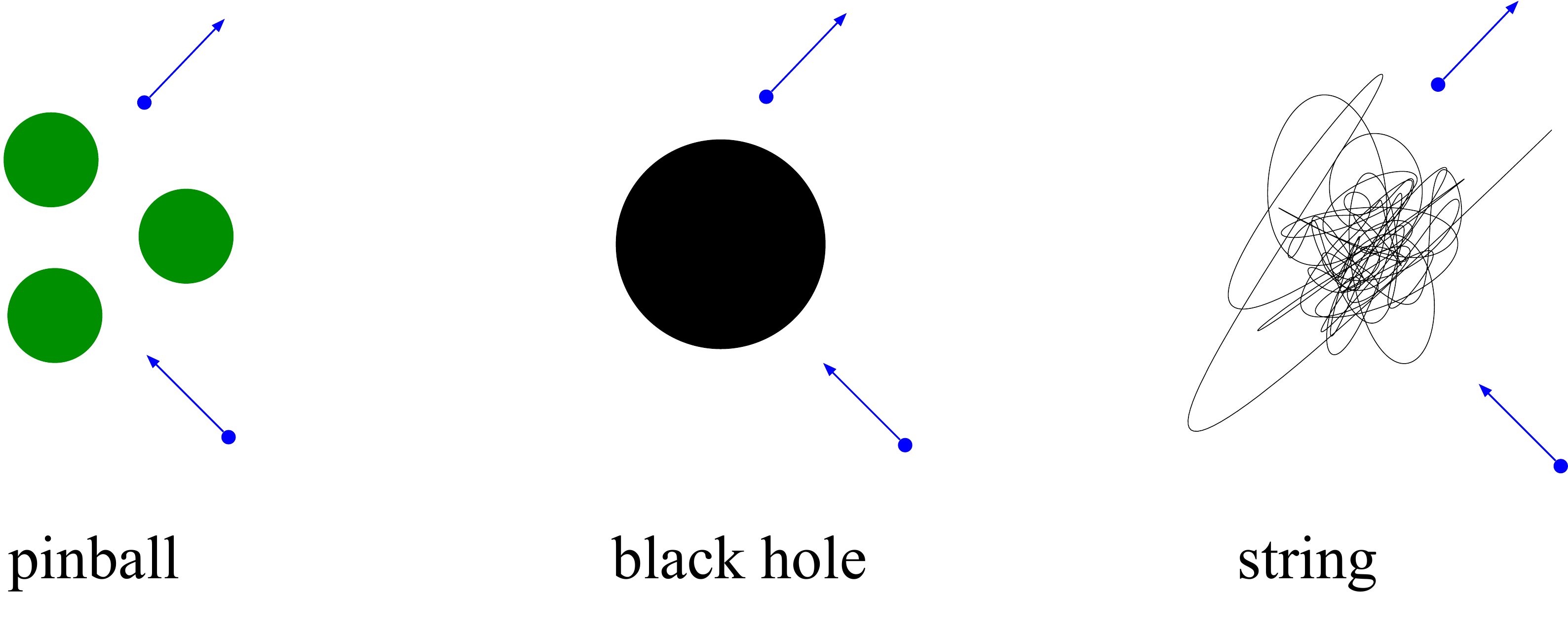}   
\caption{ (a) Pinball scattering is a prototypical example of classical chaos: the particle's outgoing angle is highly sensitive to its impact parameter \cite{Smilansky, Gaspard1,  Cvitanovich, ChaosBook, gaspard_1998}. This is a consequence of the enormous number of qualitatively different scattering trajectories. Giving the labels $1,2,3$ to the three disks, a trajectory that involves $n$ bounces between the disks before escaping is of the form e.g. $12132\cdots$. The number of such trajectories grows as $2^n$. (b) Scattering off of a black hole is also chaotic: a small perturbation of the black hole state, by for instance adding an extra soft particle, causes a large change in the state of the outgoing Hawking radiation \cite{PolchinskiC}. (c) Our goal is to study scattering off of a highly excited string, which we hope will also exhibit chaos.
 In all three of these physically very different setups, there are an exponentially large number of internal states.}  \label{IntroPic}
\end{figure}

Our proposal is  that both of these questions may have answers in the context of scattering of highly excited strings. Briefly, a generic heavy string will have a large number of excited modes; a cartoon is shown in Fig.~\ref{IntroPic} (c). We expect the scattering amplitude of a light string off of the excited string to be highly erratic as a function of the outgoing angle, or the ingoing angle, or a small change in the state of the excited string. 
Moreover, the Horowitz-Polchinski correspondence principle between black holes and strings states \cite{HorowitzPolchinski}, in effect, that some of the black hole microstates have a one-to-one mapping to excited string states, and the number of such states is enough to comprise an order-one fraction of the black hole entropy. The string coupling at the correspondence point - where the string turns into a black hole - is small,  suggesting that we may be able to study black hole chaos by studying string chaos within string perturbation theory. This is an extremely fortuitous situation wherein we have a weakly interacting system with a enormous number of almost stable resonances that can mimic the microstates of a black hole and exhibit chaos.

 Finally, we believe that the study of scattering of high energy, highly excited  strings may provide new insight into a third question, which is an old question: what is the  high energy behavior of string theory?\\[-12pt]

In the rest of the introduction we discuss these motivations and background in more detail,  overview how we will compute amplitudes with excited strings, mention what we think can be computed in future work, and give an outline of the paper. 

\subsection{Black holes, strings, and chaos}

\subsubsection*{Pinball chaos}

An iconic example of classical chaos is a ball bouncing around in a stadium. The stadium can be a rectangle with semicircular caps (Bunimovich stadium), or the inside of a square with a circle cut out (Sinai billiards), or something else. Two balls with similar starting positions and velocities will, after multiple collisions with the stadium boundaries, be at very different locations. Another example of chaos is a pinball game, sketched in Fig.~\ref{IntroPic}(a). A small ball is sent in, it bounces around between three disks, and then escapes. While trapped between the disks, it is as if it is in a stadium. The outgoing angle of the ball is a highly erratic function of the ingoing angle: an arbitrarily small range of impact parameters can lead to all possible outgoing angles \cite{ Smilansky, Gaspard1, Cvitanovich, ChaosBook, gaspard_1998}. 

The pinball game will be useful to keep in mind - it is an example of chaos manifesting itself as erratic behavior of the (one-to-one classical)  $S$-matrix, thereby providing an entry point for discussing chaos in quantum field theory and string theory, where the $S$-matrix is the natural observable.

\subsubsection*{Black hole chaos}
While it has been known since the discovery of Hawking radiation that quantum black holes have thermodynamic properties, it is only fairly recently that it was recognized how to see black hole chaos  \cite{Shenker:2013pqa, Kitaev}. The diagnostic for semiclassical quantum chaos that was applied is the exponential growth of the out-of-time-order correlator \cite{ LO, Kitaev, MSS}, which defines a quantum Lyaponuv-like exponent. Soon after, a simple argument for the Lyaponuv exponent of a black hole was given in \cite{PolchinskiC}: slightly perturbing a radiating black hole by throwing a particle into it causes the horizon to expand slightly, which due to the redshift factor, causes an exponentially large time delay in the escape of  Hawking quanta being emitted soon thereafter. The redshift is controlled by the black hole temperature $T$, which sets the Lyaponuv exponent $\lam =2\pi T/\hbar$. Black hole chaos, once mysterious, is in fact a simple geometric effect.

 One should keep in mind that these arguments are, by necessity, semiclassical. The picture of Hawking radiation as photons  ``starting off'' outside and close to the horizon, with the photons closer to the horizon escaping later, is only valid for a limited amount time (the scrambling time).~\footnote{The scrambling time for a black hole is of order $T \log T$. This is the same as the Ehrenfest time, the time after which the semiclassical approximation breaks down.} This is sufficient to establish the Lyaponuv exponent, but it is insufficient to say much else. In particular, in the setup in \cite{PolchinskiC},  one does not know what effect the extra particle that was sent in has on Hawking quanta emitted more than a scrambling time later. 
 
 Note that, for a classically chaotic system, it is simple to linearize the equations of motion around some point in phase space, $\dot{x} \approx A x$. If the matrix $A$ has eigenvalues with positive real parts then, at least in that region of phase space, the system is chaotic.~\footnote{In addition, it is assumed  that phase space is bounded. Note also, the quantum Lyaponuv exponent defined through the out-of-time-order correlator is more  like the largest positive eigenvalue of $A$ rather than  the standard classical Lyaponuv exponent, which is defined by a time average over an entire trajectory.} Of course, this says little  about  how trajectories will evolve over longer times. 
 
\subsubsection*{Beyond Lyaponuv exponents}
Diagnosing that a system is chaotic is only the start. The Lyaponuv exponent is by itself of little interest; what one would like is to compute the physical observables. For the pinball game, this is the mean escape time: if one averages over ingoing angles of the particle sent in, what is the average amount of time that the particle spends trapped between the three disks before escaping? The solution is found in  \cite{Smilansky, Gaspard1,  Cvitanovich, ChaosBook}. The three disk game is a warmup for a ball bouncing between many disks, which is a warmup for many billiard balls of the same size colliding against each other. The microscopic chaos is apparent; what one seeks are the macroscopically measured quantities - the transport coefficients. A complete understanding of chaos in the system should provide a prescription for computing the transport coefficients. 

\subsubsection*{Chaos in the $S$-matrix}

Diagnosing chaos for the black hole relied on semiclassical geometry. What are we to do in a more general context in quantum field theory and string theory when there is none? A black hole can be viewed as a long lived resonance in the scattering matrix of  quantum gravity. It is natural to propose that for a general quantum field theory, chaos is reflected in erratic behavior of the $S$-matrix under a small change in the $in$ or the $out$ state \cite{VRchaos}. To be in a chaotic regime one needs to be far from the vacuum, requiring either a very high energy collision, or an $S$-matrix involving a large number of particles. The $S$-matrix encodes far more information than the Lyaponuv exponent, but of course the $S$-matrix involving a black hole is unknown. 

In the context of quantum mechanics, in order for the $S$-matrix to exhibit chaos, one would expect a necessary condition is a large number of closely and erratically spaced resonances. To see this, note that in the case of pinball scattering, chaos is a result of the particle sometimes spending a very long time bouncing between the disks before escaping. For some amount of time the trajectory of the particle is well approximated by one of the (unstable) bound periodic orbits that are trapped within the disks. When considered quantum mechanically, these periodic orbits manifest themselves as resonances. If we consider scattering in quantum mechanics off of a potential, then if the potential has long-lived bound states,  the $S$-matrix will have poles in the complex energy plane. In particular, for a wave reflecting off of a potential, the transmission factor will have a jump at the resonance energy, and each such jump in the phase angle gives a time delay,  see e.g. \cite{fyodorov2016random,Weidenmuller:2008vb}.
 
Achieving a large number of closely spaced resonances in quantum field theory would generally require strong coupling, in order to produce a large number of bound states. This  means we would lose analytic control and be unable to compute the $S$-matrix. Indeed, this is the kind of situation which leads to black holes as intermediate states.

An evident exception is a highly excited string. A string - even a free string - has an enormous number of internal states. This suggests that weakly coupled string theory may be the rare case in which the $S$-matrix is both computable and exhibits erratic behavior. In fact, we have come full circle, due to the Horowitz-Polchinski correspondence principle between black holes and strings \cite{HorowitzPolchinski}, which we now review.

\subsubsection*{The correspondence principle}
Consider a string with zero angular momentum, in four dimensions. The string coupling $g_s$ is related to Newton's constant $G$ through $G \sim g_s^2 \al'$, where  $\al'$ is proportional to the inverse of the string tension or the string length squared. As one increases Newton's constant, the string (like any other object) will eventually collapse into a black hole. The transition is difficult to specify precisely, but it  roughly occurs  when the string is contained within its Schwarzschild radius, $2 G M$. If we take the size of the string to be of order $\sqrt {\al'}$, then the transition to a black hole occurs when the mass of the string is of order $\sqrt{\al'}/G$. The mass of a string, in terms of its excitation level $N$, is $\sqrt{N/\al'}$. Thus, the transition occurs at $g_s \sim N^{-\frac{1}{4}}$. 

This is the first nontrivial result: for a highly excited string, the string coupling at the transition is small. As \cite{HorowitzPolchinski} states, this did not have to be the case. Consequently, one can hope to use weak coupling perturbation theory to study the stringy black holes. For example, for large excitation level  $N$, and for non-interacting strings, there are exponentially (in $\sqrt{N}$) string states of mass $\sqrt{N/\al'}$. These states are degenerate and stable, but interactions will shift the masses and widths by an amount $\sim g_s^2 M \sim  g_s^2 \sqrt{N/\al'} \sim \sqrt{1/\al'}$, small compared to the mass for large $N$.

The second result is that at the transition, the entropy of the string and of the black hole are of the same order. The entropy of the string scales as $\sqrt{N}$ (this follows from the standard counting of the degeneracy of string states at level $N$, and will be reviewed in Appendix.~\ref{sec5}). The Bekenstein-Hawking entropy of the black hole scales as the horizon area divided by Newton's constant, $S \sim \frac{A}{G}$, which at the transition also scales as $\sqrt{N}$, $ S \sim G M^2 \sim g_s^2 N \sim \sqrt{N}$.

It is remarkable, and perhaps unique to string theory, that the transition occurs at weak coupling and that the entropies are of the same order. If one were to consider any other nearly static object on the verge of collapsing into a black hole - for instance, a neutron star  - then its entropy would be vastly less than  the black hole entropy, scaling as $A^{3/4}$ in Planck units, rather than $A$ \cite{Bousso:2010pm}.

In this argument, the size of the string is taken to be $\sqrt{\al'}$. This is only approximately correct: a typical excited string  would seem to be larger, due to its random walk like behavior, but the effect is partially offset by gravitational self-interactions. Defining what is meant precisely by the size of the string \cite{Mitchell:1990cu}, as well as computing the size, is challenging \cite{Horowitz:1997jc, Damour:1999aw, Chialva:2009pf}. Regardless, the argument in \cite{HorowitzPolchinski} is sufficient to establish that the entropy of a string is of the same order as the entropy of a black hole.~\footnote{We see no reason to believe that they are exactly equal \cite{Susskind:1993ws}, except in some special cases \cite{Strominger:1996sh}.}

\subsection{Scattering of highly excited strings}

The Veneziano amplitude - the tree level scattering amplitude of four tachyons - is one of the  iconic results in string theory \cite{Veneziano,Polchinski,  GSW}. A simple beta function of the Mandelstam variables, it encodes many of the remarkable properties of string theory. 
The generalization to the scattering amplitude of $N$ tachyons - the Koba-Nielsen formula \cite{Koba} - is also well-known. While the formula is compact, hidden within the integrals is a rich underlying complexity: the intermediate states of a high energy $N$-point tachyon amplitude contain highly excited strings. Indeed, the number of different strings  states at a given mass grows exponentially with the mass. 
Our goal will be to extract this structure, in a controlled way.  

We will study tree-level scattering amplitudes, not of a large number of tachyons, but of a small number of highly excited strings. As we will see, the scattering of excited strings can be regarded as extracting a subset of diagrams appearing in the scattering  of tachyons in particular kinematic configurations.

In order to study chaos in string scattering, we need to know the exclusive scattering amplitude for precisely specified excited strings; taking inclusive or averaged amplitudes would likely wash out the effects we are seeking.  As a result, many of the quantities that have been previously studied involving excited strings, such as the total decay rate of the string \cite{DaiPolchinski, Mitchell:1988qe, Wilkinson:1989tb, Sonnenschein:2017ylo, Iengo:2006gm}, the amplitude for two light strings to form a heavy string \cite{Matsuo:2008fj, Dimopoulos:2001qe}, the amplitude with an average over the excited string states \cite{Amati}, or the amplitude involving  the leading Regge trajectory \cite{Sundborg},  are inapplicable. What we need is the amplitude involving   generic string states, chosen from  the ensemble of the exponentially many states at a given mass. 

The main challenge in computing  scattering amplitudes with excited strings, at least at tree level, is actually specifying the excited string states. If, instead of a fundamental string, we had a violin string, then every state would be specified by the energy in each of the modes. However, for a fundamental string, the modes are not independent. Recall that in the Polyakov action, the worldsheet metric is a variable, which we may fix to be the flat metric. Each mode  then appears independent and free, yet the equations of motion coming from the variation of the worldsheet metric - that the worldsheet energy-momentum tensor vanishes - must  be obeyed, thereby leading to the Virasoro constrains for the modes. For the low level states, the constraints are simple to solve, but they become increasingly cumbersome at higher levels, see e.g. \cite{ Weinberg:1985tv,Manes:1988gz, Nergiz:1993gw, Hanany:2010da, Chan}. 
Fortunately, there is a more systematic and physical approach: the DDF construction \cite{DDF}. In short, the construction amounts to starting with a tachyon and then repeatedly scattering photons off of it.~\footnote{Viewing an $S$-matrix involving a highly excited string as part of an $S$-matrix with a large number of photons is in line with the proposal in \cite{VRchaos} that one look for chaos in a many-particle $S$-matrix.} The explicit form of the resulting vertex operator for an arbitrary string state, which we will need, was worked out only relatively recently \cite{Skliros}.

In this paper we concentrate on amplitudes involving tachyons and one excited string. Our focus will really be on the simplest case: the amplitude of a highly excited string  decaying into two tachyons.   We find the amplitude is highly sensitive to both the precise state of the excited string  and the tachyon scattering angle relative to the polarization of the stringy state. However, a discussion of chaos in string scattering that goes beyond this elementary observation is  left for \cite{GRchaos}.  The study of amplitudes involving more than one excited string is challenging and is largely left to future work; in \cite{GRthree} we compute the simplest amplitude, the three-point amplitude, in which all three strings are excited.~\footnote{There have been a number of previous studies of the three-point amplitude of excited strings, such as \cite{Ademollo:1974kz, GrossJevicki, Hornfeck:1987wt,  Sagnotti:2010at, Schlotterer:2010kk, Boels:2012if}, but as far as we know, none of these give a usable form for the amplitude involving  typical highly excited strings.}, \footnote{It may  be  interesting to understand how these compare with the cubic couplings \cite{Gross:2017aos, Gross:2017hcz, Rosenhaus:2018dtp} in the putative AdS dual of the SYK model. }

\subsection{Future work}
There are a number of other calculations that we believe should be possible to do.

\subsubsection*{The excited string as a random walk}

We find the scattering amplitudes through a direct computation using vertex operators. One would like a simpler and more intuitive calculation. A generic highly  excited string is often modeled as a random walk  \cite{Salomonson:1985eq,Salomonson:1988ac,Mitchell:1987hr, Kruczenski:2005pj, Mertens:2013pza, Kawamoto:2015zha, Horowitz:1997jc}, with interactions between strings a result of strings splitting and joining \cite{Polchinski:1988cn, Jackson:2004zg}.  One would like to calculate the scattering amplitude using this model. Certainly the picture of the excited string as a random walk makes it  intuitive that the amplitude should behave erratically; one would like to make this precise. 

\subsubsection*{The decaying string and black body radiation}
Amati and Russo \cite{Amati}, see also \cite{Manes:2001cs, Chialva:2006iwa, Matsuo:2009sx,Kuroki:2007aj, Iengo:2006if,  Chialva:2004xm, Chen:2005ra,  Iengo:2003ct, Penedones} studied the decay of a massive string into a photon and another massive string, averaging over the initial string states of the same mass, and summing over all outgoing heavy string states. It was found that the decay rate, as a function of the photon energy, obeys a black body spectrum with a temperature that is the Hagedorn temperature.
This is consistent with the Horowitz-Polchinski correspondence as the Hagedorn temperature is  $1 / \sqrt{\al'}$, which coincides with the Bekenstein-Hawking temperature, ${1\over GM}$, since $G \sim g_s^2 \al'$ and $M \sim \sqrt{N/\al'}$, which agree when $g_s \sim N^{-\frac{1}{4}}$. 
Using the methods presented here  allows one to compute the exclusive amplitudes of  a particular heavy string decaying into a particular heavy string and a photon \cite{GRthree}.  Averaging should of course reproduce the blackbody spectrum, but the exclusive amplitude will give much more: it will allow one to see precisely how the radiation differs for each  string microstate, the string analog of the long sought goal of extracting information from Hawking radiation.~\footnote{It was recently  proposed \cite{Penington:2019npb,Almheiri:2019psf,Almheiri:2019hni,Almheiri:2020cfm} that for black holes information can be recovered by including new replica wormhole saddles in the gravitational path integral.}

 
 

\subsubsection*{Loop corrections}

Free excited strings have an enormous degeneracy of energy levels: the number of degenerate states scales exponentially with the mass. At finite string coupling one  expects this degeneracy to be completely broken, with resulting energy level spacings that are exponentially small.

 A more tractable  starting point is finding the energy levels perturbatively in the string coupling. At leading order, one should compute the one loop string diagram correction to the two-point amplitude of a heavy string. Loop amplitudes in string theory are in principle straightforward; it is the same computation as the tree level diagram, but on a torus or cylinder, though such computations in practice are involved. One loop diagrams were studied in \cite{ Sundborg:1988tb, Sundborg:1988ai,  Okada:1989sd,  Mitchell:1989uc, Iengo:2002tf,  Chialva:2003hg, Chialva:2004ki, Gutperle:2006nb, Chialva:2009pg, Pius:2013sca, Sen:2016gqt}. These studies  average over the string states. One would like to compute the one loop amplitude for precisely defined excited string states, using the same DDF vertex operators discussed here. The imaginary part of the one loop diagram gives back the tree level amplitude  of a heavy string decaying to two heavy strings, while the real part gives the mass shift. 
  
\subsubsection*{Saddles for high energy, fixed angle scattering}
At low energies, strings behave as particles and the dynamics can be represented by a local effective field theory. At high energies (equivalently, in the tensionless limit $ T = 1/\pi \al'$, with $\al'\rightarrow \infty$), strings are very stringy. One might hope that in this  ultra high energy limit string theory simplifies. Concretely, do the string scattering amplitudes in this limit admit a simple spacetime interpretation illustrating how the strings are interacting?

Gross and Mende studied high energy, fixed angle scattering amplitudes of tachyons \cite{GrossMende1, GrossMende2, Gross:1988ue, Gross:1989ge}. They found, at each order in string perturbation theory, a particular saddle for the path integral over surfaces, and furthermore argued that this may be the dominant saddle (the saddle equations have many solutions and most are unknown, so this has never been definitively shown). Equipped with the saddle, one has a spacetime picture of how the strings interact \cite{GrossMende2}.  

Since most string states are highly excited, it seems in fact more natural to ask about the high energy, fixed angle amplitudes of generic highly excited strings.~\footnote{If one is scattering excited strings, rather than tachyons, then as long as the masses of the strings are held fixed as the scattering energies are taken to infinity, the analysis of Gross and Mende doesn't change and the same saddle solutions hold. However, if the string mass is also taken to infinity, then the saddles change. This is the limit we are interested in now: large  Mandelstam $s$ and $t$, and large excitation level $N$, while keeping the ratio fixed.} What are the equations describing the saddles in this case, and is there a dominant saddle? Concretely, in Sec.~\ref{sec62}, we will present the tree level amplitude involving one excited string and three tachyons. What are the saddles?~\footnote{As we have remarked earlier, scattering of excited strings can be viewed as picking out a subset of diagrams appearing in the many-point amplitude of light strings, suggesting that the solutions of the saddle equations for heavy string $4$-point amplitudes comprise a subset of the  solutions for $n$-point tachyon amplitude saddle equations. This may give a handle on the enormous number ($(n-3)!$ at tree level) of saddles for the $n$-point tachyon amplitude. See  \cite{Ghosh:2016fvm} for a recent discussion of the saddles at large $n$.   }

There is an intriguing connection with field theory amplitudes. The saddle equations for $n$-point tree level tachyons amplitudes \cite{GrossMende1} are the same equations as the scattering equations appearing in the CHY formula for field theory amplitudes of massless particles \cite{Cachazo:2013gna,Cachazo:2013hca,Dolan:2013isa}, see Appendix~\ref{appC}. The saddle equations we will find for highly excited string amplitudes are a generalization of the scattering equations; do they have a field theory application?

\subsection{Outline}

The paper is organized as follows. 

In Sec.~\ref{sec2} we establish notation and review tachyon scattering amplitudes. 

In Sec.~\ref{sec3} we review the explicit construction of the vertex operators which create any excited string state. 

In Sec.~\ref{sec4} we use these operators to compute the amplitude involving one excited string and any number of tachyons.  In Sec.~\ref{sec41} we discuss the special case in which the polarizations of all the photons creating the excited state  are orthogonal to each other, leading to a significant simplification of the amplitude. The amplitude with an arbitrary excited string  is discussed in Sec.~\ref{sec42}. A nontrivial check on the results that we perform is showing that  the integrand for the amplitude exhibits SL$_2$ invariance.

 In Sec.~\ref{sec6} we return to the amplitude involving one excited string, studying it in detail in Sec.~\ref{sec61} for a (typical) highly excited string decaying to two tachyons and briefly in Sec.~\ref{sec62} for an excited string decaying to three tachyons.   The reader who is uninterested in the details may go directly to Sec.~\ref{sec6}. 
 
 In Sec.~\ref{sec7} we end with a brief discussion. 

In Appendix~\ref{Virasoro} we review the construction of covariant vertex operators via the Virasoro constraints. 
In Appendix~\ref{appB} we check the normalization of the DDF operators.
In Appendix~\ref{appC} we  exhibit a  generalization of the scattering equations which determine the high energy scattering of light and massive string states.
In Appendix ~\ref{sec5} we discuss the properties of a typical excited string.

\section{Review of tachyon  scattering } \label{sec2}
In this section we review the Polyakov action for a string, the expansion of the string field into oscillator modes, and the operator product expansion (OPE) of tachyon vertex operators. We also review the tachyon scattering amplitude and show that the integrand  satisfies SL$_2$ invariance. In this paper we will be discussing $D$ dimensional open bosonic string theory for simplicity. The generalization to the superstring is straightforward, and we don't expect any qualitatively different behavior. 

\subsection*{The string action}
The location of the string, as a function of the worldsheet space and time coordinates $\sigma, t$, is given by the string field $X^{\mu}(\sigma, t)$ which satisfies the Polyakov action, 
\be
-\frac{1}{4\pi \al'} \int d \tau d\sigma\, \sqrt{-\gamma} \gamma^{a b} \d_a X^{\mu} \d_b X_{\mu}
\ee
where the indices $a,b$ range over the two worldsheet coordinates: $\sigma $ and $\tau$ (the Euclidean worldsheet time), and  the  indices $\mu$ range over the $D$ ambient spacetime coordinates. The action is diffeomorphism invariant, and we may chose $\gamma_{ab}$ to be the flat metric, however we will still have to impose the equations of motions for $\gamma_{ab}$: that the worldsheet energy-momentum tensor vanishes. We will work in conventions with $\al' = 1/2$, and focus on open strings (the analysis with closed strings is similar). 

 The expansion of an open string in terms  of oscillators is, 
\be  \label{Xmu}
X^{\mu}(\sigma, t)  = x^{\mu} + i p^{\mu} t + i \sum_{n\neq 0} \frac{1}{n} \al_{n}^{\mu} e^{- i nt} \cos n \sigma~,
\ee
where
\be \label{23}
\[\al_m^{\mu}, \al_n^{\nu} \] = m \delta_{m+n} \eta^{\mu \nu}~, \ \ \ \[ x^{\mu}, p^{\nu}\] = i \eta^{\mu \nu}~, \ \ \ \ \ \eta^{\mu \nu} = (- + + \cdots +)~,
\ee
where  $x^{\mu}$ and $p^{\mu}$ are the center of mass position and momentum of the string. The $\al_n^{\mu}$ are the creation and annihilation operators: for negative $n$, $\al_{n}^{\mu}$ excites level $-n$ of the string in the direction $\mu$. 
For an open string, the vertex operators are inserted on an endpoint of the string, which we take to be the left endpoint, $\sigma=0$. The right endpoint is at $\sigma = \pi$. At the left endpoint, 
\be \label{213}
X^{\mu}(\sigma=0, t) \equiv X^{\mu}(z) = x^{\mu} - i p^{\mu} \log z + i \sum_{n\neq 0} \frac{1}{n} \al_{n}^{\mu} z^{-n}~, \ \ \ \ \ z = e^{i t}~.
\ee

It is standard to Wick rotate the Lorentzian worldsheet time $t$ to Euclidean worldsheet time $\tau$, 
\be
\tau= - i t~, \ \ \ \ \ ds^2 = -dt^2 + d\sigma^2 = d\tau^2 + d\sigma^2~,
\ee
and to map the string worldsheet onto the upper half plane. The mapping is, 
\be
z = e^{i\sigma- \tau}~, \ \ \ \b z = e^{- i \sigma - \tau}~, \ \ \  \ \ \ \ ds^2 = \frac{d z d \b z}{|z|^2}~.
\ee
Notice that the string endpoints, $\sigma=0 $ and $\sigma= \pi$, are mapped onto the real axis: for $\sigma =0$, $z=e^{-\tau}$.
Finally, we note that the propagator is, 
\be  \label{prop}
\la X^{\mu} (z_1) X^{\nu}(z_2)\ra = - \eta^{\mu \nu} \log z_{12} -  \eta^{\mu \nu} \log \b z_{12}~, \ \ \ \ \ \la \d^{m_1} X(z_1) \d^{m_2} X(z_2)\ra =(-1)^{m_2} \frac{(m_1 {+} m_2 {-} 1)!}{z_{21}^{m_1 + m_2}}
\ee
In studying the OPE, we will only write the holomorphic piece. 

\subsection*{Tachyon amplitudes and the OPE} \label{Sec22}

Let us review the amplitude for the scattering of $n$ tachyons.  
The vertex operator for a tachyon is given by, 
\be
:e^{i p\cdot X(z)}:~,
\ee
and the amplitude is, 
\be \label{29}
\mA =\frac{1}{\text{vol}(SL_2)} \int \prod d z_i\,  \la \prod_{i=1}^{n}:e^{i p_i\cdot X(z_i)}\!:\ra~.
\ee
All that remains is to evaluate the right-hand side. Before proceeding, we review some elementary properties of the OPE, which we will need later in discussing the vertex operators for excited states. 

Recall that the purpose of normal ordering is to remove  divergent pieces which arise when two operators are brought together. For instance, 
\be
: \d \phi(z) \d \phi(z): = \lim_{w\rightarrow z} \( \d \phi(z) \d \phi(w) - \la \d \phi(z) \d \phi(w)\ra\)~.
\ee
Normal ordering is important when we need to define composite operators. The product of two operators, $A$ and $B$, can be written as, 
\be
A(z) B(w) = \sum_{n=-\infty}^N \frac{\{ AB\}_n(w)}{(z- w)^n}~, \ \ \ \ \ A(z) B(w) \sim \sum_{n=1}^N \frac{\{ AB\}_n(w)}{(z- w)^n}~,
\ee
where the second term, with the $\sim$, is the OPE which only includes  the divergent pieces in the expression. 
Thus, in the limit $z\rightarrow w$, we have a finite composite operator, along with a finite number of divergent terms which have been explicitly separated,
\be
A(z) B(w) \rightarrow : A B(w): +  \sum_{n=1}^N \frac{\{ AB\}_n(w)}{(z- w)^n}~.
\ee

When computing the OPE, we first take all possible Wick contractions, and then   Taylor expand what is left around $w$.
For instance, let us look at the composite operator formed from the free field $\phi$, with  correlation function $\la \d \phi(z) \d \phi(w)\ra = - \frac{1}{( z- w)^2}$. We have, 
\bea\nn
: (\d \phi(z))^2:  : \d \phi(w)\!: \,\  &=&\,  :\! (\d \phi(w))^3:  + 2 \d\phi(z) \la \d \phi(z) \d \phi(w)\ra\\
&\sim&  - 2 \frac{\d\phi(z)}{(z - w)^2} \sim -2 \frac{\d \phi(w)}{(z- w)^2} - 2 \frac{\d^2 \phi(w)}{z- w}~,
\eea
where in the second line we  looked at just the OPE piece. 
Let us now look at the  OPE of two tachyon vertex operators, 
\bea \nn
\hspace{-.4cm}&&:e^{i p_1\cdot X(z_1)}\!:\,  :\!e^{i p_2 \cdot X(z_2)}\!:\, \, =  \sum_{n, m} \frac{1}{n! m!} :\!(i p_1\! \cdot \!X(z_1))^n\!:\,  :\! (i p_2 \cdot X(z_2))^m\!:\\[-5pt] \nn
&=& \(1 + \la i p_1\! \cdot \!X(z_1) \,  i p_2 \cdot X(z_2)\ra + \frac{1}{2!}( \la i p_1\! \cdot\! X(z_1)\,  i p_2\! \cdot\! X(z_2)\ra )^2 + \ldots   \):e^{i p_1\cdot X(z_1)} e^{i p_2 \cdot X(z_2)}: \\
&=& e^{ - \la p_1 \cdot X(z_1) p_2 \cdot X(z_2)\ra}:e^{i p_1\cdot X(z_1)} e^{i p_2 \cdot X(z_2)}\!:\, =  |z_{12}|^{p_1 \cdot p_2}:e^{i p_1\cdot X(z_1)} e^{i p_2 \cdot X(z_2)}:~. \label{TTope}
\eea
In the first equality we Taylor expanded the exponentials. In the second equality we performed Wick contractions, in the third equality we resummed the Wick contractions, and in the last equality we made use of the propagator (\ref{prop}). 
Similarly, the OPE of multiple tachyon vertex operators is given by, 
\be
\prod_i : e^{i p_i \cdot X(z_i)}:\, = \prod_{i<j} z_{ij}^{p_i \cdot p_j}\, :e^{i\sum_i p_i \cdot X(z_i)}:~.
\ee
To compute the expectation value, we may Taylor expand $X(z_i) = X(z_1) + \ldots$, and thus pick up a delta function for momentum conservation, 
\be
\la :e^{i\sum_i p_i \cdot X(z_i)}: \ra = (2\pi)^D \delta^D(\sum_i p_i)~.
\ee

Equipped with this result, we can now evaluate the tachyon amplitude (\ref{29}) to get the Koba-Nielsen formula, 
\be \label{KN}
\mA = \frac{(2\pi)^D \delta^D(\sum_i p_i)}{\text{vol}(SL_2)} \int \prod_id z_i\, \prod_{i<j} z_{i j}^{p_i \cdot p_j} ~.
\ee
It will sometimes be convenient to write the integrand as, 
\be \label{29v2}
\prod_{i<j} z_{i j}^{p_i \cdot p_j} =e^{\mL} ~, \ \ \ \ \ \ \mL =  \sum_{1\leq i< j\leq n}  p_i \cdot p_j \log z_{i j}~.
\ee

\subsubsection*{SL$_2$ invariance }
An important property of tree-level string theory amplitudes is that the integrand exhibits SL$_2$ invariance.  While the SL$_2$ invariance of the tachyon amplitude is fairly evident, the SL$_2$ invariance of amplitudes with excited strings will be nontrivial. As warmup, here we check that the tachyon amplitude integrand is SL$_2$ invariant, by writing it in a manifestly SL$_2$ invariant form. Under an SL$_2$ transformation, 
\be
z_i \rightarrow\frac{ a z_i + b}{c z_i + d}~, \ \ \text{where } \ ad-bc = 1~,
\ee
and correspondingly the measure and difference between two points transform as, 
\be
d z_i \rightarrow \frac{1}{(c z_i + d)^2} dz_i~, \ \ \ \ \ \  z_{i j} \rightarrow \frac{z_{i j}}{(c z_i + d)(c z_j + d)}~.
\ee
Let us consider the four-point amplitude first, $n = 4$. We may use momentum conservation to eliminate $p_1 = -\sum_{i>1} p_i$ in $\mL$ (\ref{29v2}) , 
\be
\mL = p_2 \cdot p_3 \log\frac{z_{23}}{z_{12} z_{13}} + p_2 \cdot p_4 \log \frac{z_{24}}{z_{12} z_{14}} + p_3 \cdot p_4 \log\frac{z_{34}}{z_{13} z_{14}} - 2\log z_{12} z_{13} z_{14}~,
\ee
where we  used the mass-shell condition for tachyons, $p_i^2 =-m_i^2=2$. Using momentum conservation, $p_1^2 = (p_2 +p_3 + p_4)^2$, allows us to eliminate $p_2 \cdot p_3$, since $p_2 \cdot p_3 = - 2 - p_2 \cdot p_4 - p_3 \cdot p_4$, and hence, 
\be
\mL = p_2 \cdot p_4 \log \frac{ z_{24} z_{13}}{z_{14} z_{23} } + p_3 \cdot p_4 \log \frac{z_{12}z_{34} }{z_{14} z_{23}} - 2\log z_{23} z_{14}~.
\ee
We now see that the amplitude is, 
\be
\mA = \frac{1}{\text{vol}(SL_2)}\int \frac{d z_1 d z_2 d z_3 d z_4}{z_{14}^2 z_{23}^2} \, R^{\, p_2 \cdot p_4} (R-1)^{p_3 \cdot p_4}~, \ \ \ \ R=  \frac{z_{13}z_{24} }{z_{14} z_{23}}~,
\ee
which is manifestly SL$_2$ invariant.  Using SL$_2$ symmetry, three of the points can be fixed: $z_1 = \infty$, $z_2=1$, $z_3 = z$ and $z_4=0$, turning the amplitude into,  
\be \label{223}
\mA = \int_0^1 d z\, z^{p_3 \cdot p_4} (1-z)^{p_2 \cdot p_3} = \beta(p_3{\cdot}p_4 + 1, p_2 {\cdot }p_3+1)
\ee
where on the right we have the Euler beta function. 
Consider now the amplitude with $n$ tachyons. We again eliminate $p_1$ in $\mL$ (\ref{29v2}) through momentum conservation, 
\be 
\mL = \sum_{1<i<j \leq n} p_i \cdot p_j \log \frac{z_{i j}}{z_{1 i} z_{1 j}} - 2\sum_{1<i\leq n} \log z_{1 i}~.
\ee
Using momentum conservation, $p_1^2 = (\sum_{i>1} p_i)^2$, we  eliminate $p_2 \cdot p_3$,
\be
p_2 \cdot p_3 = 2 - n - \sum_{ \substack{1 <i<j\leq n\\ (i,j) \neq (2,3)}} p_i \cdot p_j~,
\ee
so that the amplitude becomes,
\be \label{227}
\mL = \sum_{ \substack{1<i<j\leq n\\ (i,j) \neq (2,3)}} p_i \cdot p_j \log \frac{z_{i j} z_{12} z_{13}}{z_{1 i} z_{1 j} z_{23}}  + \( \log\frac{z_{23}}{ z_{12} z_{13}} +  \sum_{1<i} \log \frac{ z_{12} z_{13}}{z_{1 i}^2 z_{23}} \)~.
\ee
The first term is manifestly SL$_2$ invariant. The second term,  combined with the measure, is SL$_2$ invariant.~\footnote{To see this, notice that for the argument of the $\log$ to be invariant under inversions of $z_a$, there needs to be the same powers of $z_a$ in both the numerator and denominator. For instance, $\log\frac{z_{23}}{ z_{12} z_{13}}$ is invariant under $z_2 \rightarrow 1/z_2$. From the first $\log$ in parenthesis  in (\ref{227}) there is an extra $z_1^2$ in the denominator, and from the second $\log$ an extra factor $z_i^2$ in the denominator, for $i\neq 1$; so in total, an extra $z_i^2$ for all $i$. This is just what we need to cancel off the transformation of the measure, i.e. $dz_i/z_i^2$ is invariant under inversions.   \label{foot3}}

\section{Building excited string states} \label{sec3}

In the previous section we discussed the scattering of tachyons - the lightest string states. The majority of string states are, however, excited states, and it is their scattering amplitude that we are interested in. In this section we review the construction of vertex operators for  excited string states.

The unique lightest string state is the tachyon, denoted by $|0; p\ra$, where $p$ is the center of mass momentum of the string. To build excited states, we act with the creation operator, $\al_{-m}^{\mu}$, which excites the $m$'th mode of the string, in the $\mu$ direction. An excited state  is of the form
\be \label{31}
\al_{-m_1}^{\mu_1} \al_{-m_2}^{\mu_2} \cdots \al_{-m_k}^{\mu_k} |0; p\ra~,
\ee
and has a mass,
\be
M^2 = 2(N-1) ~, \ \ \ \ \  N = \sum_{m=1}^{\infty} N_m~, \ \ \  \ \ \   N_m = \al_{-m} \cdot \al_m~,
\ee
where $N_m$ is the number operator for the number of excited modes at level $m$, and $N$ is the total level.

Not all states (\ref{31}) are allowed; we must build superpositions which are annihilated by the Virasoro generators. This is discussed in more detail in Appendix~\ref{Virasoro}. For instance, at level $N=1$, the states are of the form $\lam \cdot \al_{-1} |0;p\ra$. There are two constraints:  the polarization vector $\lam$ must be orthogonal to the momentum, $\lam \cdot p = 0$, and one can add any multiple of $p$ to $\lam$ and leave the amplitude unchanged. This leave $D-2$ independent states at level-one. These states are massless, and this result is just what we expect for  photons: in $D=4$ there are only two transverse polarizations, and changing the polarization by a vector proportional to the momentum leaves the amplitude unchanged. We will refer to the level-one states as photons (massless spin 1 particles). One can proceed to higher levels in a similar manner  ($N=2$ and $N=3$ are done in  Appendix~\ref{Virasoro}), however with increasing $N$ the process becomes increasingly more involved, making it difficult to write down the form of a state at general level $N$. 

A more physical and systematic approach comes from recognizing that all possible excited string states are already contained within $n$-point amplitudes of, for instance, tachyons or photons, in the form of intermediate states. 
To obtain the scattering amplitudes of the excited states, we simply need to pick them out.

We can do this iteratively. Suppose we have a string moving with momentum $\t p$ and created with the vertex operator $V(z) e^{i \tp\cdot X(z)}$. We will scatter a photon off of this state. The vertex operator for a photon of momentum $m\, q_{\mu}$ ($m$ is an integer and $q_{\mu}$ is a null vector) is, 
\be
i \lam \cdot \d X\, e^{i m q\cdot X(z)}~,
\ee
where the polarization $\lam^{\mu}$ is orthogonal to the momentum, $\lam \cdot q = 0$. Formally, the process of picking out the state after scattering the string with the photon consists of taking  the OPE of the photon vertex operator and the vertex operator of the initial string state. A contour integral then picks out the pole in the OPE. Explicitly, one computes,
\be
\oint \frac{dz}{2\pi i} : i \lam \cdot \d X(z)\, e^{i m q X(z)}:\, \,  :V(0)e^{i \tp\cdot X(0)}:~.
\ee
To make the procedure systematic, we define, 
\be \label{DDF2}
\lam \cdot A_m =\oint \frac{d z}{2\pi}\, \lam\cdot \d X(z)\, e^{i m q \cdot X(z)}~, \ \ \  \ \ \ q^2 = 0~, \ \ \ \ \lam \cdot q = 0~, \ \ \ |\lam|^2\neq 0
\ee 
The $A_m^{\mu}$ are the DDF operators \cite{DDF, GSW,Skliros}. It is standard \cite{GSW} to choose a coordinate system in which $q_{\mu}$ points in the $+$ direction. One can check that the transverse $A_{m}^i$ obey the commutation relations of creation and annihilation operators: they are isomorphic to the transverse components of the  operators $\al_m^{\mu}$, and describe the transverse modes of the string.
However, it is unnecessary to pick a coordinate system, so we will continue working with the covariant form (\ref{DDF2}). Notice that from the form (\ref{DDF2}) it is clear that, regardless of the coordinate system, there are $D-2$ independent $A_m^{\mu}$. The counting is identical to the counting of the number of independent photon polarizations, as discussed above. Explicitly, $\lam \cdot q=0$ gives one constraint. A second constraint is that we can add any multiple of $q$ to $\lam$ and leave the amplitude unchanged, since we just pick up a total derivative, 
\be \label{36}
\oint \frac{d z}{2\pi}\, q\cdot \d X(z)\, e^{i m q \cdot X(z)}= \frac{1}{i m} \oint \frac{d z}{2\pi} \d \, e^{i m q \cdot X(z)}=0~,
\ee

\begin{figure}
\centering
\includegraphics[width=4in]{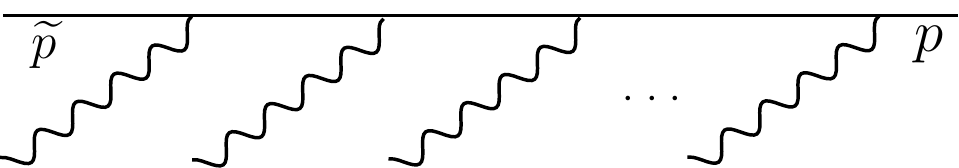}
\caption{Starting with a tachyon of momentum $\tp$, photons with momentum proportional to $q$ are scattered off of it, resulting in an excited string of momentum $p$. } \label{DDFfig}
\end{figure}

 To construct the vertex operator for an excited string state, we start with a tachyon of momentum $\tp$,  and iteratively scatter photons off of it, as shown in Fig.~\ref{DDFfig}, 
\be \label{36v2}
(\lam_k\cdot A_{-m_k}) \cdots (\lam_2 \cdot A_{-m_2}) (\lam_1\cdot A_{-m_1}) e^{i \tp\cdot X}~.
\ee
We have picked a very particular scattering configuration, in which each photon has momentum in the same direction, parallel to $q$ (this is in fact sufficient to generate any excited string state; we will elaborate on this later). The momentum of the $i$'th photon is $- m_i q$ and its polarization is $\lam_i$. 
The resulting string state has momentum $p = \tp - N q$ where $N = \sum_{i=1}^k m_i$. The mass of the state is $p^2 = 2(1- N \tp \cdot q)$, where we used that the tachyon has mass $m^2 = - \tp^2 = 2$ and the photon is massless. In order for the excited string to have the correct mass, $M^2 = -p^2 = 2(N-1)$, we need to choose $q$ such that $\tp \cdot q= 1$. \\

In what follows, we explicitly work out the form of these vertex operators, essentially reviewing the construction in \cite{Skliros,Skliros0}.~\footnote{Useful discussions of the DDF operators also include  \cite{Bianchi,DHoker}. See also  \cite{Skliros:2016fqs, Hindmarsh:2010if, Skliros:2013pka, Aldi:2019osr, Bianchi:2010es, Addazi:2020obs, Aldi:2020qfu} for further applications. }  We start in Sec.~\ref{sec31} with the $N=1$ state, $\lam\cdot A_{-1} |0\ra$, and then the two $N=2$ states: $\lam\cdot A_{-2} |0\ra$ and $(\lam_2 \cdot A_{-1})(\lam_1 \cdot A_{-1}) |0\ra$. In Sec.~\ref{sec32} we first consider the state $ \lam \cdot A_{-m_1} |0\ra$, then we look at the state $( \lam_2 \cdot A_{-m_2})(\lam_1\cdot A_{-m_1})|0\ra$, and  finally we compute the vertex operator for the general state 
(\ref{36v2}).

\subsection{Level-one and level-two states} \label{sec31}
In this section we construct the vertex operators for the states at level $N=1$ and at level $N=2$. 

\subsubsection*{Level-one}

\begin{figure}
\centering
\includegraphics[width=1.8in]{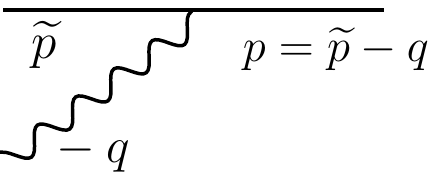}
\caption{Forming the sate $\lam \cdot A_{-1}|0\ra$ by scattering  a photon off of a tachyon.} \label{DDFlevel1}
\end{figure}

We start with the simplest case of the $N=1$ states, $\lam \cdot A_{-1} |0; p\ra$. Here we are scattering a photon of momentum $q$ off of a tachyon, see Fig.~\ref{DDFlevel1}.  
Applying (\ref{DDF2}) to the tachyon vertex operator, we need to compute,
\be  \label{37}
: \lam \cdot A_{-1} e^{i \tp \cdot X(0)}:   = \oint \frac{d z}{2\pi} : \lam \cdot \d X(z) \, e^{-i q \cdot X(z)}: \,  : e^{i \tp \cdot X(0)}:~.
\ee
To evaluate the contour integral, we must first evaluate the OPE. Taylor expanding the exponentials, 
\be
 :\lam\cdot \d X(z) \, e^{-i q \cdot X(z)}\!: \, \,  :\! e^{i \tp \cdot X(0)}: = \sum_{n, m} \frac{1}{n! m!} :  \lam \cdot \d X(z)(-i q \cdot X(z))^n\!:\, : (i \tp\cdot X(0))^m:\\ \nn
 \ee
and then Wick contracting, in a manner similar to the OPE of two tachyon vertex operators as discussed in Sec.~\ref{Sec22}, gives,
\be \label{38}
 : \!\lam\cdot \d X (z) \, e^{-i q \cdot X(z)}\!: \,\,  : e^{i \tp \cdot X(0)}:\,   = z^{ - \tp\cdot q} \(: \! \lam \cdot \d X(z)  e^{-i q \cdot X(z)} e^{i \tp \cdot X(0)}\! :\, - \frac{i \lam \cdot \tp}{z} \,:\! e^{-i q \cdot X(z)} e^{i \tp \cdot X(0)}\! : \)~.
\ee
The factor of $z^{ - \tp\cdot q}$ comes from contracting the exponentials. Since $\tp \cdot q =1$, this factor is $z^{-1}$.  Additionally, the first term in the parenthesis comes form $\d X^{\mu}(z)$ uncontracted, while the second term comes from contracting $\la \d X^{\mu}(z) \, \tp\cdot\!X(0)\ra =  - \tp^{\, \mu}/z$. The only terms that will give a contribution to the contour integral (\ref{37}) are the single poles. We should therefore Taylor expand $X(z)$ about $X(0)$ to pick out this contribution. For the first term in (\ref{38}) we can just replace $X(z)$ with $X(0)$, whereas for the second term, the extra factor of $1/z$ leads us to keep the order $z$ term in the expansion $X(z) = X(0) + z \d X(0) + \ldots$. Thus, the single pole part of the OPE of the $A_{-1}$ operator and the tachyon vertex operator is,  
\be
 :\! \lam \cdot \d X(z) \, e^{-i q \cdot X(z)}\!: \, \, :\! e^{i \tp \cdot X(0)}\!:   = \ldots + \frac{1}{z}  \( :\!\lam\cdot \d X(0)  e^{i p \cdot X(0)}\! :  - \,\lam\cdot\tp\,  :\!\,q \cdot \d X(0) e^{i p\cdot X(0)} :\)+ \ldots
\ee
where $p= \tp - q$. The integral (\ref{37}) is now immediate, giving, 
\be \label{310}
: \!\lam \cdot A_{-1}\,  e^{i \tp \cdot X}\!:\,\,   = i\,  \zeta \cdot \d X\, e^{i p\cdot X} ~, \ \ \ \  \ \zeta_{\mu}  \equiv \lam_{\mu} - (\lam \cdot \tp) q_{\mu}~.
\ee
Notice that, because $\lam \cdot q = 0$, we may equivalently express $\zeta_{\mu}$ in terms of $p$, 
\be \label{zeta}
\zeta_{\mu}  = \lam_{\mu} - (\lam \cdot p) q_{\mu}~, \ \ \ \ \  \zeta \cdot p = 0~,
\ee
where it is manifest that $\zeta_{\mu}$ is orthogonal to $p_{\mu}$. This result (\ref{310}) for the vertex operator for level-one states is exactly as  expected. It is expressed entirely in terms of the physical polarization $\zeta_{\mu}$ and the physical momentum  $p$. Moreover, as is necessary for photons, we have found that $\zeta \cdot p = 0$. Notice that the vector $q_{\mu}$, which was arbitrary, is not explicitly present in the vertex operator. It is only implicitly present, in that both $\lam_{\mu}$ and $q_{\mu}$ determine $\zeta_{\mu}$. 

\subsubsection*{Level-two}
\begin{figure}
\centering
\subfloat[]{
\includegraphics[width=2.6in]{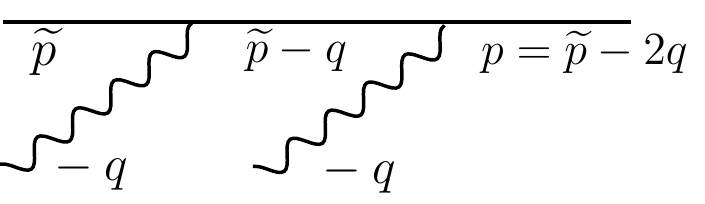}
}\ \ \ \  \ \ \ \ \ 
\subfloat[]{
\includegraphics[width=1.8in]{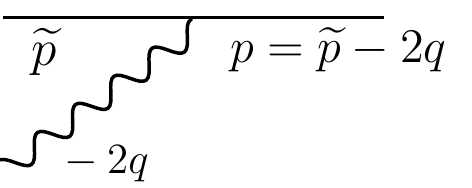}}
\caption{(a) Forming the state $(\lam_2 \cdot A_{-1})(\lam_1 \cdot A_{-1})|0\ra$. b) Forming the state $\lam \cdot A_{-2}|0\ra$. }
 \label{DDFlevel2}
\end{figure}
Let us now compute the vertex operators for  the states at level two. There are two kinds of states: those with the first mode excited twice, $A_{-1}^i A_{-1}^j |0; \tp\ra$, which corresponds to successively scattering two photons, each of momentum $q$, off of a tachyon, see Fig.~\ref{DDFlevel2} (a),  and those with the second mode excited once, $A_{-2}^i |0; \tp\ra$, which corresponds to scattering a photon of momentum $2 q$ off of a tachyon, see Fig.~\ref{DDFlevel2} (b).  We start with the latter,
\be 
: \!\lam\cdot A_{-2}\,   e^{i \tp \cdot X(0)}\!: \,\,  = \oint \frac{d z}{2\pi}\, :\! \lam\cdot\d X(z) \, e^{-2 i q \cdot X(z)}\!:\, \,  :\! e^{i \tp \cdot X(0)}\!:~.
\ee
As before, performing the OPE gives,
\be
 \hspace{-.55cm}  :\! \d X^{\mu}(z) \, e^{-2 i q \cdot X(z)}\!: \,\,  : e^{i \tp \cdot X(0)}\!:\, =\,z^{ - 2\tp\cdot q} \(:\! \d X^{\mu}(z)  e^{-2 i q \cdot X(z)} e^{i \tp \cdot X(0)}\! : -\,\,  \frac{i \tp^{\, \mu}}{z} : \! e^{-i2  q \cdot X(z)} e^{i \tp \cdot X(0)} \!: \)~.
\ee
Using that $\tp \cdot q = 1$, and Taylor expanding $X(z)$ to second order in $z$,  we get for the vertex operator, 
 \be
: \!\lam\cdot A_{-2}\,  e^{i \tp \cdot X}\!:\, \, = i :\! \lam\cdot\d^2 X\, e^{ i p \cdot X}\! :\,  +\, 2 :\! \lam\cdot \d X q \cdot \d X\, \, e^{ i p \cdot X}\! :\, 
 -\, i\lam\cdot \tp :\!q \cdot \d^2 X\, e^{ i p \cdot X}\! :  \, - \, 2 \lam\cdot\tp\, :\! (q \cdot \d X)^2 e^{ ip\cdot X} \!:
\ee
where now  $p_{\mu} = \tp_{\mu} - 2 q_{\mu}$. Expressed in terms of $\zeta$  defined earlier by (\ref{zeta}), 
we have
\be \label{315}
:\! \lam \cdot A_{-2}\,   e^{i \tp \cdot X}\! :\,\,   =\( i \zeta \cdot \d^2 X + 2 (\zeta \cdot \d X) ( q \cdot \d X) \) e^{i p \cdot X}~.
\ee
If we wish, we can define, $\zeta_{\mu \nu} = \zeta_{\mu } q_{\nu} + \zeta_{\nu} q_{\mu}$, and write this vertex operator as, 
\be
:\! \lam \cdot A_{-2}\,   e^{i \tp \cdot X}\! :\,\,  =\( i \zeta_{\mu}  \d^2 X^{\mu} + \zeta_{\mu \nu} \d X^{\mu} \d X^{\nu}\)  e^{i p \cdot X}~, \ \ \  \ \zeta_{\mu \nu}p^{\nu} = \zeta_{\mu}~, \ \ \eta^{\mu \nu} \zeta_{\mu \nu} = p \cdot \zeta =0~.
\ee
In this form, the vertex operator has no explicit $q$ dependence, except through its appearance in the polarizations. This form of the vertex operator is consistent with what one would find in the covariant construction of vertex operators, by imposing the Virasoro constraints, see Appendix~\ref{Virasoro}. 

The other states at level-two are of the form  $(\lam_2\cdot A_{-1})( \lam_1\cdot A_{-1}) |0; \tp\ra$,    see Fig.~\ref{DDFlevel2} (a), with the vertex operator,
\be 
: \lam_2 \cdot A_{-1}\,\,  \lam_1\cdot A_{-1}\,    e^{i \tp \cdot X(0)}\!:\,    =\oint\frac{d z}{2\pi} : \lam_2 \cdot \d X (z) e^{-i q \cdot X(z)}\!:\,  \oint \frac{d w}{2\pi} : \lam_1 \cdot \d X(w) \, e^{- i q \cdot X(w)}\!: \,  : e^{i \tp \cdot X(0)}:
\ee
Physically, we are starting with a tachyon of momentum $\tp$, scattering a photon of momentum $q$ off of it, placing the resulting state on-shell so as to be massless, and then scattering another photon of momentum $q$ off of it, and placing the resulting string on-shell, at mass $M^2 = 2(N-1) = 2$. 
The contour integral over $w$, corresponding to the action of the first $A_{-1}$, is just what we got previously, and gives the vertex operator for a photon. Thus, 
\be 
: \lam_2 \cdot A_{-1}\,\,  \lam_1\cdot A_{-1}\,    e^{i \tp \cdot X(0)}\!:\,    =\oint\frac{d z}{2\pi} : \lam_2 \cdot \d X (z) e^{-i q \cdot X(z)}\!:\, i  : \zeta_1 \cdot \d X(0)\, e^{i (\tp - q) \cdot X(0)}:~,
\ee
where now $\zeta_1 = \lam_1 - (\lam\cdot \tp) q$. We now just need to do the $z$ integral. Performing the OPE gives, 
\bml
\hspace{-.5cm} : \lam_2 \cdot \d X (z) e^{-i q \cdot X(z)}\!:\, : \zeta_1 \cdot \d X(0)\, e^{i (\tp - q) \cdot X(0)}\!:\, 
= z^{- (\t p - q) \cdot q} \[ : \lam_2 \cdot \d X(z)\, \zeta_1 \cdot \d X(0) e^{- i q\cdot X(z)} e^{i (\t p - q) \cdot X(0)} :\right. \\ \left. 
  -\( \frac{\zeta_1 \cdot \lam_2}{z^2} + \frac{i \lam_2 \cdot (\tp - q)}{z} \) : e^{- i q\cdot X(z)} e^{i (\t p - q) \cdot X(0)} :   \]
\end{multline}
The factor out front, $ z^{- (\t p - q)\cdot q}$, is from contracting the exponentials. Within the parenthesis, the first line comes from having no further contractions. On the second line, the first term is from contracting $ \lam_2 \cdot \d X(z)$ with $ \zeta_1 \cdot \d X(0)$, while the second term comes from contracting $\lam_2  \cdot \d X(z)$ with $(\tp - q)\cdot X(0)$ in the exponential. We will see that the second term on the second line can be combined with the term on the first line, by defining $\zeta_2$.  Proceeding, since  only the simple pole contributes to the contour integral, we may pick it out by Taylor expanding, 
\be
e^{- i q \cdot X(z)} = e^{- i q \cdot X(0)} \( 1 - z \, i q \cdot \d X(0)  - i \frac{z^2}{2} q \cdot \d^2 X(0)- \frac{z^2}{2} (q\cdot \d X(0))^2+ \ldots \)~,
\ee
and using $\tp \cdot q = 1$ and $q^2 =0$, along with $\lam_1 \cdot q = \lam_2\cdot q = 0$, which implies $\zeta_1 \cdot \lam_2 = \zeta_1 \cdot \zeta_2$~. Thus we get for the vertex operator, 
\be \label{321}
: \lam_2 \cdot A_{-1}\,\,  \lam_1\cdot A_{-1}\,    e^{i \tp \cdot X}\!:\,    = - \[ \zeta_2 \cdot \d X\,  \zeta_1 \cdot \d X + \frac{1}{2} \zeta_1 \cdot \zeta_2\( i q\cdot \d^2 X + (q\cdot \d X)^2\) \]e^{i p \cdot X}~,
\ee
where
\be
\zeta_1 = \lam_1 - (\lam_1\cdot p) q~, \ \ \ \ \zeta_2 =  \lam_2 - (\lam_2\cdot p) q~, \ \ \ \ p = \tp - 2 q~.
\ee
In the form (\ref{321}), the arbitrary vector $q$ explicitly appears. However
if we wish  we can make its appearance only implicit, by writing the vertex operator in the form, 
\be
-\frac{1}{2}\( i \xi_{\mu}  \d^2 X^{\mu} + \xi_{\mu \nu} \d X^{\mu} \d X^{\nu}\)  e^{i p \cdot X}~
\ee
where
\be \label{324}
\xi_{\mu} = \zeta_1 \cdot \zeta_2\, q_{\mu}~, \ \ \ \ \ \ \ \ \  \ \ \xi_{\mu \nu} = \zeta_1 \cdot \zeta_2\,  q_{\mu} q_{\nu} +\zeta_{1, \mu} \zeta_{2, \nu} + \zeta_{1, \nu} \zeta_{2, \mu}~.
\ee
Furthermore, these polarization vectors satisfy $\xi_{\mu \nu} p^{\nu}= \xi_{\mu}$ and  $\eta^{\mu \nu} \xi_{\mu \nu} = 2 p^{\mu}\xi_{\mu} = - \zeta_1 \cdot \zeta_2$. As a result, the Virasoro constraints that a covariant vertex operator at level-two must obey, see Eq.~\ref{B9} in Appendix \ref{Virasoro}, are obeyed.

\subsection{Level-$N$ states} \label{sec32}
In this section we construct all the vertex operators at level $N$.

\subsubsection*{A single creation operator}
\begin{figure}
\centering
\includegraphics[width=2.2in]{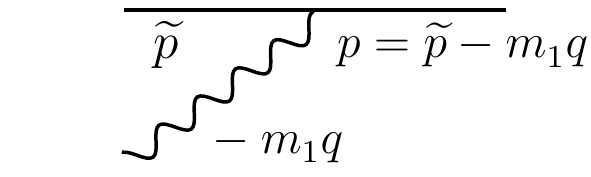}
\caption{Forming the state $\lam \cdot A_{-m_1} |0\ra$. } \label{DDFleveln}
\end{figure}
We start by constructing the vertex operator for the state $\lam \cdot A_{-m_1} |0; p\ra$, see Fig.~\ref{DDFleveln}. We did this previously for $m_1=1$ and $m_1=2$. Here we do general $m_1$. 
Applying the DDF $A_{-m_1}$  given in  (\ref{DDF2}) to the tachyon vertex operator, we must compute,
\be  
:\!\lam\cdot A_{-m_1}\, e^{i \tp \cdot X(0)}\! :\,\,= \oint \frac{dz}{2\pi} : \lam\cdot\d X(z) e^{-i m_1 q\cdot X(z)}\!: \,\, :\! e^{i \tp \cdot X(0)}:~.
\ee
To evaluate the contour integral, we need to first evaluate the OPE, 
which is given by, 
\be \label{327}
 :\!\lam\cdot A_{-m_1}\, e^{i \tp \cdot X(0)}\!:\,\, =\oint \frac{dz}{2\pi} z^{-m_1} \( : \lam\cdot \d X(z)e^{-i m_1 q\cdot X(z)} e^{i \tp \cdot X(0)}:   - \frac{i \lam\cdot \t p}{z}:e^{-i m_1 q\cdot X(z)} e^{i\tp \cdot X(0)}:\)~,
\ee
where we made use of $p \cdot \t q = 1$. \\[-5pt]

{\it  Schur polynomials}\\[-10pt]

To proceed, we will need to Taylor expand $X(z)$ about $X(0)$. More precisely, we need to Taylor expand the exponential of $X(z)$, which  will give rise to Schur polynomials. Indeed, the Schur polynomials can be defined through a Taylor expansion of the exponential of a series, 
 \be \label{SchurP}
 \exp\(\sum_{m=1}^{\infty} a_m z^m\)  = \sum_{m=0}^{\infty} S_m(a_1, \ldots, a_m) z^m~.
 \ee
 The left-hand side serves as the definition of the Schur polynomials $S_m(a_1, \ldots, a_m) $ appearing on the right-hand side. Equivalently, one may invert the expression through a contour integral, 
  \be \label{SchurP2}
 S_m(a_1, \ldots, a_m) = \oint_0 \frac{d w}{2\pi i} \frac{1}{w^{m+1}} \exp\( \sum_{s=1}^m a_s w^s\)~.
 \ee
 In our case, the series that what we have is the Taylor expansion of $X(z) = \sum_r \frac{z^r}{r!} \d^r X(0)$.  We therefore have, 
   \be  \label{329}
    e^{- i m_1 q \cdot X(z)} = \sum_{a=0}^{\infty} z^a\, S_a\( - \frac{i m_1}{r!} q\cdot \d^r X \)\, e^{ - i m_1 q\cdot X(0)}~.
 \ee
Equivalently, inverting the expression through a contour integral, 
 \be \label{Smnq}
 S_m\( - \frac{i m_1}{r!} q\cdot \d^r X \) = \oint_0\frac{d w}{2\pi i} \frac{1}{w^{m+1}} \exp\( - i m_1 q \cdot \sum_{s=1}^m \frac{w^s}{s!} \d^s X(0)\)~.
 \ee
 The sum over $s$ on the right only needs to go up to $s=m$, however one can extend it to $s=\infty$ if one wishes. 
The Schur polynomial $ S_m\( - \frac{i m_1}{r!} q\cdot \d^r X \)$ depends on $m$ variables,
\be
S_m\( - \frac{i m_1}{r!} q\cdot \d^r X \) \equiv S_m\( - i m_1 q\cdot \d X,  \frac{-i m_1}{2!} q \cdot \d^2 X, \ldots,  \frac{-i m_1}{m!} q\cdot \d^m X\)~,
\ee
and our notation on the left, with the $r$ index, is shorthand for this. By simply performing the Taylor expansion in (\ref{329}), we see that the first few Schur polynomials are,
\bea \nn
S_0( - \frac{i m_1}{r!} q\cdot \d^r X) &=& 1\\ \nn
S_1( - \frac{i m_1}{r!} q\cdot \d^r X) &=& - i m_1 q\cdot \d X \\
S_2( - \frac{i m_1}{r!} q\cdot \d^r X) &=&  - \frac{1}{2}\( i m_1\, q \cdot \d^2 X +m_1^2 (q \cdot \d X)^2 \)~.
\eea
\\
\indent Returning to our construction of the vertex operator, the expression that we need to Taylor expand in the first term in (\ref{327}) is a product $\d X(z)$ and an exponential. Correspondingly, we do a double Taylor expansion, 
 \be \label{333}
  \d X^{\mu} (z)e^{-i m_1 q\cdot X(z)}  = \sum_{a, b=0}^{\infty} \frac{z^{a+b}}{a!} \d^{a+1} X^{\mu}(0)\,  S_b\(- \frac{i m_1}{r!} q\cdot \d^r X(0)\)\, e^{ - i m_1 q\cdot X(0)}~.
  \ee
  The contour integral in (\ref{327}) then pick out the terms with $a+b = m_1-1$. The vertex operator of the state $\lam \cdot A_{-m_1} |0; \tp\ra$ is thus, 
      \be
  :  \lam \cdot A_{-m_1} \, e^{i \tp \cdot X}\!\!:\,  = \sum_{m=1}^{m_1} \frac{i}{(m-1)!} : \lam \cdot \d^m X S_{m_1-m}\(\! - \frac{i m_1}{r!} q{\cdot} \d^r X \) e^{i p \cdot X}:+  :\lam \cdot p\, S_{m_1}\( \!- \frac{i m_1}{r!} q{\cdot} \d^r X \)e^{i p \cdot X}:
  \ee
  where  $p = \tp - m_1 q$. Using a property of Schur polynomials,
  \be
S_{m_1}(a_1, \ldots, a_{m_1}) =\sum_{m=1}^{m_1} \frac{m}{m_1} a_m S_{m_1- m}(a_1, \ldots, a_{m_1})~,
\ee
we can write the vertex operator in the more compact form, 
\be \label{336}
:  \lam \cdot A_{-m_1} \, e^{i \tp \cdot X}\!:\, \,  = \sum_{m=1}^{m_1} \frac{i}{(m-1)!} \zeta \cdot \d^m X\, S_{m_1- m}\( - \frac{i m_1}{r!} q\cdot \d^r X \)\,  e^{i p \cdot X}\ \equiv \zeta \cdot P_{m_1}\, e^{i p \cdot X}
\ee
where, as before,  $\zeta_{\mu}$ is given by (\ref{zeta}), $\zeta_{\mu}  = \lam_{\mu} - (\lam \cdot p) q_{\mu}$ and $S_m$ was defined in (\ref{Smnq}). For $m_1=1$ and $m_1=2$, this gives back what we computed previously, (\ref{310}) and (\ref{315}), respectively. \\

\subsubsection*{Two creation operators}
We now turn to states that are formed by two creation operators, $\lam_2 \cdot A_{-m_2}  \lam_1\cdot A_{-m_1} |0; \tp\ra$. We have already done this for the case of $m_2 = m_1=1$, see Eq.~\ref{321}. The procedure in this more general case is similar. 

Starting with the tachyon vertex operator and acting twice with  $A_{-m}$  gives the vertex operator for these states, 
\be 
: \lam_2 \cdot A_{-m_2}\,\,  \lam_1\cdot A_{-m_1}\,    e^{i \tp \cdot X(0)}\!:\,    =\oint\frac{d z}{2\pi} : \lam_2 \cdot \d X (z) e^{-i m_2 q \cdot X(z)}\!:\,  \oint \frac{d w}{2\pi} : \lam_1 \cdot \d X(w) \, e^{- i m_1 q \cdot X(w)}\!: \,  : e^{i \tp \cdot X(0)}:
\ee
For the first application of $A_{-m_1}$ (the $w$ integral) we may use the result (\ref{336}), 
\bml
: \lam_2 \cdot A_{-m_2}\,\,  \lam_1\cdot A_{-m_1}\,    e^{i \tp \cdot X(0)}\!:\,    =\oint\frac{d z}{2\pi} : \lam_2 \cdot \d X (z) e^{-i m_2 q \cdot X(z)}\!:\,  \\
\sum_{m=1}^{m_1} : \frac{i}{(m-1)!} \zeta_1 \cdot \d^m X(0)\, S_{m_1- m}\( - \frac{i m_1}{r!} q\cdot \d^r X(0) \)\,  e^{i (\tp - m_1 q) \cdot X(0)} :
\end{multline}
where $\zeta_{1} = \lam_{1} -(\lam_1 \cdot \tp) q$. In computing the remaining OPE, the  contractions we need to account for are $\lam_2 \cdot \zeta_1 = \zeta_2 \cdot \zeta_1$ and $\lam_2 \cdot (\tp - m_1 q) = \lam_2 \cdot p$. The other possible contraction, with the arguments of the Schur polynomial, is zero because $\lam_2 \cdot q = 0$. Hence we have, 
\bml
: \lam_2 \cdot A_{-m_2}\,\,  \lam_1\cdot A_{-m_1}\,    e^{i \tp \cdot X(0)}\!:\,    =\sum_{m=1}^{m_1}  \frac{i}{(m-1)!} \oint\frac{d z}{2\pi} z^{-m_2}\\ \Big[ :  \lam_2 \cdot \d X (z) e^{-i m_2 q \cdot X(z)}  \zeta_1 \cdot \d^m X(0)\, S_{m_1- m}\( - \frac{i m_1}{r!} q\cdot \d^r X(0) \)\,\, e^{i (\tp -  m_1 q) \cdot X(0)}: \Big. \\ \Big. 
- \( \zeta_1 \cdot \zeta_2\frac{ m!  }{z^{m+1}} +\frac{i \lam_2 \cdot \t p}{z} \) :  e^{-i m_2 q \cdot X(z)} S_{m_1- m}\( - \frac{i m_1}{r!} q\cdot \d^r X(0) \)e^{i (\tp -  m_1 q) \cdot X(0)}: \Big]
\end{multline}

As before, Taylor expanding in $z$ in order to pick out the simple pole, 
we may write the result in the compact form, 
\be  \label{339}
: \lam_2 \!\cdot\! A_{-m_2}\,\,  \lam_1\!\cdot\! A_{-m_1}\,    e^{i \tp \cdot X(0)}\!:\, \,   = \( \zeta_2\! \cdot\! P_{m_2}\, \zeta_1 \!\cdot\! P_{m_1} + \zeta_1 \!\cdot\! \zeta_2\,  \, \mbS_{m_1, m_2} \)   e^{i p \cdot X}
\ee
where $\zeta\cdot P_n$ was defined in (\ref{336}), and $\zeta_i = \lam_i - (\lam_i \cdot p) q$ and $p = \tp - (m_1 + m_2) q$, and 
 \be \label{340}
 \mbS_{m_1, m_2} =-i \sum_{m=1}^{m_1} m \oint\frac{d z}{2\pi}  \frac{1}{z^{m_2 + m+1}} \sum_{a=0}^{\infty} z^a\, S_a\( - \frac{i m_2}{r!} q\cdot \d^r X(0) \)\, S_{m_1- m}\( - \frac{i m_1}{r!} q\cdot \d^r X(0) \)~.
 \ee
 Performing the contour integral sets $a= m_2 + m$, and we get, 
  \be \label{342}
 \mbS_{m_1, m_2} =\sum_{m=1}^{m_1}m\,  S_{m_1- m}\( - \frac{i m_1}{r!} q\cdot \d^r X(0) \)\, S_{m_2+m}\( - \frac{i m_2}{r!} q\cdot \d^r X(0) \)
 \ee
 One can see that in the case of $m_1=m_2 = 1$, the general result (\ref{339}) reduces to what we found previously in (\ref{321}). \\

\subsubsection*{Multiple creation operators}
One finds that the vertex operators involving three creation operators is a natural generalization of the one with two creation operators (\ref{339}) that was just found, 
\bml \label{342v2}
: \lam_3 \!\cdot\! A_{-m_3}\,\,\lam_2 \!\cdot\! A_{-m_2}\,\,  \lam_1\!\cdot\! A_{-m_1}\,    e^{i \tp \cdot X(0)}\!:\, \,   
= \( \zeta_3\! \cdot\! P_{m_3}\, \zeta_2\! \cdot\! P_{m_2}\, \zeta_1 \!\cdot\! P_{m_1} \right. \\ \left. +\, \zeta_1 \!\cdot\! \zeta_2\,  \, \mbS_{m_1, m_2} \, \zeta_3\! \cdot\! P_{m_3} + \zeta_1 \!\cdot\! \zeta_3\,  \, \mbS_{m_1, m_3} \, \zeta_2\! \cdot\! P_{m_2}+ \zeta_2 \!\cdot\! \zeta_3\,  \, \mbS_{m_2, m_3} \, \zeta_1\! \cdot\! P_{m_1}\)   e^{i p \cdot X}~.
\end{multline}
Analogously, the result for a general vertex operator is fairly clear, 
\bea \nn
\!\!\!: \lam_k \!\cdot\! A_{-m_k}\cdots \lam_2 \!\cdot\! A_{-m_2}\,\,  \lam_1\!\cdot\! A_{-m_1}\,    e^{i \tp \cdot X(0)}\!:\!\!\!&=&\!\!\!\! (\text{sum over any number of all possible Wick contractions}) e^{i p\cdot X}\\ \nn
&&\ \ \ \ \ \ \ \ \ \ \text{          where}\\ \nn
&&\text{if contract: }  \la \lam_i \cdot A_{- m_i}\, \lam_j \cdot A_{-m_j}\ra \rightarrow \zeta_i \cdot \zeta_j\, \mbS_{m_i, m_j}\\ \nn
&&\text{if don't contract: } \ \lam_i \cdot A_{- m_i}\rightarrow \zeta_i \cdot P_{m_i} ~,
\eea
where $\zeta\cdot P_m$ was given in (\ref{336}) and $\mbS_{m_1, m_2}$ was given in (\ref{342}). 

In more precise notation \cite{Skliros}, 
\be \label{344}
\!:\! \lam_k \cdot A_{-m_k}\cdots \lam_2 \cdot A_{-m_2}\,\,  \lam_1\cdot A_{-m_1}\,    e^{i \tp \cdot X(0)}\!:\, =e^{i p\cdot X}\! \sum_{a=1}^{\lfloor k/2 \rfloor} \!\sum_{\pi} \prod_{l=1}^a ( \zeta_{\pi(2l-1)} \cdot \zeta_{\pi(2 l)} )\, \mbS_{m_{\pi(2l-1)}, m_{\pi(2l)}} \! \prod_{q=2a+1}^k \! \zeta_q\cdot P_{m_{\pi(q)}}
\ee
where we are summing over $a$ (the number of contractions), $\lfloor k/2 \rfloor$ denotes all integers less than or equal to $k/2$, and  the sum over $\pi$ is over all permutations $\pi$ that give non-equivalent terms. By non-equivalent terms we mean, e.g. if $a=2$, then the sum over $l$ is from $l=1$ to $l=2$, and includes the three terms, 
\be
\zeta_1 \cdot \zeta_2\,\, \zeta_3\cdot \zeta_4\,\, \mbS_{m_1, m_2} \mbS_{m_3, m_4} + \zeta_1 \cdot \zeta_3\,\, \zeta_{2} \cdot \zeta_4\,\, \mbS_{m_1, m_3} \mbS_{m_2, m_4}+ \zeta_1 \cdot \zeta_4\,\, \zeta_{2} \cdot \zeta_3\,\, \mbS_{m_1, m_4} \mbS_{m_2, m_3}
\ee
The number of permutations of $4$  is $4!=24$, however since $\mbS_{m_1,m_2}= \mbS_{m_2,m_1}$ we must divide by a factor of $2$ for each of the two $\mbS$ that we have, and another factor of two from exchanging the two $\mbS$. Thus, in total there are $4!/2^3= 4$ distinct terms, as we clearly see above. \\

\subsubsection*{Derivation of Eq.~\ref{344}}
Let us now demonstrate that (\ref{344}) is correct. This requires only a slight generalization of the argument used in deriving the 
vertex operators with two creation operators. Let us compute the vertex operator corresponding to the action of a creation operator acting on some operator $f$, 
\be
:\! \lam\cdot A_{-m}\, f(\zeta_i \cdot \d^r X, q \cdot \d^r X)\, e^{i \t p\cdot X(0)}\!:\,\, = \oint\frac{d z}{2\pi} : \lam \cdot \d X(z)\, e^{- i m q \cdot X(z)}\! :\, \, :\, f(\zeta_i \cdot \d^r X(0), q \cdot \d^r X(0))\, e^{i \t p\cdot X(0)}\!:
\ee
where we have used the definition of $A_{-m}$. There are two possible Wick contractions: a contraction of $\lam\cdot \d X$ with $f$, and a contraction of $\lam \cdot \d X$ with $ \tp \cdot X$. Thus, 
\bml
:\! \lam\cdot A_{-m}\, f(\zeta_i \cdot \d^r X, q \cdot \d^r X)\, e^{i \t p\cdot X}\!:\,\, = \oint\frac{d z}{2\pi} \frac{1}{z^m} 
\[ :\!\lam \cdot \d X(z)\, e^{- i m q \cdot X(z)}\, f\, e^{i \t p\cdot X(0)}\!:  \right. \\ \left.  + \la \lam \cdot \d X(z)\, i \tp\! \cdot\! X(0)\ra\, :\! e^{- i m q\cdot X(z)} f e^{i \tp \cdot X(0)}\!:\,  + \la \lam \cdot \d X(z) f\ra\, e^{- i m q\cdot X(z)} e^{i \tp \cdot X(0)}\]
\end{multline}
We trivially evaluate the first correlation function on the second line,  rewriting the result as, 
\bml
\hspace{-.5cm} :\! \lam\cdot A_{-m}\, f(\zeta_i \cdot \d^r X, q \cdot \d^r X)\, e^{i \t p\cdot X}\!:\,\, =\!\! \oint\!\frac{d z}{2\pi} \frac{1}{z^m}\! \(\! :\!\lam\! \cdot \! \d X(z)\, e^{- i m q \cdot X(z)}\, f\, e^{i \t p\cdot X(0)}\!:   - \frac{i \lam{\cdot} p}{z} \!:e^{- i m q\cdot X(z)} f e^{i \tp \cdot X(0)}\!:\! \)\\
 + \oint\frac{d z}{2\pi} \frac{1}{z^m}   \la \lam \cdot \d X(z) f\ra\, e^{- i m q\cdot X(z)} e^{i \tp \cdot X(0)}~.
\end{multline}
Comparing the first line with what we had when evaluating $:\!\lam\cdot A_{-m}\,  e^{i \t p\cdot X}\!:$ in (\ref{327}), we see that it is identical, except for the extra factor of $f$. To evaluate the second line, we note that, 
\be
f = \prod_i \zeta_i \cdot P_{m_i}\, g(q \cdot \d^r X)~,
\ee
where $g$ is some function. 
Since $\lam \cdot q = 0$, the $g$ term has vanishing contraction with $\lam \cdot \d X$. Thus, 
\be
  \la \lam \cdot \d X(z) f\ra = 
\sum_j \la \lam \cdot \d X(z)\,  \zeta_j \cdot P_{n_j}(0)\ra  \prod_{i \neq j} \zeta_i \cdot P_{m_i}(0)\,\, g(q \cdot \d^r X)~.   \ee
Using (\ref{340}), we have that, after evaluating the integral over $z$, the contraction will turn into $\mbS_{m, m_j}$. Thus, 
\be
:\! \lam\cdot A_{-m}\, f(\zeta_i \cdot \d^r X, q \cdot \d^r X)\, e^{i \t p\cdot X}\!:\,\,  = \zeta \cdot P_m\, \prod_i \zeta_i \cdot P_{m_i}\, \, g(q \cdot \d^r X)\, e^{i p\cdot X} + 
\sum_j \mbS_{m, m_j}  \prod_{i \neq j} \zeta_i \cdot P_{m_i}\, \, g(q \cdot \d^r X)\,  e^{i p\cdot X}~.
\ee
So, as claimed,  we  have a sum over all Wick contractions. The result (\ref{344}) then follows. This completes the evaluation of the general DDF vertex operator. 

\subsubsection*{The generic state}
We have discussed the construction of any excited string state. However, for a string of given mass, the number of different states grows exponentially with the mass. It is useful to have a sense of what the typical state looks like. 
In $D$ space-time dimensions, there are $D-2$ independent polarization vectors and a  state can be written as ,
\be  
\prod_{k=1}^{n_1}(\lam^1_k \cdot A_{-1} )\prod_{k=1}^{n_2} (\lam^2_k\cdot A_{-2}) \cdots \prod_{k=1}^{n_r}(\lam^r_k \cdot A_{-r}) |0\ra~, \ \ \ \ \ N = \sum_{m=1}^r m n_m~.
\ee
In Appendix~\ref{sec5} we  show that, in the large $N$ limit,  the typical (or equivalently, average) occupation number $n_m$ of mode $m$ takes the form of a Bose-Einstein distribution $\la n_m\ra = \frac{1}{e^{m/T} - 1}$, with a temperature $ T=\frac{1}{\pi} \sqrt{\frac{6N}{D-2}}$.

\subsection{Completeness of operators }

The final thing that we need to show  is that the construction described above does in fact generate all the excited states. This is what we do in this section. 

Recall that, while classifying excited states in covariant gauge is cumbersome, it is straightforward in light-cone gauge. In light-cone gauge one chooses $X^+$ to be the time variable. In the classical description, this amounts to setting the oscillators coefficients $\al_n^+$ (see Eq.~\ref{Xmu}) to be zero for all $n\neq 0$.  The state is then completely determined by the specification of the $\al_n^i$ with transverse $i$, as the Virasoro constraints fix $\al_n^- $ in terms of the $\al_n^i$. The states we have been working with are formed with products of $\lam \cdot A_{-m}$, the left-hand side of (\ref{344}). Since $\lam \cdot q = 0$ and we can add any multiple of $q$ to $\lam$, see (\ref{36}), the polarization $\lambda$ has $D-2$ independent components, which is the same as the number of transverse directions in light-cone gauge.  Thus, we have the correct number of states. 

\begin{figure}
\centering
\includegraphics[width=2.5in]{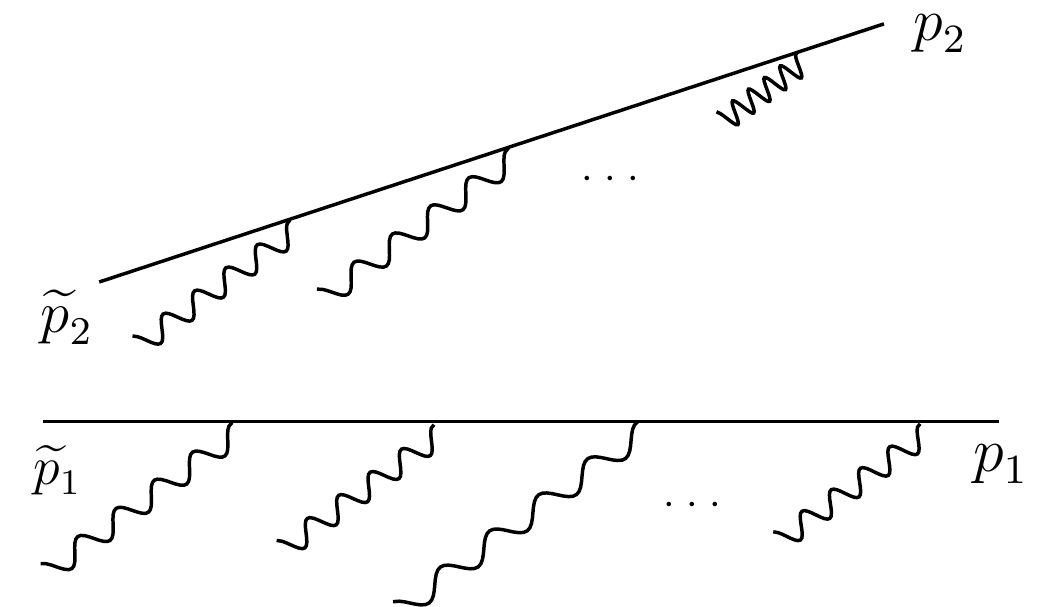}
\caption{When forming several excited string states with momenta $p_i$, we start with tachyons with momenta $\tp_i$ and add photons, all of which have momenta in the same direction.} \label{DDFfig2}
\end{figure}

We formed the general vertex operator by starting with a tachyon of momentum $\tp$ and repeatedly scattering photons off of it. All the photons had momentum proportional to some arbitrarily chosen vector $q$, which was required to satisfy $\tp \cdot q$. We need to show that this is sufficient to generate any momentum $p$. This is easy to see. Let us work in light-cone coordinates, and go to the frame in which $q$ points in the minus direction, $ q = (q^+, q^-, q^i) = (0, q^-, \vec{0})$. Since $q$ is null, $q^2 =0$, this is possible. The tachyon has momentum $\tp = (\tp^{\, +}, \tp^{\, -}, \tp^{\, i})$, which is arbitrary, subject to the mass-shell condition $\tp^{\, 2} = 2$. Imposing that $q \cdot \tp = 1$ allows us to fix the magnitude of $q$, 
\be
q = - \frac{1}{\tp^{\, +}} (0,1,\vec{0})~, \ \ \  \ \ \ \tp = (\tp^{\, +}, \tp^{\, -}, \tp^{\, i})~, \ \ \ \ \tp^{\, 2}= 2~.
\ee
We can view $\tp^{\, +}, \tp^{\, i}$ as fully specifying $\tp$, with the remaining component, $\tp^{\, -}$, fixed through the mass shell condition; the components $\tp^{\, +}, \tp^{\, i}$ can be anything. By adding an arbitrary multiple of $q$ to $\tp$, we get a vector in which the minus component is now also arbitrary, so we have formed an arbitrary $D$-dimensional vector. If we take the particular vector $p = \tp - N q$, then $p^2 = \tp^{\, 2} - 2 N \tp\cdot q = -2(N-1)$. So, we have shown that we can achieve any excited string state momentum $p$ by scattering photons off of a tachyon, where all the photons have momenta in the same direction. Moreover, if we are forming multiple heavy string states, with different momenta, photons in the same direction are added to all of them. In other words, if we have two heavy strings with momenta $p_1$ and $p_2$, then we would form them by starting with tachyons of momenta $\tp_1$ and $\tp_2$ and scattering photons off of them with momenta that are multiples of $q_1 = - \frac{1}{\tp_1^{\, +}} (0,1,\vec{0})$ and $q_2 = - \frac{1}{\tp_2^{\, +}} (0,1,\vec{0})$, respectively. See Fig.~\ref{DDFfig2}.

\subsection{Summary}
We end with a summary of the results for the vertex operator of an excited string state. \\

The vertex operator for a tachyon of momentum $p$ is $e^{i p\cdot X}$. The vertex operator for an excited string of momentum $p$ is some polynomial of derivatives of $X$, $\d^{k} X^{\mu}$, multiplying $e^{i p\cdot X}$. 
Schematically, 
\be \label{Vgen}
V= e^{i p \cdot X}
 \sum_{\{m_r\}}  c_{\{m_r\}} \prod_r \d^{m_r} X^{\mu_r}~,
 \ee
 with some coefficients $c_{\{m_r\}}$, which can be thought of as the polarizations. At level $N$, corresponding to a string of mass $M^2 = 2(N-1)$, we include all terms with a total of $N$ derivatives. For instance, at level two, the vertex operator is of the form, 
 $\( i \xi_{\mu}  \d^2 X^{\mu} + \xi_{\mu \nu} \d X^{\mu} \d X^{\nu}\)  e^{i p \cdot X}$. Imposing the Virasoro constraints gives sets of constraints that the polarizations must satisfy. For the first few levels these constraints are worked out in Appendix~\ref{Virasoro}. For general level $N$, the vertex operator is a superposition of a large number of terms, with a large number of nontrivial constraints among the coefficients. Writing the operator in this form is neither trivial nor intuitive.

 The DDF construction of vertex operators provides a more physical and systematic procedure. The coefficients $c_{\{m_r\}}$ mentioned above are all expressed in terms of a null vector $q$ which satisfies $p\cdot q =1$,~\footnote{The appearance of an arbitrarily chosen vector $q$ in the expression for the vertex operator may initially seem odd, but it is no different than the appearance of an arbitrary polarization vector in the vertex operator. Expressed in covariant form, the $q$ are part of defining the polarization tensors. See e.g. (\ref{324}).}  and polarizations $\lam_i$ which have nonzero norm and are orthogonal to $q$,  $\lam_i \cdot q=0$. The vertex operator is expressed as a polynomial involving $q\cdot \d^m X$ and $ \zeta_i \cdot \d^m X$ where $\zeta_i^{\mu} = \lam_i^{\mu} - (p\cdot q) q^{\mu}$. The precise form of the vertex operator is given in (\ref{344}), with $\zeta\cdot P_n$ (a function of  $q\cdot \d^m X$ and $ \zeta_i \cdot \d^m X$) given in (\ref{336}) and $\mbS_{n_1, n_2}$ (a function of $q\cdot \d^m X$) given in (\ref{342}). 
 
 Physically, the vertex operators are formed by starting with a tachyon of momentum $\tp$ and scattering photons off of it (i.e. repeatedly taking the OPE with photon vertex operators) where the action of $\lam \cdot A_{-m}$ corresponds to a photon with polarization $\lam$ and momentum $ -m q$ where $q\cdot \tp = 1$. The $\zeta_{\mu}$ can be viewed as the polarization vectors of the excited string. Doing the scattering/OPE procedure once gives a vertex operator of the form (\ref{336}). Doing the procedure twice gives the vertex operator (\ref{339}), doing it three times gives (\ref{342v2}), and doing it $k$ times gives (\ref{344}).

\section{Amplitude with one excited string: formalism} \label{sec4}

In this section we  compute the scattering amplitude involving one arbitrary excited string state and any number of tachyons. 

The excited state is created with the DDF vertex operator, as described in the previous section: the state $\prod_k (\lam_k \cdot A_{-m_k}) |0;\tp\ra$ can be thought of as having been built up by successively scattering photons of polarization $\lam_k$ and momenta $-m_k q$  off of a tachyon, where $q$ is some chosen null vector satisfying $q\cdot \tp=1$ (and the photon's polarization is orthogonal to its momentum, $ \lam\cdot q = 0$).

 We start in Sec.~\ref{sec41} with the special case in which all the DDF photons have the same polarization which squares to zero, $\lam^2=0$ (this can be achieved by, for instance, having transverse circular polarization). In this special case the equations for the amplitude simplify. In Sec.~\ref{sec42} we compute the amplitude in the general case, with arbitrary polarizations for the DDF photons. \\

Let $V(z_1)$ be the vertex operator for the the excited string with momentum $p_1$. All the other strings are tachyons with momentum $p_i$, worldsheet coordinate $z_i$, and vertex operator $e^{i p_i \cdot X(z_i)}$. There are $n$ tachyons, with index $i$ running from $2$ to $n+1$.  The amplitude is then,
\be \label{41}
\mA = \frac{1}{\text{vol}(SL_2)} \int d z\, \la V(z_1) \prod_{i\neq 1} e^{i p_i \cdot X(z_i)}\ra~, \ \ \ dz\equiv \prod_i d z_i~.
\ee
We wrote down the general vertex operator, in schematic form, in (\ref{Vgen}).  The vertex operator is given by a sum of products of derivatives of $X$. Evaluating the correlation function appearing in the amplitude is straightforward, once one recognizes that the Wick contractions can be perform successively, for each $X$. Namely, one has that, 
\be
\la \d^n X^{\mu}(z_1)   \prod_{i\neq 1} e^{i p_i \cdot X(z_i)} \cdots \ra = \la   \prod_{i\neq 1} e^{i p_i \cdot X(z_i)} \cdots \ra \sum_{i \neq 1} \la \d^n X^{\mu}(z_1)  (i p_i \cdot X(z_i)\ra~,
\ee
as one can see by using the Taylor expansion of the exponential. Using the form of the vertex operator (\ref{Vgen}) we have that the correlation function is, 
\be
\la  V(z_1) \prod_{i\neq 1} e^{i p_i \cdot X(z_i)}\ra = \prod_{i<j} z_{i j}^{p_i \cdot p_j} \sum_{\{n_r\}}  c_{\{n_r\}} \prod_r ( \sum_{i\neq1} \la \d^{n_r} X^{\mu_r}  (i p_i \cdot X(z_i)\ra )~.
 \ee
Using that the Wick contraction is given by, 
\be
\la \d^k X^{\mu} (z_1) X^{\nu} (z_2)\ra = \eta^{\mu \nu}\frac{(k-1)!}{z_{21}^k}~,
\ee
we have for the amplitude (\ref{41}), 
\be
\mA = \frac{1}{\text{vol}(SL_2)} \int d z\,  \prod_{i<j} z_{i j}^{p_i \cdot p_j}\ \sum_{\{n_r\}}  c_{\{n_r\}} \prod_r \( \sum_{i\neq1} i p_i^{\mu_r} \frac{(n_r - 1)!}{z_{i 1}^{n_r}} \)~.
\ee

This warmup was useful as a way to see how to perform the Wick contractions. We saw that going from the vertex operator to the amplitude essentially amounts to replacing,
\be
 \d^m X^{\mu}(z_1) \rightarrow i (m-1)!\sum_{i \neq 1} \frac{p_i^{\mu}}{z_{i 1}^m}~.
 \ee
We now move on to using the actual vertex operator, rather than  just its schematic form. 

\subsection{A special choice of polarizations} \label{sec41}
In this section we give the explicit form of the amplitude, for the  simple case  in which all the DDF photons forming the the excited state have the same polarization $\lam$, 
\be \label{829}
(\lam \cdot A_{-1})^{n_1} (\lam\cdot A_{-2})^{n_2} \cdots (\lam \cdot A_{-k})^{n_k} |0\ra~, \ \ \ \ \ N = \sum_{i=1}^k i n_i~.
\ee
Furthermore, we take  the polarization to square to zero, $\lam^2 = 0$.~\footnote{The polarization is always required to satisfy $|\lam|^2 = 1$; we are taking $\lam^2=0$.}  An example of such a polarization is, for instance, circular polarization with two nonzero components, $\frac{1}{\sqrt{2}}(1, \pm i)$. (It is actually not important if all the photons have the same polarization: to get a simplification, what we really need is for the polarizations of all the photons to square to zero and be orthogonal, and the simplest way to achieve that is with one $\lam$ for which $\lam^2=0$.) With $\lam^2 = 0$, the form of the vertex operator corresponding to the state (\ref{829}) is relatively simple, and 
was found earlier in (\ref{336}), 
\be \label{336v2}
:  \lam \cdot A_{-n} \, e^{i \tp \cdot X}\!:\,\, = \sum_{m=1}^n \frac{i}{(m-1)!} \zeta \cdot \d^m X\, S_{n- m}\( - \frac{i n}{r!} q\cdot \d^r X \)\,  e^{i p \cdot X}~,
\ee
where the Schur polynomial that appears was defined in (\ref{Smnq}). 
Using the above results, the amplitude for the scattering of the state (\ref{829}) with tachyons is, 
\be \label{48}
\mA = \frac{1}{\text{vol} (SL_2)}\int dz\,  \exp\(\mL\)~,
\ee
where 
\be \label{930}
\mL = \sum_{i< j} p_i \cdot p_j \log z_{i j} + \sum_{m= 1}^k n_m \log (-\Sigma_{m})~,
\ee
and
\be \label{931}
\Sigma_{n} =  \sum_{m=1}^{n} \,\( \sum_{j\neq 1} \frac{p_j \cdot \zeta}{z_{j 1}^m} \) S_{n-m}\( n\sum_{i\neq 1} \frac{p_i \cdot q}{s\, z_{i 1}^s}\)~,
\ee
where $p_1$ is the momentum of the excited string and $p_i$, with $i \geq 2$,  are the momenta of the tachyons. The amplitude also has a momentum conservation delta function which we suppressed, $(2\pi)^D  \delta(\sum_i p_i) $. 
The result is expressed in term of Schur polynomials, however in this form it appears unnecessarily complicated. In particular, we may use the contour integral representation of the Schur polynomial given in (\ref{SchurP2}) to write this as, 
\be \label{SchurInt}
S_m\( n\sum_{i\neq 1} \frac{p_i \cdot q}{s\, z_{i 1}^s}\) = \oint \frac{dw}{2\pi i} \frac{1}{w^{m+1}}  \prod_{i\neq 1} \(1 - \frac{w}{z_{i 1}}\)^{- n p_i \cdot q}~,
\ee
where inside the integral representation of the Schur polynomial we extended the sum to infinity and performed the sum, 
\be
\sum_{s=1}^{\infty} \sum_{i\neq 1} \frac{n p_i \cdot q}{s\, z_{i 1}^s} w^s = -\sum_{i \neq 1} n\,  p_i\! \cdot \!q\,  \log\( 1 - \frac{w}{z_{i 1}}\)~.
\ee
If we wish, we may now perform the sum over $m$ appearing in $\Sigma_n$,~\footnote{In performing the sum, we obtain a factor of $\frac{1 - (z_{i1}/w)^n}{w - z_{i1}}$. We discard the $1$ in the numerator, since its contribution will vanish after the contour integral is done.}
\be \label{413}
\Sigma_n = -\oint \frac{d w}{2\pi i }\frac{1}{w^n} \(\sum_{j\neq 1} \frac{p_j \cdot \zeta}{w- z_{j1}} \) \prod_{i\neq1} \(1 - \frac{w}{z_{i 1}}\)^{- n p_i \cdot q}~.
\ee

The form of the amplitude as written is acceptable, but it is not optimal. In particular, the integrand in the amplitude must have SL$_2$ invariance of the $z_i$, as we discussed earlier in Sec.~\ref{Sec22}. In the current form, the SL$_2$ invariance is not manifest. In order to make it manifest, we should rewrite the amplitude in terms of SL$_2$-invariant cross-ratios. As we will momentarily show, upon doing this  we find the amplitude to be (\ref{48}) with~\footnote{The last term in parenthesis in (\ref{414}) is subdominant at large $N$.} 
\be \label{414}
\mL = \sum_{1<i<j} p_i \cdot p_j \log \frac{z_{i j} z_{1 a} z_{1b}}{z_{1i} z_{1 j} z_{ab}}+  \sum_{m= 1}^k n_m \log (-\Sigma_{m}') + \( \sum_{i \neq 1} \log \frac{z_{1a} z_{1b}}{z_{1 i}^2 z_{ab}} + \log \frac{z_{ab}}{z_{1a}z_{1b}}\)~,
\ee
where $a$ and $b$ are any distinct indices (chosen out of the  indices labeling the tachyons) and,
\be \label{Sigmanp}
\Sigma_n' =  \sum_{m=1}^{n} \(\sum_{j \neq 1, a} p_j \cdot \zeta\, R_j^m\) S_{n-m} \( \frac{n}{s }\sum_{i \neq 1,a}p_i \cdot q\, R_i^s\)~, \ \  \ \ \ \ \ \ R_i = \frac{z_{a i}z_{1b}}{z_{1 i} z_{a b}}~.
\ee
Equivalently, using the integral representation of the Schur polynomial, 
\be \label{963}
\Sigma_n' = \sum_{j\neq 1, a} p_j \cdot \zeta\, R_j \oint \frac{d w}{2\pi i} \frac{w^{-n}}{1 - w R_j} \prod_{i\neq 1, a} \(1 - w R_i\)^{-n\, p_i \cdot q}~.
\ee
Notice that $R_a = 0$ and $R_b = 1$.

\subsubsection*{Derivation}
Let us derive the form of the amplitude given in (\ref{414}).  Starting with the ``tachyon'' piece of $\mL$ in (\ref{930}), we use momentum conservation to eliminate $p_1$, 
\be
\sum_{i <j } p_i \cdot p_j \log z_{i j} = \sum_{1<i<j} p_i \cdot p_j \log \frac{z_{i j}}{z_{1 i} z_{1 j}}- 2 \sum_{i \neq 1} \log z_{1 i}~.
\ee
Using momentum conservation again, $p_1^2 = (\sum_{i \neq 1} p_i)^2$, as well as the mass-shell condition, we have that
\be
-2(N-1) = p_1^2 = ( \sum_{i\neq1 } p_i)^2= 2 \sum_{1<i<j} p_i \cdot p_j+ 2n ~,
\ee
where $n$ is the number of tachyons. 
We may solve for  $p_a \cdot p_b$, with $a$ and $b$ of our choosing, 
\be
p_a \cdot p_b = -\!\!\!\! \sum_{\substack{1<i<j\\ (i,j) \neq (a,b)}}\!\!\! p_i \cdot p_j - (N-1 +n)~.
\ee
Using this to eliminate $p_a \cdot p_b$, we get,
\be \label{952}
\sum_{i <j } p_i \cdot p_j \log z_{i j} =\sum_{\substack{1<i<j\\ (i,j) \neq (a,b)}}p_i \cdot p_j \log \frac{z_{i j} z_{1 a} z_{1b}}{z_{1i} z_{1 j} z_{ab}}  +  \( \sum_{i \neq 1} \log \frac{z_{1a} z_{1b}}{z_{1 i}^2 z_{ab}} + \log \frac{z_{ab}}{z_{1a}z_{1b}}\)- N \log \frac{z_{ab}}{z_{1a } z_{1 b}} ~.
\ee
The first term is manifestly SL$_2$ invariant, the second term combines with the measure to give something SL$_2$ invariant (see footnote \ref{foot3}), and  the final term will be absorbed into the $\Sigma_n$. In particular, we define, 
\be \label{422}
\Sigma_n' =\(\frac{z_{ab}}{z_{1a } z_{1 b}}\)^{-n} \Sigma_n~.
\ee
Our goal now is to write  $\Sigma_n'$ in a form in which it is manifestly SL$_2$ invariant. To do this, we note that as a result of momentum conservation and $p_1\cdot \zeta = 0$, we may eliminate $p_a \cdot \zeta = - \sum_{i \neq 1, a} p_i \cdot \zeta$. 
Applying this to $\Sigma_n$ given in (\ref{413}) we get, 
\be
\Sigma_n = \sum_{j \neq 1, a}p_j \cdot \zeta \frac{ z_{a j}}{z_{j1} z_{a 1}}\oint \frac{d w}{2\pi i }\frac{1}{w^n} \frac{1}{1 - \frac{w}{z_{j1}}}\frac{1}{1- \frac{w}{z_{a 1}}}\prod_{i\neq 1} \(1 - \frac{w}{z_{i 1}}\)^{- n p_i \cdot q}~.
\ee
Next, we change integration variables to $w\rightarrow z_{a1}/w$, and use $\sum_{i \neq 1} p_i \cdot q = - 1$ (a consequence of momentum conservation and that $p_1 \cdot q = 1$)  to get, 
\be
\Sigma_n =- \sum_{j \neq 1, a}p_j \cdot \zeta \frac{ z_{a j}}{z_{j1} z_{a 1}}\frac{1}{z_{a1}^{n-1}}\oint \frac{d w}{2\pi i } \frac{1}{w - \frac{z_{a1}}{z_{j1}}}\frac{1}{w-1}\prod_{i\neq 1} \(w- \frac{z_{a1}}{z_{i 1}}\)^{- n p_i \cdot q}~.
\ee
We see that the term involving $w$ to a power disappeared - a result of SL$_2$ invariance. Finally, we change variables $w\rightarrow 1 + \frac{z_{ab}}{z_{b1} w}$, and obtain the claimed result, (\ref{963}).

\subsection{General polarization} \label{sec42}
We take the most general state, with the DDF photons taking any allowed polarizations $\lam_i \cdot q = 0$, 
\be \label{425}
\lam_k \cdot A_{-m_k}\cdots \lam_2 \cdot A_{-m_2}\,\,  \lam_1\cdot A_{-m_1} |0\ra~, \ \ \ \ \ N= \sum m_i~.
\ee
The vertex operator for this state was given in (\ref{344}). The amplitude involving this state and any number of tachyons is given by, 
\be 
\mA = \frac{1}{\text{vol} (SL_2)}\int dz\,  \exp\(\mL\)~,\\[-5pt]
\ee
where 
\be  \label{427}
\mL = \sum_{i< j} p_i \cdot p_j \log z_{i j} +\log\(\sum_{\rho=1}^{\lfloor k/2 \rfloor} \sum_{\pi}\, \prod_{l=1}^{\rho}( \zeta_{\pi(2l-1)} \cdot \zeta_{\pi(2 l)} )\, \mbS_{m_{\pi(2l-1)}, m_{\pi(2l)}} \prod_{q=2\rho+1}^k(-\Sigma_{m_{\pi{q}}}(\zeta_q) )\)~. 
\ee
where $\Sigma_m(\zeta)$ was defined in (\ref{931}) and $\mbS_{n, m}$ now refers  not to the operator  in (\ref{342}), but rather its contraction with $e^{i \sum p_i \cdot X(z_i)}$, 
 \be
\mathbb{S}_{m, n} = \sum_{r =1}^n r\, S_{m+r}\(m \sum_{i\neq 1} \frac{p_i \cdot q}{s\, z_{i 1}^s}\) S_{n-r}\(n \sum_{i\neq 1} \frac{p_i \cdot q}{s\, z_{i 1}^s}\)~.
\ee
As in Sec.~\ref{sec41}, we would like to rewrite the amplitude in an SL$_2$ invariant form. Doing this gives, 
\bml  \label{429}
\mL = \sum_{1<i<j} p_i \cdot p_j \log \frac{z_{i j} z_{1 a} z_{1b}}{z_{1i} z_{1 j} z_{ab}} + \( \sum_{i \neq 1, a, b} \log \frac{z_{1a} z_{1b}}{z_{1 i}^2 z_{ab}} + \log \frac{z_{ab}}{z_{1a}z_{1b}}\) \\
+\log\(\sum_{\rho=1}^{\lfloor k/2 \rfloor} \sum_{\pi}\, \prod_{l=1}^{\rho} ( \zeta_{\pi(2l-1)} \cdot \zeta_{\pi(2 l)} )\, \mbS_{m_{\pi(2l-1)}, m_{\pi(2l)}}' \prod_{q=2\rho+1}^k(-\Sigma'_{m_{\pi{q}}}(\zeta_q)) \)~,
\end{multline}
where $a$ and $b$ are any distinct indices (chosen out of the $n$ indices labeling the tachyons) and $\Sigma_n'$ was defined in terms of $\Sigma_n$ in (\ref{422}) and explicitly given in (\ref{963}), while $\mbS_{m, n}'$ is defined by
\be
\mbS_{m, n}' =\(\frac{z_{ab}}{z_{1a } z_{1 b}}\)^{-n-m}\mbS_{m, n}~.
\ee

We still need to write $\mbS_{m, n}'$  in a form which makes the SL$_2$ invariance manifest. 
By using the integral representation of the Schur polynomial (\ref{SchurInt}), performing the sum over $r$, and doing a change of integration variables, analogous to what we did in the derivation of $\Sigma_n'$, we get, 
\be \label{431}
\mathbb{S}_{m, n}'  =  \oint\frac{d w}{2\pi i} \oint\frac{d v}{2\pi i} \frac{1}{(v-w)^2} \frac{1}{w^m v^n} \prod_{i\neq 1,a} \(1 - w R_i\)^{-m p_i \cdot q}  \prod_{j \neq 1,a} \(1 - v R_j\)^{-n p_j \cdot q}~,
\ee
where the cross-ratio of the points, $R_i$, was given in (\ref{Sigmanp}). We may equivalently write this in terms of the Schur polynomials, 
\be \label{432}
\mathbb{S}_{m, n}' =\sum_{r =1}^n r\, S_{m+r}\(m \sum_{i\neq 1,a} \frac{p_i \cdot q}{s} R_i^s\) S_{n-r}\(n \sum_{i\neq 1,a } \frac{p_i \cdot q}{s} R_i^s\)~.
\ee

\subsection*{Summary}
Let us summarize where we currently stand. We have computed the amplitude involving the most general excited state and any number of tachyons. The result is (\ref{429}), and the various terms appearing in this expression are defined in this section and in the previous one. The expression is clearly involved. 
The amplitude simplifies in the special case when the dot product of all the polarizations vanishes, in which case (\ref{429}) turns into the expression (\ref{414}) found in Sec.~\ref{sec41}. 

 Another special case is an excited state involving only one excited mode $(\lam \cdot A_{-k})^n$. In this case the combinatorial sum in (\ref{429}) simplifies and we find,
\be
 \mL = \sum_{1<i<j} p_i \cdot p_j \log \frac{z_{i j} z_{1 a} z_{1b}}{z_{1i} z_{1 j} z_{ab}} + \( \sum_{i \neq 1, a, b} \log \frac{z_{1a} z_{1b}}{z_{1 i}^2 z_{ab}} + \log \frac{z_{ab}}{z_{1a}z_{1b}}\)  + 
n! \sum_{r=0}^{\lfloor n/2\rfloor}  \frac{(\Sigma_k')^{n- 2 r}}{(n-2r)!}\frac{1}{r!} \( \frac{\lam \cdot \lam}{2}\, \mbS'_{k,k}\)^r 
\ee
A special case of this would be exciting the mode $k=1$, the leading Regge trajectory. 

This is as far as we can go without specifying the number of tachyons. In  Sec.~\ref{sec6} we will specialize to amplitudes with one excited string and two tachyons, and one excited string and three tachyons, and give more explicit expressions.

\section{Amplitude with one excited string: properties} \label{sec6}
\subsection{Two tachyons and one excited string} \label{sec61}
\begin{figure}
\centering
\subfloat[]{
\includegraphics[width=1.4in]{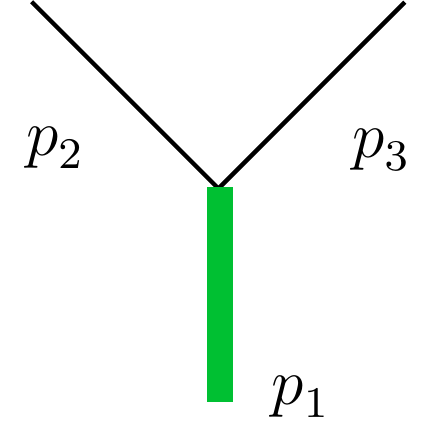} } \ \ \ \ \ \ \  \ \ \ \ \ \ \ \ \ \  \  \ \ \ \ 
\subfloat[]{
\includegraphics[width=1.8in]{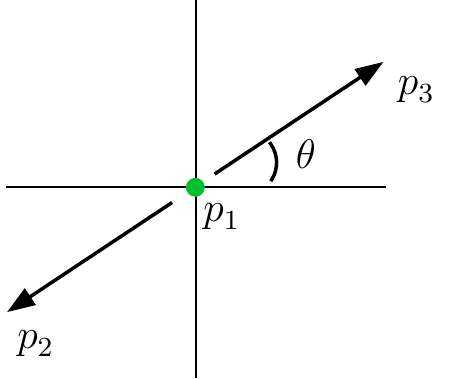} } \\[50pt]
\subfloat[]{
\includegraphics[width=2.7in]{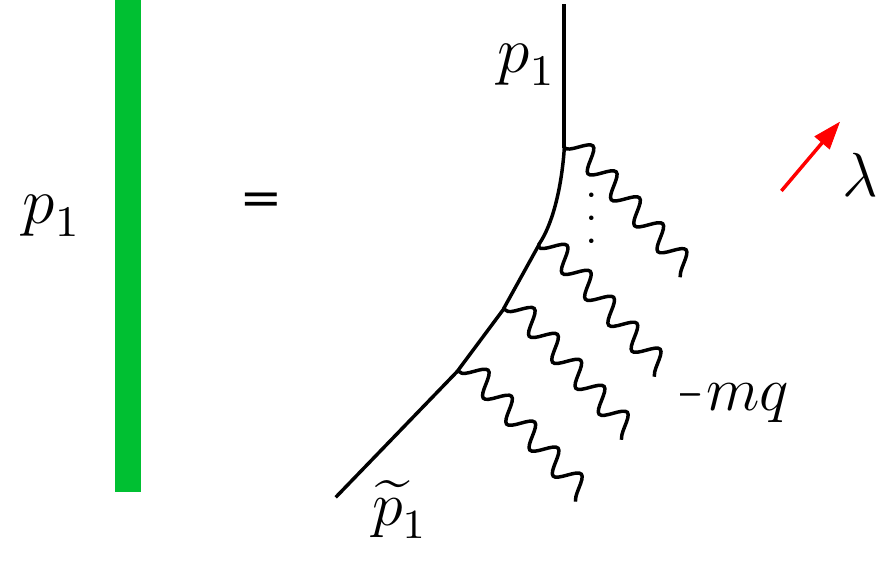} }  \ \ \ \ \ \ 
\subfloat[]{
\includegraphics[width=2.9in]{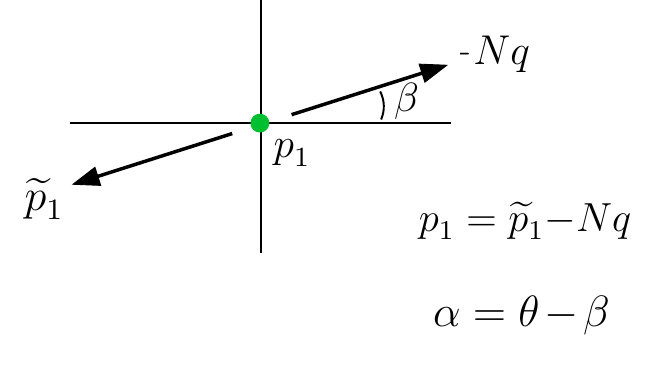} } 
\caption{(a) A spacetime diagram of heavy (highly excited) string of momentum $p_1$ decaying into two tachyons of momenta $p_2$ and $p_3$. (b) The spatial plane of the decay process.  The tachyons leave at angle $\theta$, see  (\ref{htt}). (c) The heavy string state (\ref{61}), with momentum $p_1$, is in turn formed by repeatedly scattering DDF photons of momenta $-m q$  a total $n_m$ times (for $m = 1,2,3, \ldots$) off of an initial tachyon of momentum $\t p_1$. The photons have polarization $\lam$ which is orthogonal to $q$, and are sent in at angle $\beta$, see (\ref{64q}) and (\ref{661}). A spacetime diagram is shown in (c), and the spatial plane is shown in (d) .  The amplitude depends on the difference $\al$ between the angles, $\al = \theta - \beta$.  }\label{Heavdecay}
\end{figure}

In this section we compute the amplitude for an excited string to decay into two tachyons, as shown in Fig.~\ref{Heavdecay}. 

 As was discussed in Sec.~\ref{sec3}, the excited string is formed by repeatedly scattering photons of momenta $-m q$ off of an initial tachyon, where $q$ is a null vector of our choosing, provided $p \cdot q = 1$, where $p$ is the momentum of the string. Here $m$ is an integer, and to create a state in which mode $m$ of the string is excited $n_m$ times, we send in $n_m$ photons with momentum $-m q$. The polarizations $\lambda$ of these DDF photons are transverse to their momentum, $\lam\cdot q = 0$.  For simplicity, we first discuss the case in which all the photons have the same polarization $\lam$.  (The general case will be discussed later; the chaos features are unchanged.)
With one polarization, the excited state is uniquely specified by the occupation numbers $\{n_m\}$ of each of the transverse string modes $m = 1, 2,3 \ldots$. 
In terms of the DDF creation operators ($A_{-m}^{\mu}$ for mode $m$), the state is,
\be \label{61}
\prod_{m=1}^{\infty} (\lam \cdot A_{-m})^{n_m}  |0\ra~, \ \ \ \ \ N = \sum_{m=1}^{\infty} m \, n_m~.
\ee

We will find that the amplitude for the excited string (\ref{61}) to decay to two tachyons is proportional to,
\be \label{62}
\mA  \propto \prod_{m=1}^{\infty} \( p_3 \cdot \zeta\,  P_m(p_3\cdot q)\)^{n_m} ~, \  \ \ \ \  p_3 \cdot \zeta =- \sqrt{\frac{N}{2}} \sin \al~, \ \ \ \  p_3 \cdot q= - \cos^2 \frac{\al}{2}~, \ \ \  P_m(a) = \frac{(1+ m\, a)_{m-1}}{(m-1)!}~,
\ee
where $\al$ is the relative angle between the direction of the tachyons and the photons creating the string, and $(a)_m$ is the Pochhammer symbol. 
This  expression is valid in the approximation that the state is highly excited,  $N\gg 1$, and the angle $\al$ is not close to $0$ or $\pi$. 
 This expression captures the angular dependence $\al$ of the amplitude; the amplitude also contains a state-dependent normalization prefactor, for which we do not have a simple expression.
We will spend the rest of this section deriving and studying the result (\ref{62}).

\subsubsection*{Kinematics}

Let us look at the kinematics for the process of a heavy string decaying into two tachyons. The kinematics will occur in a three spacetime dimensional plane, and we choose the frame in which the heavy string is at rest. The heavy string is taken to have momentum $p_1$, while the two tachyons have momenta  $p_2$ and $p_3$. Conservation of momentum is $p_1 +p_2 +p_3 = 0$. To slightly simplify the presentation, we take the heavy string to be very massive $N\gg1$, so that we may approximate the mass $M^2 = 2(N-1) \approx 2N$.
The kinematics is, 
\bea  \label{htt}
p_1 &=&\sqrt{2N} \, (1,0,0)\\ \nn
p_2 &=& - \frac{\sqrt{2N}}{2}(1,\sin \theta,  \cos \theta)  \\ \nn
p_3 &=& - \frac{\sqrt{2N}}{2}(1,- \sin \theta, - \cos \theta)~. \ \ \ \ 
\eea

As discussed in Sec.~\ref{sec3}, the heavy (excited) string is formed by repeatedly scattering photons with momentum proportional to $q$ (which we refer to as DDF photons) off of a tachyon. The vector $q$ must satisfy $p_1\cdot q = 1$, in order to produce a heavy string of the correct mass. Combined with the requirement that the photon momentum be null, $q^2=0$, this fixes $q$ to take the form, 
\be \label{64q}
q = -\frac{1}{\sqrt{2N}}(1,\sin \beta,  \cos \beta)~,
\ee
where $\beta$ is an arbitrary angle which we are free to choose.
A dot product we will need later on, between $q$ and the momenta of the tachyons, is, 
\be \label{63}
p_2 \cdot q = - \sin^2 \frac{\al}{2}~, \  \ \ \ p_3 \cdot q = - \cos^2\frac{\al}{2}~, \ \ \  \ \ \ 
\al = \theta - \beta~.
\ee

The polarization $\lambda$ of the DDF photon must be orthogonal to its momentum, $\lambda \cdot q = 0$. In three dimensions,  the polarization is therefore, 
\be \label{661}
\lam = (0, - \cos \beta, \sin \beta)~.
\ee
If we are in dimension $D$  that is four or higher, we can let the polarization have arbitrary components  $\lam^i$ in the higher dimensions,  $i= 3, \ldots, D$.  A special case is circular polarization
\be \label{RL}
\lam = \frac{1}{\sqrt{2}}(0, - \cos \beta, \sin \beta, \pm i)~,
\ee
which has $\lam^2 = 0$. 
As usual, we can add a multiple of $q$ to $\lam$,  and leave the amplitude unchanged. Recalling that $\zeta_{\mu} = \lam_{\mu} - (\lam \cdot p_1) q_{\mu}$, since $\lam \cdot p_1 = 0$, we have $\zeta_{\mu} = \lam_{\mu}$. The dot product of the momenta with the polarization is thus, 
\be \label{66}
p_2 \cdot \zeta = -p_3 \cdot \zeta = \sqrt{\frac{N}{2}} \sin \al ~.
\ee

The amplitude can only be a function of: $p_i \cdot p_j$ (all of which are constants), $p_i \cdot q$ and $p_i \cdot \zeta$. We see that neither of these depend on $\theta$ or $\beta$ individually, but only on their difference $\al = \theta - \beta$, and so the amplitude will only depend on $\al$. This had to be the case: since the choice of $q$, and hence $\beta$, is arbitrary and any particular $q$ is enough to get a complete basis of excited states, it could not have been the case that different $\beta$ lead to physically different amplitudes. We see that a change of $\beta$ is just a coordinate rotation.

\subsubsection*{The amplitude}
We now turn to the amplitude for a heavy string to decay into two tachyons. In Sec.~\ref{sec4} we discussed the amplitude involving one heavy string and any number of tachyons. We can easily specialize this to two tachyons. 

The formulas for the amplitude were considerably simpler in the case in which the polarization $\lam$ of the DDF photons is null, $\lam^2 = 0$, as discussed in Sec.~\ref{sec41}. This can be achieved in dimensions four or higher by taking  the polarizations to  either all be right-handed circularly polarized, or all left-handed circularly polarized, see (\ref{RL}). We will start with this case.

The amplitude found in Sec.~\ref{sec41} is, 
\be \label{67}
\mA =\mN \prod_{m=1}^{\infty} \(\Sigma_m' \)^{n_m}~, 
\ee
where $\Sigma_m'$ was given in (\ref{963}) and $\mN$ is an irrelevant normalization factor. Applying that formula to  two tachyons, we take $a =2$, $b=3$ and find $R_3 = 1$, and 
\be
\Sigma_n' = p_3 \cdot \zeta \oint \frac{d w}{2\pi i}\frac{w^{-n}}{(1 - w )^{1 + n p_3 \cdot q}} = p_3 \cdot \zeta \frac{(1+ n p_3 \cdot q)_{n-1}}{(n-1)!}~.
\ee
The amplitude is thus,
\be\label{69}
\mA = \mN (p_3 \cdot \zeta)^J \prod_{m=1}^{\infty} P_m(p_3\cdot q)^{n_m}~,
\ee
where $J$ is the spin $J = \sum_m n_m$, and $p_3\cdot\zeta$ was given in (\ref{66}), $p_3 \cdot q$ was given in (\ref{63}) and we defined $P_m(a)$ in terms of the Pochhammer symbol, 
\be \label{Pma}
P_m(a) = \frac{(1+ m\, a)_{m-1}}{(m-1)!}~, \ \ \ \ \ \ (a)_{m} = a(a{+}1) \cdots (a{+}m{-}1)~.
\ee
For some low values of $m$, the explicit form of $P_m(a)$ is, 
\be \label{PmaE}
P_1(a) = 1~, \ \  \ \ P_2(a) = 1+2a~, \ \ \ \ P_3(a) = \frac{(1+3a)(2+3a)}{2}~, \ \ \ P_4(a) = \frac{(1+4a)(2+4a)(3+4a)}{6}~.
\ee

Let us now consider the case of general polarization $\lam$. We will find the amplitude is essentially proportional to (\ref{69}), in a way in which we will elaborate on. 
In Sec.~\ref{sec42} we discussed the case of general polarization, and found  that the amplitude is given by (\ref{67}) multiplied by a function of,
\be \label{614}
 \frac{\zeta^2\,  \mbS_{i j}'}{\Sigma_i'\, \Sigma_j'} 
\ee
for various $i$ and $j$. The precise function is found by considering all possible Wick contractions involving the creation operators forming the state. The function will be discussed  in \cite{GRthree}, but we actually don't need to know it. The reason is that in the large $N$ limit the ratio (\ref{614}) is just a number, independent of the angle $\al$. Let us compute this ratio. First, we note that if we specialize the formula (\ref{432}) for $\mbS'_{i j}$ to the case of two tachyons,  we can take the indices to be $a = 2$ and $i=3$ and it reduces to, 
\be
\mathbb{S}_{m, n}' = \sum_{r=1}^n r S_{m+r} \( m \frac{p_3 \cdot q}{r}\) S_{n-r}\( n \frac{p_3\cdot q}{r}\)~.
\ee
Upon performing the sum, 
\be \label{428}
\mathbb{S}_{m, n}' =  \sum_{r=1}^n r \frac{(m p_3 \cdot q)_{m+r}}{(m+r)!} \frac{(n p_3 \cdot q)_{n-r}}{(n-r)!} = \frac{m n}{m+n} p_3{ \cdot} q\, (1+ p_3 {\cdot} q) \frac{(1+ n\,  p_3 \cdot q)_{n-1}}{(n-1)!} \frac{(1 + m\,  p_3 \cdot q)_{m-1}}{(m-1)!}~.
\ee
As a result the ratio that we need is,
\be
\frac{\mbS_{i j}'}{\Sigma_i'\, \Sigma_j'} = \frac{i j}{i+j} \frac{p_3 \cdot q (1 + p_3 \cdot q) }{(p_3 \cdot \zeta)^2}~. 
\ee
The kinematic terms appearing here were given earlier, see (\ref{63}) and (\ref{66}), which gives, 
\be  \label{620}
 \frac{p_3 \cdot q (1 + p_3 \cdot q) }{(p_3 \cdot \zeta)^2}=- \frac{1}{2N}~.
\ee
Actually, we should be slightly careful - the kinematics we have been using involved taking the large $N$ limit. If we do this more carefully, at finite $N$. we get, 
\be \label{621}
 \frac{p_3 \cdot q (1 + p_3 \cdot q) }{(p_3 \cdot \zeta)^2} =-\frac{1}{2} \frac{3 + N -\frac{4}{\sin^2 \al}}{ N^2 + 2N -3}\approx  \frac{-1}{2N} + \frac{1}{2N^2}\( \frac{4}{\sin^2 \al}-1\) +\mO(\frac{1}{N^3})~.
 \ee
 Thus, as long as we are away from $\al=0, \pi$, the leading term is a good approximation. Therefore, in the large $N$ limit and away from $\al = 0, \pi$ we have, 
 \be
 \frac{\mbS_{i j}'}{\Sigma_i' \Sigma_j'}  \approx -\frac{1}{2N} \frac{i j}{i+j}~.
 \ee
 The amplitude will contain a function of these variables, which we don't know. However, since the variables are independent of $\al$, the function will be independent of $\al$.

Thus, away from $\al = 0$ and $\al =\pi$, the amplitude is proportional to (\ref{69}), 
\be \label{Ampp}
\mA  \propto(p_3 \cdot \zeta)^J \prod_{m=1}^{\infty} P_m(p_3\cdot q)^{n_m}~,
\ee
where $P_m(a)$ was defined in (\ref{Pma}), $p_3 \cdot q$ was given in (\ref{63}), and $p_3\cdot \zeta$ was given in (\ref{66}).
We may equally well write the amplitude with $p_3 \cdot \zeta$ under the product, or the amplitude expressed in terms of $p_2 \cdot q$ and $p_2 \cdot \zeta$, 
\be
\mA  \propto \prod_m \( p_3 \cdot \zeta\,  P_m(p_3\cdot q)\)^{n_m} =(-1)^N (p_2 \cdot \zeta)^J \prod_m P_m(p_2\cdot q)^{n_m}~,
\ee
where in the second equality we used that $P_m(-1-a) = (-1)^{m-1}P_m(a)$ combined with $p_3 \cdot q = -1 - p_2 \cdot q$, as well as $p_2 \cdot \zeta = - p_3 \cdot \zeta$.\\

The amplitude (\ref{Ampp}) is the result (\ref{62}) that was advertised at the beginning of the section as the amplitude for the heavy state (\ref{61}) to decay to two tachyons. Perhaps the most familiar heavy state is the leading Regge trajectory $ (\lam \cdot A_{-1})^N |0\ra$. Since $P_1(a) = 1$ (\ref{PmaE}), the amplitude is simply, 
\be
\mA \sim (p_3 \cdot \zeta)^J \sim (\sin \al)^J, \ \ \ \ J = N~, \ \ \ \ \ \text{for the state      }\ \ \    (\lam \cdot A_{-1})^N |0\ra~.
\ee
As would be expected, the amplitude is simple. 
A slightly more general simple state is one in which only one mode is excited, 
\be
(\lam \cdot A_{-k})^n |0\ra~, \ \ \ \ \ \ N = n k~, \ \ \ \ \ \mA \propto \(p_3 \cdot \zeta\,  P_k(p_3 \cdot q) \)^n 
\ee
The amplitude is again a smooth function of the angle $\al$. Of course, this state is still special, so this is to be expected. We will soon show that, in contrast, for generic states the amplitude is erratic. 

\subsubsection*{General polarization}

The excited string state that we have so far discussed, (\ref{61}), with one polarization is the general  state in three dimension. 
In $D$ space-time dimensions, there are $D-2$ independent polarization vectors and the most general state is given by,
\be  \label{51}
\prod_{k=1}^{n_1}(\lam^1_k \cdot A_{-1} )\prod_{k=1}^{n_2} (\lam^2_k\cdot A_{-2}) \cdots \prod_{k=1}^{n_r}(\lam^r_k \cdot A_{-r}) |0\ra~, \ \ \ \ \ N = \sum_{m=1}^r m n_m~.
\ee
If all the polarization vectors are orthogonal to each and null then the amplitude is, 
\be \label{627}
\mA \propto \prod_{k=1}^{n_1}(p_3\cdot \zeta^1_k)  \prod_{k=1}^{n_2} (p_3 \cdot \zeta^2_k) \cdots \prod_{k=1}^{n_r}(p_3 \cdot \zeta^r_k )     \prod_{m=1}^{\infty} P_m(p_3\cdot q)^{n_m}~,
\ee
up to a normalization constant, where $\zeta_{\mu, k}^i = \lam_{\mu, k}^i - (\lam^i_k \cdot p_1) q_{\mu}$. If there is only one polarization vector, this reduces to   (\ref{Ampp}).  In the case the polarization vectors are not orthogonal to each other, (\ref{627}) will be multiplied by a function of 
\be \label{zikjl}
 \frac{\zeta^i_k\cdot \zeta^j_l\,  \mbS_{i j}'}{\Sigma_i'\, \Sigma_j'}~.
\ee
As we saw earlier in (\ref{620}), the ratio $ \mbS_{i j}'/(\Sigma_i'\, \Sigma_j')$ is independent of $\al$. So this function of (\ref{zikjl}) will only be a function of the dot products of the polarization vectors. Thus, the $\al$ dependence of the amplitude is captured by (\ref{627}), whereas the dependence on the relative polarization angles is not something we explicitly know. 
The essential point is that the factor that will give rise to chaos is the $\al$ dependent term $\prod_{m=1}^{\infty} P_m(p_3\cdot q)^{n_m}$ that appears in the amplitude. This term is independent of the choices of polarization. We will therefore continue our discussion of the amplitude by focusing on the amplitude (\ref{Ampp}) for the case  of one polarization vector.

\subsubsection*{Oscillations in $P_m(a)$}
\begin{figure}
\centering
\includegraphics[width=2.5in]{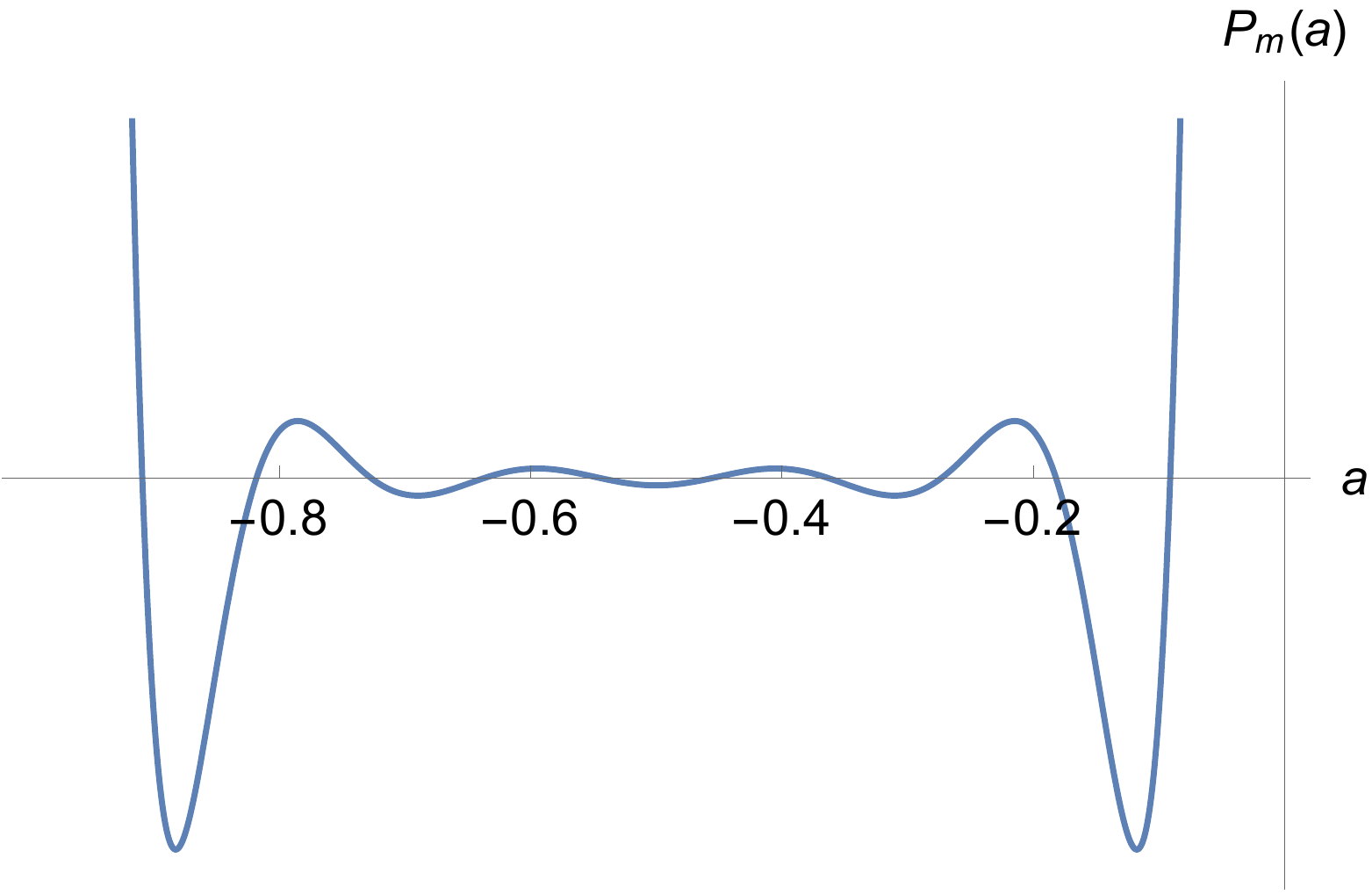}
\caption{A plot of $P_m(a)$ (\ref{Pma}) with $m=11$. At large $m$ $P_m(a)$ is approximated by (\ref{1735}).} \label{Pmaplot}
\end{figure}
The essential ingredient in the amplitude is the function which we called $P_m(a)$ in (\ref{Pma}). We notice that $P_m(a)$ is oscillatory for $-1<a<0$, and monotonic for $a$ outside this range. For us $a = p_3\cdot q$ and is in the oscillatory range $-1<a<0$, at large $N$. A plot of $P_m(a)$ for $m=11$ is shown  in Fig.~\ref{Pmaplot}.  The larger $m$, the more oscillatory $P_m(a)$ is. 
Indeed, we can simplify $P_m(a)$ at large $m$ through use of Stirling's approximation. For   $-1<a<0$, we get, 
\be \label{1735}
P_m(a) = - \sin (\pi m a) \sqrt{\frac{2}{m \pi}}\frac{(1+a)^{m(1+a) - \frac{1}{2}}}{(-a)^{ma + \frac{1}{2}}}~, \ \ \ \ \  m |a| \gg 1~.
\ee

\subsubsection*{An erratic amplitude}

Each $P_m(a)$ comprising the amplitude (\ref{Ampp}) is a smooth function of $a$ that has $m-1$ zeros as $a$ ranges from $-1$ to $0$. In the limit of large $m$, as we said, $P_m(a)$ can be approximated by (\ref{1735}), and the zeros lie at $ a= \frac{1}{m}, \frac{2}{m}, \ldots, \frac{m-1}{m}$. Each $P_m(a)$ by itself is a fairly regular function.

 However, with a product of $P_m(a)$ over many $m$ we have a much more interesting function. Taking the amplitude (\ref{Ampp}) and using the approximation (\ref{1735}), the amplitude takes the form,
 \be
 \mA\sim (\sin \al)^J \exp\(- \sum_m n_m \log m\)\frac{(1+p_3 \cdot q)^{N(1 + p_3 \cdot q) - \frac{J}{2}}}{(- p_3 \cdot q)^{N p_3 \cdot q + \frac{J}{2}}} \prod_m (\sin( \pi m p_3 \cdot q))^{n_m}~, \ \ \ \ m\gg 1~.
 \ee
 The kinematic factor $p_3 \cdot q$ was given in (\ref{63}),   $ p_3 \cdot q= - \cos^2 \frac{\al}{2}$. The erratic nature of the amplitude comes from the last term, the product of the sine factors. 
 Consider taking the limit of $N$ goes to infinity, and taking our excited string state to have a nonzero occupation number for each mode. The amplitude then has a zero  for every angle $\al$ at which $\cos^2 \frac{\al}{2}$ is a rational number.
 
For any particular (generic, highly  excited) state, the amplitude appears to be erratic, in both the change in the outgoing tachyons or the ingoing excited string. In particular, a small change in the angle $\theta$ of the detector for the tachyons potentially leads to a large change in the amplitude (recall that $\al = \theta - \beta$). Likewise, the amplitude has the same sensitivity to the ingoing state, under a small change in the angle $\beta$ of the DDF photons that form the excited string. In addition, the amplitude is sensitive to the precise occupation levels of the initial string: suppose we take an initial string in which a large number of different modes have nonzero occupation number. Let one of these modes be $m$, so that $n_m$ is nonzero. Let $r$ be a nearby mode for which $n_r$ is zero. It is a fairly minor change in the state to change $n_m$ to zero and $n_r$ to something nonzero. The ratio of these two amplitudes contains $P_r(p_3 \cdot q)/P_m(p_3 \cdot q)$, a function which has both zeros and poles.

\subsubsection*{An example}
\begin{figure}[t]
\centering
\subfloat[]{
\includegraphics[width=2.8in]{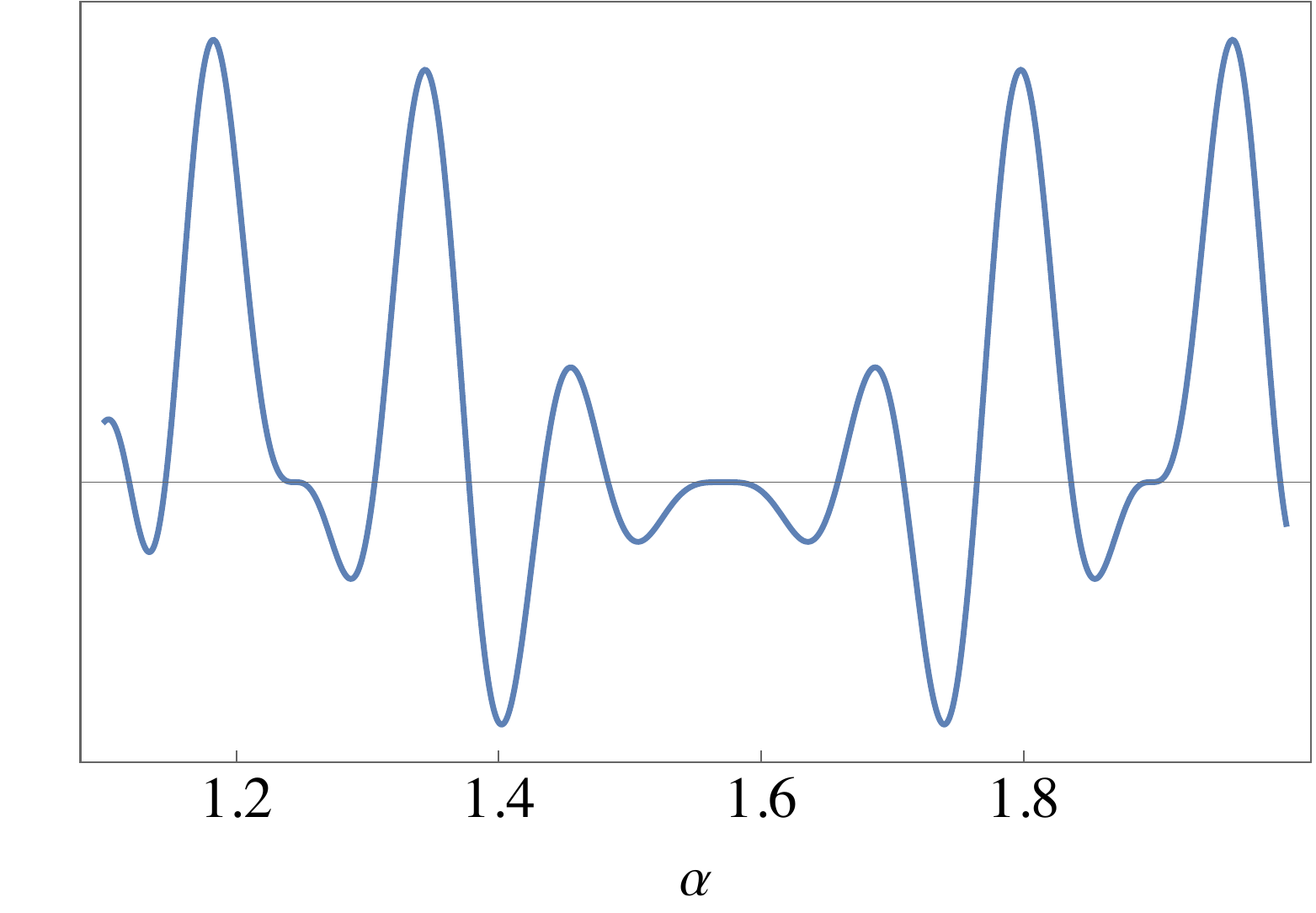}
} \ \ \ \  \ \ \ 
\subfloat[]{
\includegraphics[width=2.8in]{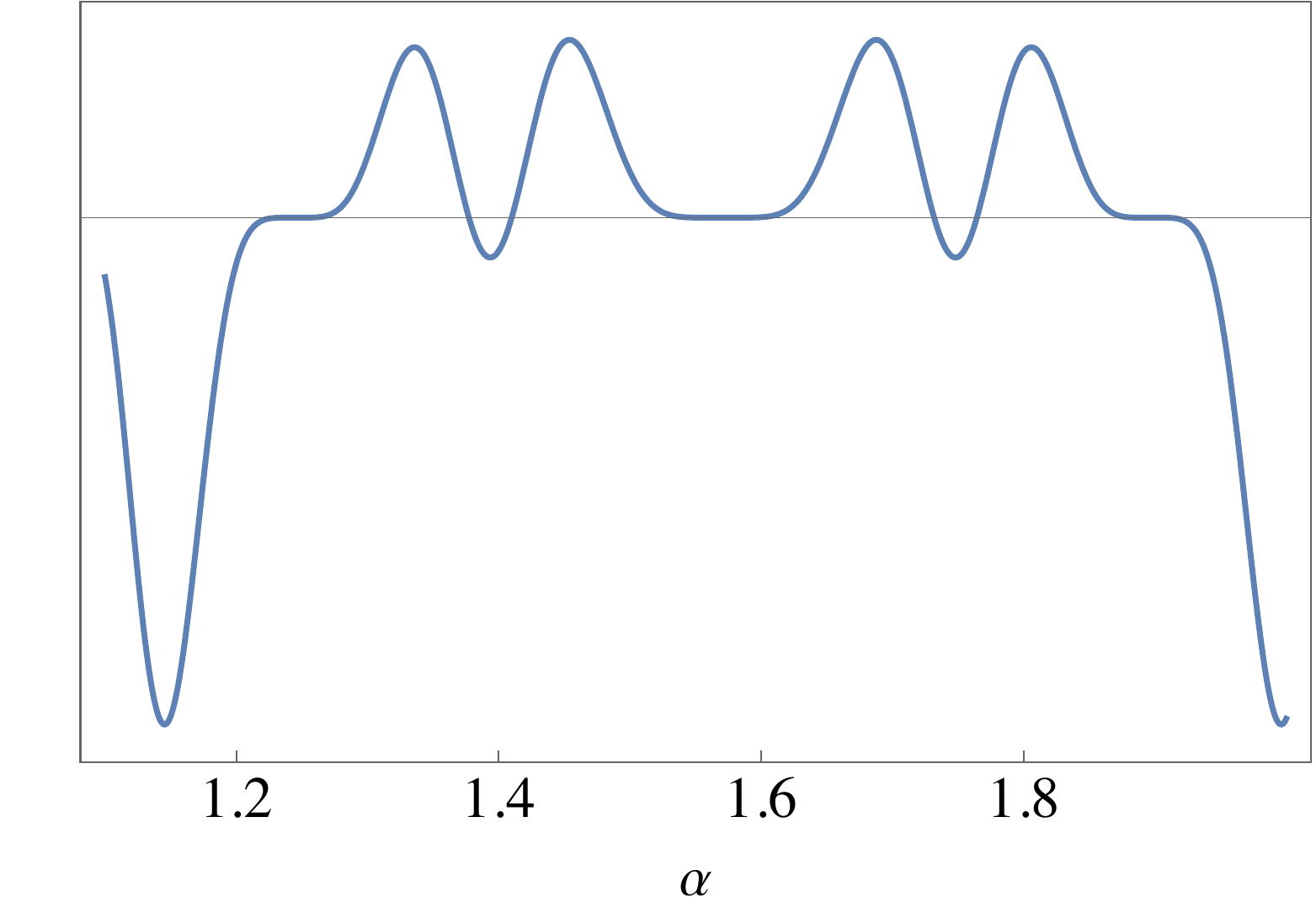}
}
\caption{A plot of the amplitude for an excited string (at level $N=50$, in a particular state we specify) to decay to two tachyons, as a function of the angle $\al$, as given by (\ref{Ampp}). (a) The amplitude for the state with occupation numbers $n_{11}= n_7=n_5=1$, $n_4 = 4$, $n_3=3$, $n_1=2$. (b) We slightly change the state to have $n_{12}=n_6=1$ and $n_{11}=n_7=0$, and other occupation numbers unaffected. It is very different from plot (a). } \label{Amp1}
\end{figure}
 
 Let us look at the amplitude (\ref{Ampp}) for some particular excited string state. As an example, we take the excited string to be at level $N=50$, and take some randomly chosen state at this level. For instance, 
 \be
 (\lam \cdot A_{-11})(\lam \cdot A_{-7})(\lam \cdot A_{-5})(\lam\cdot A_{-4})^4(\lam\cdot A_{-3})^3(\lam \cdot A_{-1})^2 |0\ra~.
 \ee
 We plot the amplitude (\ref{Ampp}) as a function of angle $\al$ in Fig.~\ref{Amp1} (a). Even at this relatively small value of $N$, the amplitude exhibits nontrivial behavior. If we change to a different state, also at level $50$, the amplitude looks very different, as shown in Fig.~\ref{Amp1} (b).

\subsubsection*{A detector measuring erratic behavior}

Let us now combine everything to discuss our setup. We start with a highly excited string in a generic state. At finite string coupling $g_s$, the excited string will decay into other strings. For small $g_s$, we can look at just the leading order process, of the string decaying into two strings. 

For simplicity, we focus on the case in which the string decays into two tachyons. This is not the generic decay channel, and not the dominant decay channel, however it is the simplest to calculate. Operationally, we can arrange to have a detector which only detects tachyons. There is a finite amplitude for the initial string to decay into two tachyons, and so the detector will occasionally click.  

We are interested in the behavior of the amplitude as a function of the outgoing angle of the tachyons. 
If the initial excited string state is one of the basis states we discussed, then the amplitude is highly erratic as a function of the angle, and is extremely sensitive to the precise state of the heavy string.  As we move our detector around (varying $\theta$), its detection of tachyons will vary erratically. 

A natural way of forming the initial highly excited string is through a scattering process of light strings. This is encoded in the DDF construction of an excited string that we have been using, in which photons are repeatedly scattered off of an initial tachyon. A  small change in the excited string state then corresponds to a small change in one of the momenta of the many photons that formed the excited string. Note also that the erratic behavior in the amplitude arrises in the limit in which the string is in a generic highly excited state (large $N$); a large number of photons are sent in  to form it. This is in line with the proposal of \cite{VRchaos}, that chaos can be seen in the erratic behavior of the scattering amplitude of many particles, under a change in the momentum of one of the particles.

This erratic behavior is not seen in the imaginary part of the high energy, large $s$, Veneziano scattering amplitude of two tachyons, even though this  is given by a sum,  over all states of mass $\sqrt{s}$,  of the square of these amplitudes. Most of the terms in the sum will be erratic, but the chaotic, erratic behavior is washed out in the sum.

\subsection{Three tachyons and one excited string} \label{sec62}

In this section we discuss the amplitude involving an excited string and three tachyons.
The amplitude involving an excited string and any number of tachyons was discussed in Sec.~\ref{sec4}. In this section we specialize the formulas found there to three tachyons.

\subsubsection*{Special polarization}
We start with the case considered in Sec.~\ref{sec41}: an excited state, 
\be \label{829v2}
(\lam \cdot A_{-1})^{n_1} (\lam\cdot A_{-2})^{n_2} \cdots (\lam \cdot A_{-k})^{n_k} |0\ra~, \ \ \ \ \ N = \sum_{i=1}^k i n_i~,
\ee
in which the polarizations $\lam$ of all the DDF photons forming the excited state are the same and $\lam^2$=0. We found the amplitude is, 
\be \label{48v2}
\mA = \frac{1}{\text{vol} (SL_2)}\int dz\,  \exp\(\mL\)~,
\ee
where $\mL$ was given by (\ref{414}) which, after taking $n= 3$ (three tachyons), and $a=2, b=3$, becomes, 
\be
\mL =  p_2 \cdot p_4 \log R + p_3 \cdot p_4 \log(R-1) - \log (z_{14}^2 z_{23}^2) + \sum_{m=1}^k n_m \log(- \Sigma_m')~, \ \ \ \ \ R \equiv R_4 = \frac{z_{13} z_{24}}{z_{14}z_{23}}~,
\ee
and $\Sigma_n'$ given in (\ref{963}) becomes, 
\be
\Sigma_n' = p_3{ \cdot} \zeta \oint \frac{d w}{2\pi i} \frac{w^{-n}}{(1-w)^{1 + n p_3 \cdot q} (1 -w R)^{n p_4 \cdot q}} + p_4{\cdot} \zeta  \oint \frac{d w}{2\pi i} \frac{R\, w^{-n}}{(1 - w)^{n p_3 \cdot q}(1  - w R)^{1 + n p_4 \cdot q}}~,
\ee
where we used that $R_3 = 1$. Performing the integrals gives, 
 \be  \label{Snp}
 \Sigma_n' =   p_3 {\cdot } \zeta 
\sum_{m=0}^{n-1} \frac{(1+ n\, p_3 \cdot q)_{n-1-m}}{(n{-}1{-}m)!} \frac{(n\, p_4 \cdot q)_m}{m!}R^m  + p_4{\cdot} \zeta 
\sum_{m=0}^{n-1} \frac{(1+ n\, p_4 \cdot q)_{n-1-m}}{(n{-}1{-}m)!} \frac{(n\, p_3 \cdot q)_m}{m!}R^{n-m}~.
\ee
We can use SL$_2$ symmetry to fix three of the four $z_i$ in the integral for the amplitude (\ref{48v2}) (see the discussion above (\ref{223})) to the values $z_1 = \infty$, $z_2=1$, $z_3 = z$ and $z_4=0$, so that $R=(1-z)^{-1}$. The amplitude becomes, 
\be \label{630}
\mA = \int_0^1 d z \, z^{p_3 \cdot p_4} (1-z)^{p_2 \cdot p_3}\, \prod_m ( - \Sigma_m')^{n_m}~.
\ee
This expression for the amplitude, with $\Sigma_n'$ given above in (\ref{Snp}),  is fully explicit and relatively compact. We  see that the scaling with the polarization  is $\zeta^J$, where $J= \sum_m n_m$ is the spin, which is correct. If we wished, we could do a multinomial expansion of  $(-\Sigma_m')^{n_m}$ in powers of $(1-z)$, and then perform the $z$ integrals, thereby obtaining the amplitude as a sum of beta functions. However, such an expression wouldn't be any more enlightening than the current form. 

Notice that if we take the excited state to be a tachyon, setting $n_m =0$ for all $m$, then we recover the Veneziano amplitude. As is well known, for high energy, fixed angle scattering of tachyons (large Mandelstam $s$ and $t$), one can get the amplitude by taking the saddle with respect to $z$ \cite{GrossMende1, GrossMende2}. If the excited state were only moderately excited, so that $N$ remains finite as $s, t \rightarrow \infty$, then the same saddle would hold here. However, the situation we are studying is one in which $N$ is of the same order as $s$ and $t$, and so one can not neglect the 
$( - \Sigma_m')^{n_m}$ in (\ref{630}) when finding the saddle. We leave a discussion of the saddle to future work. 

\subsubsection*{Special polarization and special kinematics}
There is one special case in which the amplitude simplifies.
Looking at $\Sigma_n'$  (\ref{Snp}) we notice that it  simplifies significantly if $q\cdot p_4 = \zeta \cdot p_4 = 0$, 
\be
\Sigma_n'  = p_3 {\cdot} \zeta \frac{(1+n p_3 \cdot q)_{n-1}}{(n-1)!} =  p_3 {\cdot} \zeta \, P_n(p_3 \cdot q)~, \ \ \ \ \ \text{for} \  \  \ \ \ q\cdot p_4 = \zeta \cdot p_4 = 0~.
\ee
The dependence on $R$ has disappeared, and  the amplitude becomes,
\be 
\mA = \prod_m ( - \Sigma_m')^{n_m} \int_0^1 d z \, z^{p_3 \cdot p_4} (1-z)^{p_2 \cdot p_3} = \prod_m \(-   p_3 {\cdot} \zeta \, P_m(p_3 \cdot q)\)^{n_m} \beta(p_3{\cdot}p_4 + 1, p_2 {\cdot }p_3+1)~,
\ee
where the beta function is just the tachyon amplitude, see (\ref{223}). Comparing with the three-point amplitude in (\ref{69}), the four-point amplitude - for our special kinematic configuration - is the three-point heavy-tachyon-tachyon amplitude multiplied by the four-point tachyon amplitude. 

The conditions that we needed, $q\cdot p_4 = \zeta \cdot p_4 = 0$, can equivalently be written as $q\cdot p_4 = \lam \cdot p_4 = 0$, because $\zeta_{\mu} = \lam_{\mu} - (\lam \cdot p_1) q_{\mu}$. So the momentum of the tachyon is orthogonal to the momenta  of the DDF photons that created the excited state (which are proportional to $q$), and orthogonal to their polarization. Since $q$ is null this means that in the high energy limit, in which tachyons are approximately massless, $p_4$ is approximately some multiple of $q$. Also, the condition $p_4 \cdot\lam =0$ follows automatically, since $q \cdot \lam =0$.

\subsubsection*{General polarization}
Finally, we look at the case considered in Sec.~\ref{sec42} of an excited string in a general state, 
\be 
\lam_k \cdot A_{-m_k}\cdots \lam_2 \cdot A_{-m_2}\,\,  \lam_1\cdot A_{-m_1} |0\ra~, \ \ \ \ \ N= \sum_i  m_i~.
\ee
The amplitude is given by,
\be 
\mA = \frac{1}{\text{vol} (SL_2)}\int dz\,  \exp\(\mL\)~,
\ee
where $\mL$ was given by (\ref{429}) which, after taking $n = 3$ (three tachyons), and $a=2, b=3$, becomes, 
\bml  
\mL =   p_2 \cdot p_4 \log R + p_3 \cdot p_4 \log(R-1) - \log (z_{14}^2 z_{23}^2)  \\
+\log\(\sum_{\rho=1}^{\lfloor k/2 \rfloor} \sum_{\pi}\, \prod_{l=1}^{\rho} ( \zeta_{\pi(2l-1)} \cdot \zeta_{\pi(2 l)} )\, \mbS_{m_{\pi(2l-1)}, m_{\pi(2l)}}' \prod_{q=2\rho+1}^k(-\Sigma'_{m_{\pi{q}}}(\zeta_q)) \)~,
\end{multline}
where $ R \equiv R_4 = \frac{z_{13} z_{24}}{z_{14}z_{23}}$, and $\Sigma_m'(\zeta)$ was given by (\ref{Snp}), and $\mbS_{m, n}'$  in (\ref{432}) becomes,
\be \label{642}
\mathbb{S}_{m, n}' =\sum_{r =1}^n r\ \oint \frac{dw}{2\pi i} \frac{w^{-m-r-1}}{(1-w)^{m p_3 \cdot q}(1 - w R)^{m p_4 \cdot q}}\oint \frac{dv}{2\pi i} \frac{v^{-n+r-1}}{(1-v)^{n p_3 \cdot q}(1 - v R)^{n p_4 \cdot q}}~.
\ee
If we wish, we can evaluate the integrals over $w$ and $v$; each will give a sum, as in (\ref{Snp}). We may instead do the sum over $r$ to get,
\be \
\mathbb{S}_{m, n}' = \oint \frac{dw}{2\pi i} \oint \frac{dv}{2\pi i} \frac{1}{(v - w)^2 w^m v^n}\frac{1}{(1-w)^{m p_3 \cdot q}(1 - w R)^{m p_4 \cdot q}}\frac{1}{(1-v)^{n p_3 \cdot q}(1 - v R)^{n p_4 \cdot q}}~.
\ee
Notice that if we take $p_4 \cdot q =0$ in (\ref{642}), then $S_{m, n}'$ becomes identical to the $S_{m, n}'$ found in the three-point amplitude (\ref{428}).

\section{Discussion} \label{sec7}

In this paper we looked at scattering amplitudes involving highly excited strings. A general excited string state was formed by the DDF construction, of repeatedly scattering photons off of an initial tachyon. 
We studied in detail the amplitude for the decay of a highly excited string into two tachyons. The result is compact and rich, and summarized in the beginning of Sec.~\ref{sec6}. We found that the amplitude involving generic excited strings (in contrast to the more commonly studied special excited states, such as the leading Regge trajectory) is highly sensitive to the precise excited string state (the microstate) and erratic as a function of the relative angle between the outgoing tachyons and the photons used to create the state. 
We interpret this as chaos in the scattering amplitude. 

Although our computation was for one excited string decaying into two tachyons, it seems fairly clear that the effect is general: a scattering amplitude involving any number of generic highly excited strings  should be chaotic. The next simplest case to consider is a highly excited string decaying into another highly excited string by emitting a low energy tachyon or photon. This is more challenging to calculate than the case studied here, 
and we hope to report on it soon.

An important question is what is the dominant decay channel of a highly excited string. If a highly excited string is to behave like a conventional thermodynamic system, then the dominant decay channel should consist of the string gradually and sequentially emitting low energy tachyons/photons. One might hope that this amplitude can be found by gluing together the three-string amplitudes of an excited string decaying into another excited string and a photon. This process would be an analog of a black hole decaying by gradual emission of Hawking radiation.

String theory is known to have a number of unique and extraordinary properties. The remarkable simplicity and complexity of scattering amplitudes of highly excited strings, exhibiting  chaos at weak coupling,  is yet one more.

\section*{Acknowledgements}
The work of DG is supported  by NSF grant 1125915. The work of VR is supported by NSF grant PHY-1911298 and the Sivian Fund of the IAS.

\appendix

\section{Covariant vertex operators} \label{Virasoro}
In the main body of the text we found the vertex operators for excited string states through the DDF construction. 
In this appendix we review the more familiar construction of covariant vertex operators, as found by imposing the Virasoro constraints. This works well for light string states, but becomes increasingly unwieldy for heavy string states. 

Recall that the open string field $X^{\mu}$ can be expanded in terms of modes, 
\be
\d X^{\mu}(z) = - i \sum_n \frac{\al_n^{\mu} }{z^{n +1}}~, \ \  \ \ \ \ \ \al_{-n}^{\mu}  = \frac{i   }{(n-1)!}\d^n X^{\mu}(0)~.
\ee
An excited string state is obtained by acting with the creation operators $\al_{-n}^{\mu}$ on the vacuum. The relation above allows us to translate between states and their corresponding vertex operators, 
\be
\al_{-n_1}^{\mu_1} \al_{-n_2}^{\mu_2} \cdots \al_{-n_k}^{\mu_k} |0; p\ra\, \,   \leftrightarrow\, \,  \, \d^{n_1} \!X^{\mu_1}\, \d^{n_2}\! X^{\nu_2} \cdots \d^{n_k}\! X^{\mu_k}\, \,e^{i p \cdot X}~.
\ee
Only particular superpositions of these are allowed states.
In order for a state to be physical, it must  both be annihilated by the Virasoro generators, 
\be
 L_m |\phi\ra = 0~, \ \ m>0~, \ \ \ \ \ \ \ \ L_m = \frac{1}{2} \sum_{n=-\infty}^{\infty} \al_{m- n} \cdot \al_n~,
\ee
and it must be on-shell, $L_0 |\phi\ra = |\phi\ra$, which corresponds to $M^2 = - 2 + 2 \sum_{n=1}^{\infty} \al_{-n} \cdot \al_n$. 
Note that $\al_0^{\mu} = p^{\mu}$. 
We will show explicitly how this works for states at levels one, two, and three.

\subsection*{$N=1$}
At level $N=1$ the states are massless and take the form, 
\be \label{532}
\zeta \cdot \al_{-1} |0; p\ra~ \ \ \ \ \ \leftrightarrow \ \ \ \  \zeta \cdot \d X\, e^{i p \cdot X}~.
\ee
Acting with $L_1$ gives, 
\be \label{533}
0 = L_1\, \zeta \cdot \al_{-1} |0; p\ra  = \zeta \cdot p\, |0; p\ra ~,
\ee
where we used that the relevant part of $L_1$ is $L_1 = \al_0 \cdot \al_1 + \ldots$, and that the modes satisfy the commutation relations $\[\al_m^{\mu}, \al_n^{\nu}\] = m\, \eta^{\mu \nu} \delta_{m n}$. 
We get the constraint that $\zeta \cdot p =0$. A second constraint is that we may add any multiple of $p$ to $\zeta$, while leaving the amplitude involving the vertex operator unaffected. To see this, note that if we take $\zeta = p$, then the vertex operator is a total derivative of the tachyon vertex operator
\be
p \cdot \d X \, e^{i p \cdot X}= -i \d\, e^{i p \cdot X}~.
\ee
Thus we have two constraints, and correspondingly a basis of  $D{-}2$ states at level-one in dimension $D$.  

Explicitly, suppose we are in three dimensions and take the momentum to be $p =(1,0,1)$. Letting the polarization have components $\zeta_{\mu} = (\zeta_0, \zeta_1, \zeta_2)$, the constraint $p \cdot \zeta = 0$ gives $\zeta_2 = \zeta_0$. Being able to add any multiple of $p$ to $\zeta$ lets us further set $\zeta_0 = 0$. Thus, the polarization is proportional to $(0,1,0)$.

\subsection*{$N=2$}
At  level $N=2$ the states have mass $M^2 = 2$ and take the form, 
\be
\( \zeta \cdot \al_{-2} - \zeta_{\mu \nu} \al_{-1}^{\mu} \al_{-1}^{\nu}\) |0; p\ra, \ \ \ \ \ \leftrightarrow \ \ \ \  \(i \zeta_{\mu} \d^2 X^{\mu} + \zeta_{\mu \nu} \d X^{\mu} \d X^{\nu}\) e^{i p \cdot X}~.
\ee
Requiring that the state be annihilated under the action of  $L_1$ and $L_2$ gives, respectively, 
\be \label{B9}
\zeta_{\mu} - p^{\nu} \zeta_{\mu \nu} = 0~, \ \ \ \ \eta^{\mu \nu} \zeta_{\mu \nu} -2 p \cdot \zeta = 0~,
\ee
where we used that the relevant part of $L_1$ is $L_1 =   \al_0{ \cdot} \al_1 + \al_{-1} {\cdot} \al_{2}+ \ldots$ and that the relevant part of $L_2$ is $L_2= \al_2 {\cdot} \al_0 + \frac{1}{2} \al_1 {\cdot} \al_1+ \ldots$. Finally, the norm of the state is $
2\( \zeta_{\mu} \zeta^{\mu\, *}+ \zeta_{\mu \nu} \zeta^{\mu\, \nu *}\)$. Since $\zeta_{\mu \nu}$ is a symmetric tensor, this gives $D(D+1)/2 $ components. The vector $\zeta_{\mu}$ is fixed via (\ref{B9}) in terms of $\zeta_{\mu \nu}$, and the second equation in (\ref{B9}) give one constraint. Thus we have a total of $\frac{(D-2)(D+1)}{2}$ independent states, which is the dimension of the symmetric traceless representation of $SO(D-1)$.

\subsection*{$N=3$}
At level $N=3$  the state  takes the form, 
\be \nn
\( 2 \zeta \cdot \al_{-3} - t_{\mu \nu} \al_{-2}^{\mu} \al_{-1}^{\nu} - \zeta_{\mu \nu \rho}\al_{-1}^{\mu} \al_{-1}^{\nu} \al_{-1}^{\rho}\) |0; p\ra \leftrightarrow \(i \zeta_{\mu} \d^3 X^{\mu}  + t_{\mu \nu} \d^2 X^{\mu} \d X^{\nu} +i  \zeta_{\mu \nu \rho} \d X^{\mu} \d X^{\nu} \d X^{\rho} \) e^{i p \cdot X}~.
\ee
Requiring that the state be annihilated by $L_1$, $L_2$ and $L_3$ gives the constraints, respectively,  
\bea
6 \zeta_{\mu} - t_{\mu \nu} p^{\nu} &=& 0~, \ \ \ \ \  \nn 
 t_{\mu \nu} + t_{\nu \mu}+ 3 \zeta_{\mu \nu \rho} p^{\rho} = 0 \\ \nn
 6 \zeta_{\mu} -2 t_{\nu \mu} p^{\nu} - 3 \eta^{\nu \rho} \zeta_{\mu \nu \rho} &=& 0\\ 
 3 \zeta \cdot p - t_{\mu \nu} \eta^{\mu \nu}&=&0~,
 \eea
 where in the first line, corresponding to annihilation by $L_1$, we have two constraints, coming from the requirement that the coefficients of both $\al_{-2}^{\mu} |0\ra$ and $\al_{-1}^{\mu} \al_{-1}^{\nu}|0\ra$ vanish. 
 
 Proceeding to higher $N$ in this fashion is tedious.  For some discussion see e.g. \cite{Weinberg:1985tv, Manes:1988gz,  Nergiz:1993gw}.

\section{Normalization of DDF operators} \label{appB}
   In order to gain experience working with the DDF vertex operators, it is useful to check that they are correctly normalized. 
 \subsubsection*{Single creation operators}
Since  the DDF operators $A_{-m}$ obey the commutation relations of creation operators, (\ref{23}), we have that the overlap of two states $\lam_1 \cdot A_{-m_1}|0\ra$ and $\lam_2 \cdot A_{-m_2}|0\ra$ is given by, 
  \be \label{Norme}
\la 0| \lam_1^*\cdot A_{m_1}  \lam_2 \cdot A_{-m_2}|0\ra = \lam_1^*\cdot \lam_2\, m_1 \delta_{m_1, m_2}~.
\ee
We would like to get the same thing by computing the two point function of the corresponding DDF vertex operators given in (\ref{336}),
\be \label{338}
\la V_1^* V_2\ra \equiv \la (:  \lam_1 \cdot A_{-m_1} \, e^{i \tp \cdot X(z_1)}\!:)^* \, :  \lam_2 \cdot A_{-m_2} \, e^{i \tp \cdot X(z_2)}\!:\ra~. 
\ee
We use  (\ref{336}) for the vertex operator (note that taking  the complex conjugate is equivalent to (\ref{336}) with $q\rightarrow - q$ and $p\rightarrow -p$). We get that (\ref{338}) is equal to,
\be
\la V_1^* V_2\ra= \zeta_1^*\cdot \zeta_2 \sum_{a_1, a_2} \frac{\la \d^{a_1} X(z_1) \d^{a_2} X(z_2)\ra}{(a_1-1)!(a_2-1)!}\,  
S_{m_1- a_1}\( - \frac{m_1}{r z_{21}^r}\)  S_{m_2- a_2} \( - \frac{m_2}{r z_{12}^r}\)  \la e^{-i p\cdot X(z_1)} e^{ i p\cdot X(z_2)}\ra~.
\ee
We perform the Wick contractions using (\ref{prop}), as well as (\ref{TTope}) which gives, $\la e^{-i p\cdot X(z_1)} e^{ i p\cdot X(z_2)}\ra = z_{12}^{-p^2} $, and we use for the Schur polynomial, 
 \be \label{B4}
 S_k\(- \frac{m_1}{r z_{21}^r}\) = \oint \frac{d w}{2\pi i} \frac{1}{w^{k+1}} \(1 - \frac{w}{z_{21}}\)^{m_1} = \binom{m_1}{k}\frac{(-1)^k}{z_{21}^{k}}~,
 \ee
 to get
 \be
  \la V_1^* V_2\ra =\zeta_1^*\cdot \zeta_2 \frac{1}{z_{12}^{m_1 + m_2 + p^2}}\sum_{a_1, a_2}\frac{(-1)^{m_2-a_1- a_2 } (a_1+a_2-1)!}{(a_1-1)!(a_2-1)!}  \binom{m_1}{a_1}\binom{m_2}{a_2} ~.
  \ee
  Evaluating the sum gives,~\footnote{For  $a_1\leq m_1$, one can evaluate the sum by using,  
  \be \nn
  \sum_{a_2} \frac{ (-1)^{m_2-a_2} (a_1{+}a_2{-}1)!}{(a_2-1)!} \binom{m_2}{a_2}   = \( - \d_z\)^{a_1} \sum_{a_2}  \binom{m_2}{a_2} \frac{(-1)^{m_2-a_2}}{z^{a_2}}\Big|_{z=1} 
  = (-1)^{m_2+a_1} \d_z^{a_1}\(\frac{z{-}1}{z}\)^{m_2}\Big|_{z=1} = m_2! \, \delta_{m_2, a_1}
  \ee}
  \be \label{343}
    \la V_1^* V_2\ra  = \zeta_1^*\cdot \zeta_2 \, m_1 \, \frac{ (-1)^{m_1}}{z_{12}^2} \, \delta_{m_1, m_2}~,
    \ee
    where we used that $p^2 = -2(m_1 - 1)$. 
    Since $\zeta_1^*\cdot \zeta_2 = \lam_1^* \cdot \lam_2$, this matches what we expected (\ref{Norme}). 

\subsubsection*{Two creation operators}

Let us check the normalization of a state created with two creation operators, $\lam_2 \cdot A_{-m_2} \lam_1 \cdot A_{-m_1}|0\ra$. From the commutation relations of creation operators  (\ref{23}), we have, 
\be \label{B8}
\la 0|\lam_1^* \cdot A_{m_1} \lam_2^* \cdot A_{m_2}\ \lam_2 \cdot A_{-m_2} \lam_1 \cdot A_{-m_1}|0\ra = |\lam_1|^2 |\lam_2^2| m_1 m_2 + |\lam_1 \cdot \lam_2^*|^2 m_1 \delta_{m_1, m_2}~.
\ee
Now let us get the same result by computing the two-point function of the vertex operator (\ref{339})  at $z_2$ and its complex conjugate at $z_1$, 
\be  \label{351}
\la  \[\( \zeta_2\! \cdot\! P_{m_2}\, \zeta_1 \!\cdot\! P_{m_1} + \zeta_1 \!\cdot\! \zeta_2\,  \, \mbS_{m_1, m_2} \)   e^{i p \cdot X(z_1)}\]^*  \( \zeta_2\! \cdot\! P_{m_2}\, \zeta_1 \!\cdot\! P_{m_1} + \zeta_1 \!\cdot\! \zeta_2\,  \, \mbS_{m_1, m_2} \)   e^{i p \cdot X(z_2)}\ra~.
\ee
We note that since $\mbS_{m_1, m_2}$ contains only terms of the form $q\cdot \d^r X(z)$, and since $q^2 = 0$  and $q\cdot \zeta_i =0$, we can only contract the $\mbS_{m_1, m_2}$ with the exponential. Thus the $ \mbS_{m_1, m_2}(z_1)$ that we have on the left becomes, 
\be
 \mbS_{m_1, m_2}(z_1) = 
\sum_{m=1}^{m_1}m\,  S_{m_1- m}\(- \frac{m_1}{r z_{21}^r}\)\, S_{m_2+m}\( - \frac{m_2}{r z_{21}^r}\)~.
\ee
Using the contour integral representation of the Schur polynomial (\ref{B4})   we see that $S_{m_2+m}\( - \frac{m_2}{r z_{21}^r}\)=0$ for $m\geq 1$. Hence (\ref{351}) becomes, 
\be  
\la  \[\( \zeta_2\! \cdot\! P_{m_2}\, \zeta_1 \!\cdot\! P_{m_1} \)   e^{i p \cdot X(z_1)}\]^*  \( \zeta_2\! \cdot\! P_{m_2}\, \zeta_1 \!\cdot\! P_{m_1}  \)   e^{i p \cdot X(z_2)}\ra~.
\ee
We may now contract $(\zeta_2\cdot\! P_{m_2})^*(z_1)$ with either $\zeta_2\! \cdot\! P_{m_2}(z_2)$ or with $\zeta_1 \!\cdot\! P_{m_1} (z_1)$. Using (\ref{343}) we thus find, 
\be
\(|\zeta_1|^2 |\zeta_2^2| m_1  m_2  + |\zeta_1 \cdot \zeta_2^*|^2 m_1^2 \delta_{m_1, m_2}  \)\frac{(-1)^{m_1+m_2}}{z_{12}^2}~.
\ee
This matches (\ref{B8}).\\[-10pt]

The generalization to a state created with multiple creation operators is clear.

\section{A generalization of the scattering equations} \label{appC}

The bosonic closed (open)  $n$ string scattering amplitudes, to lowest order in $g_s$ , are given by saddles in the integration over the positions ($z_i$) of the vertex operators  on the sphere (upper half plane), when we take all kinematic invariants to be large or, equivalently, $\al' \to \infty$
\be  \label{ampl}
\mA = \frac{1}{\text{vol} (SL_2)}\int \prod_{i=1}^n dz_i \,  \exp \mL(z)~,\ \ \ \ \  ~ \mL(z) = 2 \al'\sum_{0 \leq i< j \leq n}^np_i\cdot p_j \log z_{ij}~.
\ee
The $z_i's$ are determined, up to SL$_2$ transformations, by the  ``scattering equations'':
\be \label{C2}
\sum_{j\neq i} \frac {p_i\cdot p_j}{z_i-z_j}=0, ~ {\rm for} ~i= 1\ldots n ~ 
\ee
which determine the saddles of (\ref{ampl}).
Note that in this limit we can take the tachyons to be massless, as $p_i^2= 1/\al'$.
These equations are invariant under  SL$_2$ transformations of the $z_i$ and thus yield $n-3$ equations. They have $(n-3)!$ solutions.

Although these equations determine high energy string scattering they are also key ingredients in the CHY formulas \cite{Cachazo:2013gna,Cachazo:2013hca} for massless particle scattering in field theory, and have been the subject of much investigation recently. The scattering of excited strings will be governed by the same saddle as long as one keeps the mass of the excited strings  small compared to the energies and momentum transfers, leading to the many relations between scattering amplitudes discussed in \cite{GrossMende1, GrossMende2, Gross:1988ue, Gross:1989ge}. However, if we consider strings with masses of order the energies,  the saddles will be shifted. This will lead to interesting generalizations of the scattering equations. In this section we will exhibit some of the simplest generalizations. 

Consider  the simple case discussed in Sec.~\ref{sec41}  of the scattering  of one heavy string  in which all the DDF photons forming the the excited state have the same polarization $\lam$, orthogonal to its momentum $p_h$ (in this section we let $p_h$ denote the momentum of the heavy string) and with $\lam^2=0$. The analog of  (\ref{ampl}) was given in equations (\ref{48}), (\ref{930}) and (\ref{931}). Specializing further, consider the case where the excited string has  $n_1=N$, $ n_{i >1}=0 $, namely the exited string lies on the leading Regge trajectory, has momentum $p_h$ (where $p_h^2=-2N$) and polarization $\xi$, ($ \xi \cdot p_h =0$). In this case, for $n$  tachyons  (which we can take to be massless)  of momenta $p_i$, the action is given by,
\be \label{48prime}
\mL(z) = \sum_{1\leq i< j\leq n} p_i \cdot p_j \log z_{i j} + \sum_{i=1}^n p_i \cdot p_h\log( z_i-z_h) +N  \log (\Sigma_1)~,\ \ \ \ 
~\Sigma_1 =   \sum_{j=1}^n \frac{p_j \cdot \zeta}{z_j-z_h}  ~.
\ee
The saddles are given by $\frac{\partial \mL(z_i, z_h)}{\partial z_i}=\frac{\partial \mL(z_i, z_h)}{\partial z_h}=0$,
\bea
\sum_{j\neq i} \frac {p_i\cdot p_j}{z_i-z_j}+ \frac {p_i\cdot p_h}{z_i-z_h}-\frac{N}{\Sigma_1}\frac{p_i \cdot \zeta}{(z_i-z_h)^2}&=&0, ~\ \ \ \ \ {\rm for} ~i= 1, \ldots, n ~ ,\\
 \sum_{j=1}^n \frac{p_j \cdot p_h}{ z_h-z_j} + \frac{N}{\Sigma_1} \sum_{j=1}^n \frac{p_j \cdot \zeta}{(z_h-z_j)^2} &=&0~.
\eea
These equations for $z_i$ and $z_h$ are invariant under M\"obius transformations, $\{z\to \frac{az+b}{cz+d},~ ad-cb=1\}$. They are equivalent to requiring that the vector field $E^\mu(z)$
\be
 E^\mu(z) \equiv \sum_{j=1}^n \frac{p^\mu_j}{z-z_j} + \frac{p^\mu_h}{z-z_h} + \frac{N\xi^\mu}{\Sigma_1} \frac{1}{(z-z_h)^2} 
\ee
is null, $ E^\mu(z) E_\mu(z)=0.$  $E^\mu(z)$ is the electric field created by charges $(p^\mu_i  ,p^\mu_h)$ at positions $(z_i , z_h)$, as well a dipole
$\frac{N\xi^\mu}{\Sigma_1} $ at $z_h$.

Similar generalizations of the scattering equations can easily be derived by considering the critical points of the Lagrangians given in 
(\ref{930}), as well as (\ref {427}).

\section{A typical string state} \label{sec5}

In the main body of the paper we  discussed scattering amplitudes of excited string states. However, for a string of given mass, the number of different states grows exponentially with the mass. Clearly, we need to have a better sense of which states in particular we would like to study. We are interested in the generic state. In $D$ space-time dimensions, there are $D-2$ independent polarization vectors.  A state is created by acting with the DDF creation operators, $A_{-m}^{\mu}$, which excite mode $m$ in direction $\mu$. We let $n_m$ denote the number of times mode $m$ is excited. The modes are excited in direction (polarization) $\lam_k^m$, where $k=1, 2,\ldots, n_k$. A state at level $N$, having mass $M^2 = 2 (N-1)$,  therefore takes the form, 
\be  
\prod_{k=1}^{n_1}(\lam^1_k \cdot A_{-1} )\prod_{k=1}^{n_2} (\lam^2_k\cdot A_{-2}) \cdots \prod_{k=1}^{n_r}(\lam^r_k \cdot A_{-r}) |0\ra~, \ \ \ \ \ N = \sum_{m=1}^r m n_m~.
\ee
The state is specified by the polarizations $\{\lam_m^k\}$ and the occupation numbers $\{n_m\}$. 

The question of what a typical state looks like is more precisely: for a given $N$, what is the typical value of the occupation number $n_m$ of mode $m$? In this section we will show that, in the large $N$ limit,  the typical (or equivalently, average) occupation number takes the form of a Bose-Einstein distribution $\la n_m\ra = \frac{1}{e^{m/T} - 1}$, with a temperature $ T= \frac{\sqrt{6N(D-2)}}{\pi}$.

 \subsection*{A typical Young diagram}
 \begin{figure}
 \centering
 \includegraphics[width=1.8in]{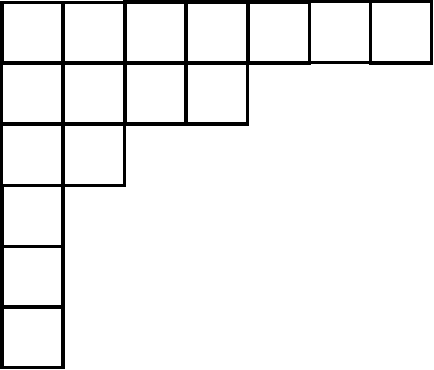}
 \caption{A Young diagram corresponding to the state in (\ref{egstate}).} \label{Young1}
 \end{figure}
Let us reformulate the problem of finding the expected occupation numbers $n_m$ as one of finding the generic partition. We associate every state with a partition, and take $D=3$ so that the polarization is unique. An example of a state and the corresponding partition is, 
\be \label{egstate}
(\lam\cdot A_{-7})(\lam\cdot A_{-4})(\lam\cdot A_{-2})(\lam\cdot A_{-1})(\lam \cdot A_{-1})(\lam \cdot A_{-1})  |0\ra~, \ \ \ \ \ \ \{7,4,2,1,1,1\}~,
 \ee
which is just one of the $231$ partitions of $N=16$. A general state/partition is, 
 \bea  \nn
 (\lam \cdot A_{- l_1} )(\lam \cdot A_{-l_2}) \cdots (\lam \cdot A_{-l_r}) |0\ra~, \ \ \ \ \ && \{l_1, l_2, \ldots, l_r\} \\
&& l_1 \geq l_2 \geq \cdots l_r >0~, \ \ \ \ \sum_i l_i = N~.
 \eea
 Notice that we have ordered the $l_i$ in nonincreasing order, in order to not count the same state twice. We may associate each partition of $N$ with a unique Young diagram. For each $A_{-l}$ we draw a row of $l$ boxes, and the rows are arranged with nonincreasing $l$ as one moves down. The Young diagram corresponding to the state (\ref{egstate}) is shown in Fig.~\ref{Young1}. 
 
\begin{figure}
 \centering
 \includegraphics[width=2in]{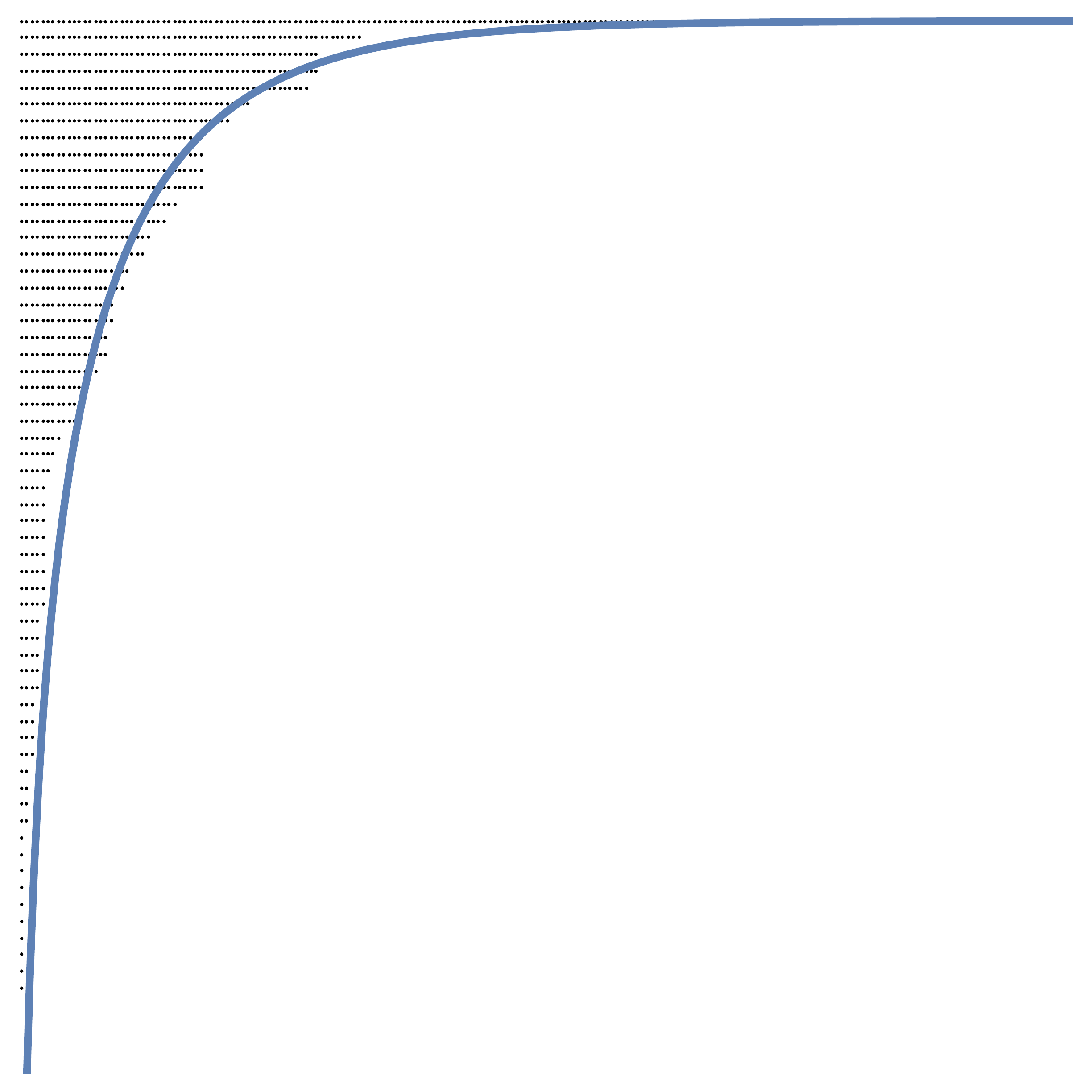}
 \caption{A Young diagram corresponding to a random partition of $N=1000$. The solid line is $-y(x)$ in (\ref{55}).} \label{RandomYoung}
 \end{figure}

Our question of what a generic  large $N$ state looks like has become a question of what a generic Young diagram looks like (assuming that any individual Young diagram is equally likely). The answer to this is well known, see e.g. \cite{Vershik2, Vershik}: the number of elements of the partition that have a $l\geq x$ is given by $y(x)$, 
\be \label{55}
y(x) = -\frac{1}{\beta} \log(1 - e^{- \beta x})~~, \ \ \ \ \ \beta = \frac{\pi}{\sqrt{6N}}~.
\ee
The shape of the Young diagram is given by the curve $y(x)$, as is sketched in Fig.~\ref{RandomYoung}. One can check that $y(x)$ is properly  normalized (at large N), $N = \int_1^{N} d x\, y(x)$.  

What we  are really interested in is the number of times mode $m$ is  excited - labelled by $n_m$ in (\ref{51}) -  for a typical state. Clearly, $y(x)$ is the number of creation operators with an $m\geq x$, 
\be
y(x) = \sum_{m=x}^N n_m~.
\ee
To express $n_m$ in terms of $y(x)$, let us use the notation $n_m \equiv n(m)$ and think of $m$ as a continuous variable. Then, 
\be
y(x) = \int_x^{N} d m\, n(m)~, \ \ \ \ \Rightarrow \, \, n(x) = - y'(x) = \frac{1}{e^{ \frac{\pi}{\sqrt{6 N}} x}-1}~.
\ee
This is just the Bose-Einstein distribution, with a temperature, 
\be 
T= \frac{\sqrt{ 6N}}{\pi}~.
\ee

Actually, we could have derived this result from first principles, without relying on knowledge of the mathematical result (\ref{55}) for $y(x)$. 
The argument is a straightforward extension of the derivation in  e.g. \cite{GSW} for the density of states, $\Omega(N)$ (the number of string states at level $N$).
 Recall that $\Omega(N)$ is, and we now work in arbitrary dimension $D$. The polarizations are picked from a basis of $D-2$ independent polarization vectors, and   the number of choices of  $\lam^r_1, \lam^r_2,...,\lam^r_{n_r}$ is  given by $ C_r \equiv {D-2 + n_r -1 \choose n_r} =(-1)^{n_r} {2-D\choose n_r}$. 
\bea\nn
\Omega(N) = \sum_{\{n_k\}} \delta(N - \sum_k k n_k) \prod_kC_k 
&=&\sum_{\{n_k\}} \int d\beta e^{\beta N}  \prod_k e^{-\beta k n_k}(-1)^{n_k} {2{-}D\choose n_k}  \\ 
&=& \int d \beta\,  e^{\beta N} \prod_k \(1 - e^{- \beta k}\)^{2-D} ~,
\eea
where we represented the delta function as an integral and performed the sum over the occupation number $n_k$. 
Next, we  rewrite the integrand as,
\be
 \prod_k \(1 - e^{- \beta k}\)^{2-D} = \exp\( - \sum_{k=1}^{\infty} (D-2)\log(1 - e^{-\beta k})\) = \exp\( (D-2) \sum_{n, k= 1}^{\infty} \frac{e^{-\beta n k}}{n} \)~,
\ee
where in the last step we expanded the $\log$. 
We are going to look for the saddle in the large $N$ limit. Thus, we should take ${\beta}$ to be close to $0$.   Expanding the term being summed, 
\be
 \exp\( \sum_{n=1}^{\infty}\frac{1}{n} \frac{D-2}{e^{\beta n} - 1}\) \sim  \exp\( \frac{D-2}{\beta} \sum_{n=1}^{\infty}\frac{1}{n^2} \) = \exp\(\frac{\pi^2 (D-2)}{6\beta}\)~,
 \ee
and so we find the number of states at level $N\gg 1$ is, 
 \be \label{513}
 \Omega(N) \sim \int d\beta \exp\( \beta N + \frac{\pi^2 (D-2)}{6\beta}\)~   \sim \exp\(2 \pi \sqrt{\frac{N(D-2)}{6}} \)~,
 \ee
 where  the saddle is at $ \beta = {\pi} \sqrt \frac{D-2}{6N} $. The result is of course the Hardy-Ramanujan formula for the asymptotic number of partitions of $N$.

Our interest is in $n_k$ for a typical state, or equivalently, the expectation value of $n_k$ averaged over all states at level $N$, 
 \be
 \la n_k\ra = \frac{\sum_{n_m} n_k\,  \delta( \sum m n_m - N)\prod_mC_m}{\Omega(N)}~.
 \ee
 We write the numerator in the same way as we did for $\Omega(N)$,  
 \bea\nn
 \la n_k\ra &=& \frac{1}{\Omega(N)} \int d\beta \sum_{\{n_m\}} n_k \exp\(  \beta( N- \sum_m m n_m )\) \prod_m C_m \\
 &=& \frac{1}{\Omega(N)}\int d\beta\, e^{\beta N} \frac{1}{e^{\beta k} - 1} \prod_m  \( \frac{1}{1 - e^{-\beta m}} \)^{D-2}~.
 \eea
 The saddle will clearly be at the same $\beta$ as it was for $\Omega(N)$, and so, 
 \be \label{516}
  \la n_k\ra = \frac{1}{e^{\beta k} - 1} ~, \ \ \ \ \   \beta = {\pi} \sqrt \frac{D-2}{6N} ~.
 \ee

Actually, this Bose-Einstein distribution for $\la n_k\ra$ was guaranteed, once we knew the density of states. We can think of our setup as a statistical mechanical system consisting of harmonic oscillators, one for each integer frequency. Working in the microcanonical ensemble, and thinking of $N$ as the energy, from $\Omega(N)$ in (\ref{513}) we identify the entropy, and correspondingly the temperature, as, 
\be
S(N) =  \pi \sqrt{\frac{2N(D-2)}{3}}~, \ \ \ \ \ \ \ \frac{1}{T} \equiv \frac{\d S}{\d N} =  {\pi} \sqrt \frac{D-2}{6N}~,
\ee
implying the Bose-Einstein distribution for $n_k$  found in (\ref{516}). Note that this ``temperature'' is not the same as that of a string in the canonical ensemble, as the mass of the string is $M=\sqrt{\frac{N}{\al '}}$, which yields the Hagedorn temperature
\be \frac{1}{T_H} \equiv \frac{\d S}{\d M}  ={\pi} \sqrt \frac{2 \al ' (D-2)}{3} ~.
\ee
\\[-10pt]

Let us now look at some properties of $y(m)$.  We see that the spin (the total number of creation operators) is,
\be \label{D17}
J \approx y(1)\approx T \log T + \frac{1}{2}~.
\ee
More generally,  
for an $m$ of order $1$,
$
y(m) = - T \log(1 - e^{-m/T}) \approx T \log\frac{T}{m}$ for $N\gg 1$. 
However, as one can see from the plot of $y(m)$ in Fig.~\ref{RandomYoung}, most of the contribution to the area (the total ``energy'' $N$) comes from large $m$.  In particular, we will show that most of the contribution to $N$ comes from those creation operators with an $m$ of order $m\sim T\sim \sqrt{N}$. 
Let us look at the number of creation operators that have an $m$ greater than $N^{\gamma}$, for some power $0<\gamma<1$, 
\be
y(N^{\gamma}) = - T \log \( 1- e^{- \frac{N^{\gamma}}{T}}\)~.
\ee
We see that if $\gamma>1/2$, this is zero in the large $N$ limit. 
We may compute the expectation value of $m$, 
\be \label{518}
\la m\ra= \frac{1}{y(1)} \int_1^{\infty} dm\, m\, n(m) = \frac{T^2}{y(1)} \int_{1/T}^{\infty} dx \frac{x}{e^x - 1} =\frac{T}{\log T} \frac{\pi^2}{6} , \ \ \ \ \ N\gg 1~.
\ee
This result is consistent with there being of order $T$ creation operators with an $m$ of order $T$. 
More precisely, we may look at the creation operators with an $m$ in the range $\al T< m<\beta T$ (with $\al$ and $\beta$ constant in the large $N$ limit), and see the fraction of the total energy $N$ they make up,
\be
\int_{\al T}^{\beta T} dm\, y(m) = - T^2 \int_{\al}^{\beta} dx\, \log(1-e^{-x})~.
\ee
The right hand side is of order $N$. For instance, if we take $\al = .01$ and $\beta = 10$, then the integral gives $.965 N$. Thus, nearly all the energy is in this range. The number of creation operators in this range is, 
\be
y(\al T) - y(\beta T) = - T \log\( \frac{1- e^{-\al}}{1 - e^{-\beta}}\)~.
\ee
This is smaller than the total number of creation operators, $J$, by a factor of $\log T\sim \log N$. The reason is that there are many creation operators with small $m$, which don't contribute much to the energy. Indeed the number of $A_{-1}$'s is, 
\be
n(1) = \frac{1}{e^{1/T} - 1} \approx T~.
\ee
So there are of order $T$ creation operators with an $m$ of order one (and hence an energy contribution of order $T$), and of order $T$ creation operators with an $m$ of order $T$ (and hence an energy contribution of order $T^2 \sim N$). 

Finally, we note that the Bose-Einstein distribution gives an $\la n_k\ra$ that is of order-one for $k$ of order $T$. Clearly, for $k$ much larger than $T$, the expectation value $\la n_k\ra \ll 1$ is no longer a good indicator of the value of $n_k$ for a typical state. However, as we saw, most $k$ are parametrically less than $T$, by a factor of $\log T$  (see (\ref{518})), so for most $k$, $n_k$ is large. \\

So far we have discussed the generic state for fixed $N$. One could also ask about the states occurring at fixed $N$ and fixed $J$. The number of states as a function of both $N$ and $J$ is, 
\bea
\Omega(N,J)\!\! &=&\!\! \sum_{\{n_k\}} \delta(N - \sum_k k n_k)  \delta(J - \sum_k n_k)\prod_kC_k \\ \nn
&=&  \sum_{\{n_k\}} \int d \beta d\mu \, e^{\beta N+ \mu J}  \prod_k e^{-\(\beta k +\mu\) n_k}(-1)^{n_k} {2{-}D\choose n_k} 
=\! \int \!d \beta d\mu\,  e^{\beta N+ \mu J} \prod_k \! \(1 - e^{- \beta k-\mu}\)^{2-D}~.
\eea
The integrals over $\beta$ and $\mu$ run over the imaginary axis. We will deform the contour to pick up the saddle which lies on the real axis. 
Repeating the previous calculation,
\be
 \prod_k \(1 - e^{- \beta k-\mu}\)^{2-D} = \exp\( \sum_{n=1}^{\infty}\frac{e^{-\mu n}}{n} \frac{D-2}{e^{\beta n} - 1}\) \sim \exp\(\frac{ (D-2)}{\beta} {\rm Li}_2(e^{-\mu}) \)~,
 \ee
 where the ${\rm Li}_2$ is the polylogarithm: $ {\rm Li}_2(e^{-\mu}) =\sum_{n=1}^{\infty}\frac{e^{-\mu n}}{n^2} = \pi^2/6 +\mu \log (\mu) + O(\mu^2)$. The density of states at the saddles is given by, 
 \be  \label{D25}
  \Omega(N, J) \sim \exp\( \beta N + \mu J +\frac{ (D-2)}{\beta} {\rm Li}_2(e^{-\mu}) \)~,
 \ee
where for large $N$ and $J$, the saddle is given by:
\be  
N= \frac{D-2}{\beta^2}\,  {\rm Li}_2(e^{-\mu})~, \ \ \ \ \ \ \ J=-\frac{D-2}{\beta} \partial_\mu {\rm Li}_2(e^{-\mu})=-\frac{D-2}{\beta}\log(1-e^{-\mu})~.
\ee
To eliminate $\beta$ and $\mu$ in $\Omega(N, J)$, one can take the second equation and divide by the square root of the first equation, 
\be \label{D27}
\frac{J}{\sqrt{N}} = -\sqrt{D-2}\,  \frac{\log(1-e^{-\mu})}{\sqrt  {\rm Li_2(e^{-\mu})}}~,
\ee
and solve for  $\mu$ in terms of $\frac{J}{\sqrt{N}} $, and then $\beta$ is given by $\beta^2 N= (D-2)\rm Li_2(e^{-\mu})$. 
We now have the density of states for fixed $N$ and $J$. 

Finally, let us vary  $J$ and see for which $J$ there are the most states. It is easiest to express the exponent in (\ref{D25}) in terms of $\mu$. This gives, 
\be
\sqrt{N}\sqrt{D-2}\( 2 \sqrt{ {\rm Li}_2(e^{-\mu})} -\frac{\mu}{\sqrt{ {\rm Li}_2(e^{-\mu})}} \log(1 - e^{- \mu})\)~.
\ee
One might have expected this to have a local maximum at some value of $\mu$, but in fact this function is monotonic, increasing as $\mu$ goes to zero. There is a limit to how small $\mu$ can be however, since for it to be legitimate to  use the saddle point the terms in the exponent, such as $\mu J$, must be large. This limits $\mu$ to be greater than of order $1/\sqrt{N}$. Taking this $\mu$ and using (\ref{D27}) gives a $J$ of order $T \log T$, where $T= \frac{1}{\pi} \sqrt \frac{6N}{D-2}$. This is just what  we found earlier  in (\ref{D17}).


\bibliographystyle{utphys}

\end{document}